\documentclass[12pt]{iopart}
\usepackage{iopams}  
\usepackage{graphicx}
\usepackage{ulem} 
\usepackage{color} 

\newcommand{\BellAmp}{N_B}
\newcommand{\beq}{\begin{eqnarray}}
\newcommand{\eeq}{\end{eqnarray}}
\renewcommand{\vec}[1]{\mathbf{#1}}
\newcommand{\rel}{relativistic}
\newcommand{\syn}{synchrotron}
\newcount\listnorom
\newcommand\listromanDE{\global\advance \listnorom by 1
{\lowercase\expandafter{(\romannumeral\listnorom)}\ }}

\def\be{\begin{eqnarray}}
\def\ee{\end{eqnarray}}
 
\def\lsim{\;\raise0.3ex\hbox{$<$\kern-0.75em\raise-1.1ex\hbox{$\sim$}}\;}
\def\gsim{\;\raise0.3ex\hbox{$>$\kern-0.75em\raise-1.1ex\hbox{$\sim$}}\;}

\def\lsim{\;\raise0.3ex\hbox{$<$\kern-0.75em\raise-1.1ex\hbox{$\sim$}}\;}
\def\gsim{\;\raise0.3ex\hbox{$>$\kern-0.75em\raise-1.1ex\hbox{$\sim$}}\;}

\def\kms{\rm ~km~s^{-1}}

\def\etal{{ et al. }}

\def \kms {\rm ~km~s^{-1}}

\def\arcsec{\hbox{$^{\prime\prime}$}}

\def \chan{{\sl Chandra }}

\def \hcm {\hbox {\ifmmode $ atom cm$^{-2}\else atom cm$^{-2}$\fi}}

\def \arcsec {\hbox{$^{\prime\prime}$} }

\def\approxgt{\mathrel{\hbox{\rlap{\lower.55ex \hbox {$\sim$}}
        \kern-.3em \raise.4ex \hbox{$>$}}}}
\def\approxlt{\mathrel{\hbox{\rlap{\lower.55ex \hbox {$\sim$}}
        \kern-.3em \raise.4ex \hbox{$<$}}}}

\def \RXJ1713 {RXJ1713.72--3946 }

\begin{document}

\title{The microphysics of collisionless shock waves}
\author{A Marcowith$^1$, A Bret$^{2,3}$, A Bykov$^{4,5,6}$, M E Dieckman$^7$, L O'C Drury$^8$, B Lemb\`ege$^9$, M Lemoine$^{10}$, G Morlino$^{11,12}$, G Murphy$^{13}$, G Pelletier$^{14}$, I Plotnikov$^{14,15,16}$, B Reville$^{17}$, M Riquelme$^{18}$, L Sironi$^{19}$, A Stockem Novo$^{20}$}

\address{$^1$Laboratoire Univers et Particules de Montpellier CNRS/Universit\'e de Montpellier, Place E. Bataillon, 34095 Montpellier, France.}
\address{$^2$ETSI Industriales, Universidad de Castilla-La Mancha, 13071 Ciudad Real, Spain.}
\address{$^3$Instituto de Investigaciones EnergŽticas y Aplicaciones Industriales, Campus Universitario de Ciudad Real, 13071 Ciudad Real, Spain.}
\address{$^4$A.F. Ioffe Institute for Physics and Technology, 194021, St. Petersburg, Russia.}
\address{$^5$St. Petersburg State Politecnical University, Russia.} 
\address{$^6$International Space Science Institute, Bern, Switzerland.}
\address{$^7$Department of Science and Technology (ITN), Linkšpings University, Campus Norrkšping, SE-60174 Norrkšping, Sweden.}
\address{$^8$School of Cosmic Physics, Dublin Institute for Advanced Studies, 31 Fitzwilliam Place, Dublin 2, Ireland.}
\address{$^9$LATMOS - CNRS - UVSQ - IPSL,  11 Bd. d'Alembert , 78280, Guyancourt, France.}
\address{$^{10}$Institut d'Astrophysique de Paris, CNRS - UPMC, 98 bis boulevard Arago, F-75014 Paris,France.}
\address{$^{11}$INFN, Gran Sasso Science Institute, viale F. Crispi 7, 67100 LÕAquila, Italy.}
\address{$^{12}$INAF, Osservatorio Astrofisico di Arcetri, L.go E. Fermi, 5, 50125 Firenze, Italy.}
\address{$^{13}$Niels Bohr International Academy, Niels Bohr Institute, Blegdamsvej 17 Copenhagen 2100 Denmark.}
\address{$^{14}$Institut de Plan\'etologie et d'Astrophysique de Grenoble (IPAG) UMR 5274 F-38041 Grenoble, France.}
\address{$^{15}$Universit\'e de Toulouse, UPS-OMP, IRAP, Toulouse, France.}
\address{$^{16}$CNRS, Institut de Recherche en Astrophysique et Plan\'etologie 9 Av. colonel Roche, BP 44346, F-31028 Toulouse cedex 4, France.}
\address{$^{17}$Department of Physics and Astronomy Queen's University Belfast University Road Belfast BT7 1NN United Kingdom.}
\address{$^{18}$Department of Physics (FCFM) - University of Chile, Santiago, Chile.}
\address{$^{19}$Harvard-Smithsonian Center for Astrophysics, 60 Garden Street, Cambridge, MA 02138, USA.}
\address{$^{20}$ Institut fur Theoretische Physik, Lehrstuhl IV: Weltraum und Astrophysik, Ruhr Universitat, Bochum, Germany }
\ead{Alexandre.Marcowith@umontpellier.fr}

\begin{abstract}
Collisionless shocks, that is shocks mediated by electromagnetic processes, are customary in space physics and in astrophysics. They are to be found in a great variety of objects and environments: magnetospheric and heliospheric shocks, supernova remnants, pulsar winds and their nebul\ae, active galactic nuclei, gamma-ray bursts and clusters of galaxies shock waves. Collisionless shock microphysics enters at different stages of shock formation, shock dynamics and particle energization and/or acceleration. It turns out that the shock phenomenon is a multi-scale non-linear problem in time and space. It is complexified by the impact due to high-energy cosmic rays in astrophysical environments. This review adresses the physics of shock formation, shock dynamics and particle acceleration based on a close examination of available multi-wavelength or in-situ observations, analytical and numerical developments. A particular emphasize is made on the different instabilities triggered during the shock formation and in association with particle acceleration processes with regards to the properties of the background upstream medium. It appears that among the most important parameters the background magnetic field through the magnetization and its obliquity is the dominant one. The shock velocity that can reach relativistic speeds has also a strong impact over the development of the micro-instabilities and the fate of particle acceleration. Recent developments of laboratory shock experiments has started to bring some new insights in the physics of space plasma and astrophysical shock waves. A special section is dedicated to new laser plasma experiments probing shock physics. 
\end{abstract}

\maketitle

\section{Introduction}
\label{S:Intro}
Collisionless shocks are shock wave systems in which interactions between the different components are mediated by electromagnetic forces. These interactions involve a great variety of electromagnetic fluctuations that are triggered by a great variety of sources of free energy, all of them being at some stage connected with the existence of fast supersonic plasma flows that develop in interplanetary or interstellar environments. As they benefit from in-situ measurements, shocks occurring in planetary magnetospheres or in the heliosphere are a prime source of information about shock formation and dynamics on one hand and plasma instabilities and particle energization one the other hand. Recent developments in diagnostic tools in numerical simulations in laser plasma experiments have also brought fresh insights into the microphysics of shocks. Besides space and laser plasma physics, the shocks that develop in astrophysical environments do have their own peculiarities. Astrophysical shock waves are ubiquitous in the sources of high-energy (supra-thermal) particles such as supernova remnants (SNRs), galaxy clusters (GCs), active galactic nuclei (AGN) or gamma-ray bursts (GRBs). In the former two cases, the shock waves are non-relativistic (NR), while in the latter two cases, they can be ultra-relativistic (UR). Yet a common feature is the emission of non-thermal power-law spectra of high-energy radiation, which is usually observed as synchrotron or Inverse Compton (IC) photons emitted by in-situ accelerated electrons. Theoretical studies as well as observational findings have now begun to bring to light the complex relationship that exists between the accelerated (non-thermal) particles, the dynamics and structure of the shock wave, the surrounding magnetized turbulence and the efficiency of particle injection, indeed, the very nature of the acceleration process. \\
One way to power such radiation and produce high-energy particles is to generate a high level of magnetic fluctuations on both sides of the shock front to boost the efficiency of diffusive shock acceleration (DSA). The process was first analyzed in the late 70s in a series of papers \cite{Krymsky77, Axford77, Bell78a, Blandford78} (see \cite{Drury83} for a review). It involves repeated scattering of particles by resonant \footnote{The resonant interaction between the electromagnetic fluctuations and the particle gyro-motion involves the matching of the fluctuation wavelength and the particle Larmor radius.} magnetic fluctuations back and forth across the shock front. The particles gaining a constant relative amount of energy at each shock crossing cycle may reach very high energies if they are kept confined over long enough time-scales or if the acceleration time-scale is shorter than any loss or escape time-scale. The review will discuss for large parts the efficiency of DSA in the above astrophysical shock configurations. It became clear soon after these seminal papers that DSA is very efficient in producing high-energy tails which may contribute to up to 30\% of the kinetic gas energy. Particle acceleration then should modify the shock profile meaning that DSA is an intrinsic non-linear process. Multiple solutions for the particle distribution have been obtained solving a system of coupled fluid and kinetic equations \cite{BerezhkoEllison99}. Distinguishing between the different solutions can only be achieved through the shock microphysics and requires a deeper understanding of the origin and the role played by electromagnetic fluctuations in the shock environment. The nature of these, unfortunately, remains highly hypothetical. If the level of turbulent fluctuations is of the same order of as the mean magnetic field, as was first envisaged \cite{McKenzie82}, interestingly both observational and theoretical works have recently uncovered possibilities to produce strong magnetic field amplification (MFA) with magnetic field strengths several orders of magnitude above the standard interstellar values. In SNRs such a high level of fluctuations should help in confining particles over longer time-scales and hence producing energies that may even reach the so-called cosmic ray (CR) ankle beyond $10^{17}$ eV \cite{Ptuskin10}. The detection of X-ray filaments in several objects by the X-ray satellite {\it Chandra} (see \S \ref{S:NRobs}, \cite{Vink12} and references therein for a review) has further constrained the magnetic field at the very edge of the sources. Limited by the angular resolution of the instruments the derived magnetic fields are only lower limits and can reach up to several hundred of micro Gauss in the youngest objects such as Tycho or Kepler. The filaments in another young SNR RX J1713-3946.5 show also yearly variations \cite{Uchiyama07} that have received different explanations (see \S \ref{S:NRobs} and \ref{S:NRth}). Finally, peculiar features similar to stripes have been observed in the images of Tycho SNR that may be related to the physical processes at the origin of the filaments (see \cite{Eriksen11} and \S \ref{S:NRth}). The DSA model has been adapted with some success in the limit of low Mach number GC shocks \cite{Ensslin98, Miniati01}. The particle (electron and proton) injection in the shock process in low Mach number shocks is discussed in \S \ref{S:PICNR}. In GRBs only indirect hints of MFA have been provided notably by the analysis in \cite{LiWaxman06} of early X-ray afterglows even if this result does not preclude the possibility of the relativistic shock wave propagating into a magnetized wind (see also \S \ref{S:Robs} and \cite{LiZhao11} for a recent analysis of the data of gamma-ray satellite Fermi). Since early 2000 with the seminal work of \cite{Lucek00, Bell01} a lot of scenarios for MFA and particle acceleration to high energies have been proposed (see recent reviews by \cite{Schure12} and \cite{Bykov12} and \S \ref{S:NRth} and \ref{S:RTh}). One can roughly divide the MFA scenarios into three categories all related to the presence of energetic particles in the shock precursor \footnote{The shock precursor stands here for the region upstream the shock front populated by the most energetic particles. Notice that the terms defining the different parts of the shock structure are defined in \S \ref{S:M-HS}.}. These instabilities can be classified depending on the source of free energy that can destabilize them. A first type of instability is a kinetic instability generated by the resonant interaction of particles with plasma waves \cite{Skilling75, Bell78a, McKenzie82}. A second kind is produced by the return plasma current compensating the energetic particle current in the upstream medium \cite{Bell04}. Finally a third kind of instability is produced by the gradient of energetic particles that destabilizes compressible modes \cite{Drury86, Kangetal92, MalkovDiamond09}. The detailed nature of these instabilities and their connection to DSA will be the central topic of this review. Yet another problem especially in the second case above is to produce waves at a scale comparable to the particle gyroradius (see \cite{Schure12} and \S \ref{S:NRth}). In effect, the generation of long wavelength perturbations are essential in confining high energy particles the latter have to be injected from the thermal plasma.This injection problem is very essential in the regulation of the DSA process \cite{Blasietal05}. It is also interesting as the process is now being investigated by means of numerical simulations (see \S \ref{S:NRsim} and \ref{S:Rsim}) in particular by particle-in-cell and hybrid technics \cite{Riquelme10, Sironi11, Caprioli14a}. Recent simulation efforts have now begun to uncover the nature of the instabilities that mediate the particle energization close to the shock front. The energization of the incoming upstream plasma and the shock formation process is a central subject in magnetospheric and space plasma physics; we postpone its discussion to section \ref{S:M-HS}.\\
The above discussion shows that the interpretation of current and forthcoming high resolution astrophysical data will provide us with further insight into the microphysics of collisionless shock waves, whether non-relativistic or ultra-relativistic. Accordingly, it is important to have a trans-disciplinary approach in order to study the physics of high-energy radiation from shock waves, starting from the microphysics of the shock wave itself, then discussing the development of numerical tools dedicated to these studies and finally invoking recent progresses made in laser plasma experiments. This work will be mostly dedicated to the Fermi acceleration process at collisionless shocks; However sections \ref{S:M-HS} and \ref{S:RTh} will address the physics of other particle acceleration mechanisms. Special attention will be paid to the efficiency of the Fermi acceleration process with respect to the shock velocity and to the physical parameters of the environment (e.g. magnetization, magnetic field obliquity).\\

The review is organized into following different sections. Section \ref{S:Voc} introduces a basic common vocabulary used all over the report. Section \ref{S:NRS} discusses the microphysics of NR shocks: section \ref{S:M-HS} reviews recent progresses made in our understanding of magnetospheric and heliospheric shocks. This section also includes a repository of plasma instabilities that develop up- and downstream of the shock. Section \ref{S:NRobs} reviews observations that support particle acceleration and MFA in astrophysics while section \ref{S:NRth} discusses plasma instabilities that likely are connected to the two phenomen\ae. Section \ref{S:NRsim} reviews recent results on the microphysics of shock waves from numerical simulations. Section \ref{S:NRcr} connects the above results with the origin of high-energy cosmic rays. Section \ref{S:RS} discusses the microphysics of mildly relativistic and relativistic shocks: section \ref{S:Robs} summarizes the observations that support MFA in GRBs. Section \ref{S:RTh} discusses plasma instabilities that are relevant in the relativistic shock case. Section \ref{S:Rsim} reviews the recent findings on the microphysics of shock waves due to numerical simulations. The special case of striped pulsar winds is addressed in section \ref{S:Rpuls}. Section \ref{S:Ruhecr} discusses the link between shock microphysics and the origin of ultra high-energy cosmic rays. Section \ref{S:LS} reviews the recent developments in the laboratory experiments of shock formation and particle acceleration and radiation. Finally section \ref{S:Sum} summarizes the most relevant points in the review and concludes it. 

\section{Definitions}
\label{S:Voc}
This section introduces a common vocabulary valid for both magnetospherical and astrophysical contexts and for both non- and ultra-relativistic shock velocity limits. Once again this report only addresses the case of shock propagating in a collisionless plasma. For any further inquiry about basic concepts in collisionless shock physics the reader is directed to \cite{Treumann09} for further details.

\subsection{Notations used in the review}
\label{S:Def}
All quantities in this review are expressed in gaussian units. 

\begin{itemize}

\item $B_0$ is the background large scale magnetic field strength and $\delta B$ is the fluctuating component. 

\item $\ell$ is the  coherence length of the turbulent fluctuations.

\item $R_g = p c/Z eB$ is the gyro-radius of a particle of momentum $p$ and charge $Z e$ in a magnetic field of strength $B$.

\item $V_a = B/\sqrt{4\pi \rho}$ is the Alfv\'en velocity in a plasma of density $\rho$ and magnetic field strength $B$.

\item The electron plasma skin depth is $d_e=c/\omega_{ce}$, where $\omega_{ce}= e B/ m c$ is the electron cyclotron frequency. We will also use $\omega_{pe} =\sqrt{4\pi n e^2/m_e}$, the electron plasma frequency.  

\item The ion inertial length is $l_i=c/\omega_{pi}$.

\end{itemize}

\subsection{Shock classification}
There are two basic types of shocks:
\begin{enumerate}

\item Electrostatic shocks: Electrostatic shocks are sustained by the ambipolar electric field that is linked to the density gradient between a non-magnetized downstream plasma and the upstream plasma. In a simple approach, there is no jump in magnetic field. However, certain intrinsic mechanisms (such as microinstabilities) can generate some induced magnetic field. A detailed presentation of electrostatic shocks is given in \S \ref{S:Rsim}.

\item Magnetized shocks: Most of the shocks in geophysics and astrophysics do carry a mean magnetic field. It should not be confused with highly magnetized shocks where the magnetization parameter is high (see next).

\end{enumerate}

\subsection{Shock sub-structures}
The detailed structure of a (super-critical, see \S \ref{S:M-SF}) magnetospheric shock can be decomposed into several parts: the shock foot, the shock ramp and the overshoot-undershoot (see \S \ref{S:M-SF}).  The {\bf foot} can be defined as a bump in the magnetic field and pressure located upstream of the shock ramp and which results from the local accumulation of gyrating ions during their reflection against the ramp. The {\bf ramp} is the steepest part of the magnetic field/density gradient within the shock front. The {\bf overshoot-undershoot} are related to the reflection and subsequent gyration of ions which about the shock ramp. Foot and overshoot-undershoot parts are signatures of a noticeable number of reflected ions, i.e. of supercritical shocks. These signatures are almost all but absent in so-called subcritical shocks.\\
In astrophysics the shock front itself is usually described as the location where magneto-hydrodynamical quantities (density, pressure, temperature ...) have a jump corresponding to the Rankine-Hugoniot conservation conditions. We identify this location as the magneto-hydrodynamic shock (or MHD) front which is different for instance from the shock ramp. Other structures can also be present in astrophysical shock waves. The energetic particle {\bf precursor} is a structure ahead of the MHD shock occupied by energetic particles diffusing in the upstream medium. In some cases a radiative precursor can exist produced by the ionizing radiation emitted at the shock front (see \S \ref{S:NRobsopt}). Astrophysical shocks differ from magnetospherical shocks as they can extend over several orders of magnitude in spatial scales. 
 
\subsection{Reference frames}
In shock physics one considers usually three different rest-frames. The {\bf upstream rest-frame} (URF) is the frame where the upstream medium is at rest with respect to the (MHD) shock front. The shock front in this frame is approaching the upstream medium with a velocity $V_{\rm sh}$ or a Lorentz factor $\gamma_{\rm sh}=(1-(V_{\rm sh}/c)^2)^{-1/2}$ in the relativistic case. The {\bf downstream rest-frame} (DRF) is the frame where the downstream medium is at rest with respect to the shock front. The {\bf shock} (front) {\bf rest-frame} (SRF) is the frame moving with the MHD discontinuity. One may for practical purposes also define frames moving with magnetic irregularities for instance at the origin of the scattering of energetic particles in the Fermi diffusive shock acceleration process (see \S \ref{S:RThPart}). More detailed discussions about shock rest frames in astrophysics may also be found in \cite{Begelman90, Kirk94}. Hereafter the subscripts $u$ and $d$ will refer to quantities determined in the upstream and downstream medium respectively.

\subsection{Shock orientation}
In the case shocks are magnetized one can classify them under two main sub-classes:
\begin{enumerate}
\item {\bf Perpendicular or quasi-perpendicular shocks}: Perpendicular shocks \cite{Auer62,Forslund84} move through a plasma with a magnetic field vector, which is oriented strictly perpendicularly to the shock normal. For a quasi-perpendicular shock, the angle $\theta$ between the shock normal and the magnetic field is typically $45^\circ < \theta < 90^\circ$.

\item {\bf Parallel or quasi-parallel shocks}: Parallel shocks move through a plasma with a magnetic field vector which is oriented strictly parallel to the shock normal. The angle $\theta$ between the shock normal and the magnetic field is typically $0^\circ < \theta < 45^\circ$ for a quasi-parallel shock.

\end{enumerate}
In the context of oblique shocks (neither perpendicular nor parallel) it is possible to define the point of intersection $I$ between the field lines and the shock front. This point propagates at a velocity $V_{int}$ in the SRF. The shock is called {\bf sub-luminal} if $V_{int} < c$, and is {\bf super-luminal} if $V_{int} > c$. In the former case it is always possible to define a Lorentz transformation and find a frame where $I$ is at rest and the flow velocity is parallel to the magnetic field line in the upstream and downstream media. This frame is the de Hoffman-Teller frame. In this frame the convective electric field $-\vec{V} \wedge \vec{B}$ carried by the flow of velocity $\vec{V}$ and magnetic field $\vec{B}$ vanishes \cite{Kirk94}. In the latter case no such transformation is possible \cite{Begelman90}. This case is most relevant to relativistic shocks.

\subsection{Particle acceleration processes}
\begin{enumerate}

\item{{\bf Diffusive shock acceleration} (DSA).}
In diffusive shock acceleration particles gain energy by scattering off magnetic disturbances present in the upstream and downstream media. The difference of velocity propagation of the scattering centers induces a systematic energy gain at each shock crossing \cite{Krymsky77, Axford77, Bell78a, Blandford78, Bell78b}

\item{{\bf Shock drift acceleration} (SDA).}
In shock drift acceleration particles gain energy as their guiding centers move along the convective electric field due to the drift effects of the magnetic field gradient or the curvature of the shock front \cite{Decker85, Begelman90, Chalov01, Caprioli15}. 

\item{{\bf Shock surfing acceleration} (SSA).}
In shock surfing acceleration particles are reflected by the shock potential, and then return to the shock front due to the upstream Lorentz force. During this process, particles are trapped at the shock front and accelerated by the convective electric field \cite{Sagdeev66, Katsouleas83a, Zank96, Lee96}.

\end{enumerate}

\subsection{The magnetization parameter}
The magnetization of a given medium is the ratio of the Poynting flux to the particle energy flux, namely $\sigma =B^2/4\pi \rho c^2$ for a medium with a magnetic field $B$, a mass density $\rho=n \times m$ composed of particles of mass $m$ and proper density $n$. One may write $\sigma = V_a^2/c^2$. In case the flow is moving at relativistic velocity the magnetization parameter becomes $\sigma = B^2/4\pi \rho \Gamma U c^2$ for a medium with a four velocity $U$ and Lorentz factor $\Gamma=\sqrt{1+U^2}$\cite{Kennel84}. Typical values range from $\sigma \sim 10^{-9/-10}$ in the interstellar medium, to $\sigma \sim 10^{-4}$ in massive stellar winds and $\sigma \sim 0.1$ in pulsar winds. 

\section{Non-relativistic shock waves}
\label{S:NRS}

\subsection{Microscopic processes in non-relativistic shocks}
\label{S:M-HS}
This section discusses processes at the very base of shock formation and dynamics. It also addresses the question of particle energization connected to the development of instabilities in the different parts of a collisionless shock front. This section gives a special emphasis to planetary and solar magnetospheric shocks and also to the heliospheric shock which all have benefited from recent in-situ measurements.

\subsubsection{Shock formation}
\label{S:M-SF}
\begin{figure*}[t]
\centering
\includegraphics[width=1.5\textwidth]{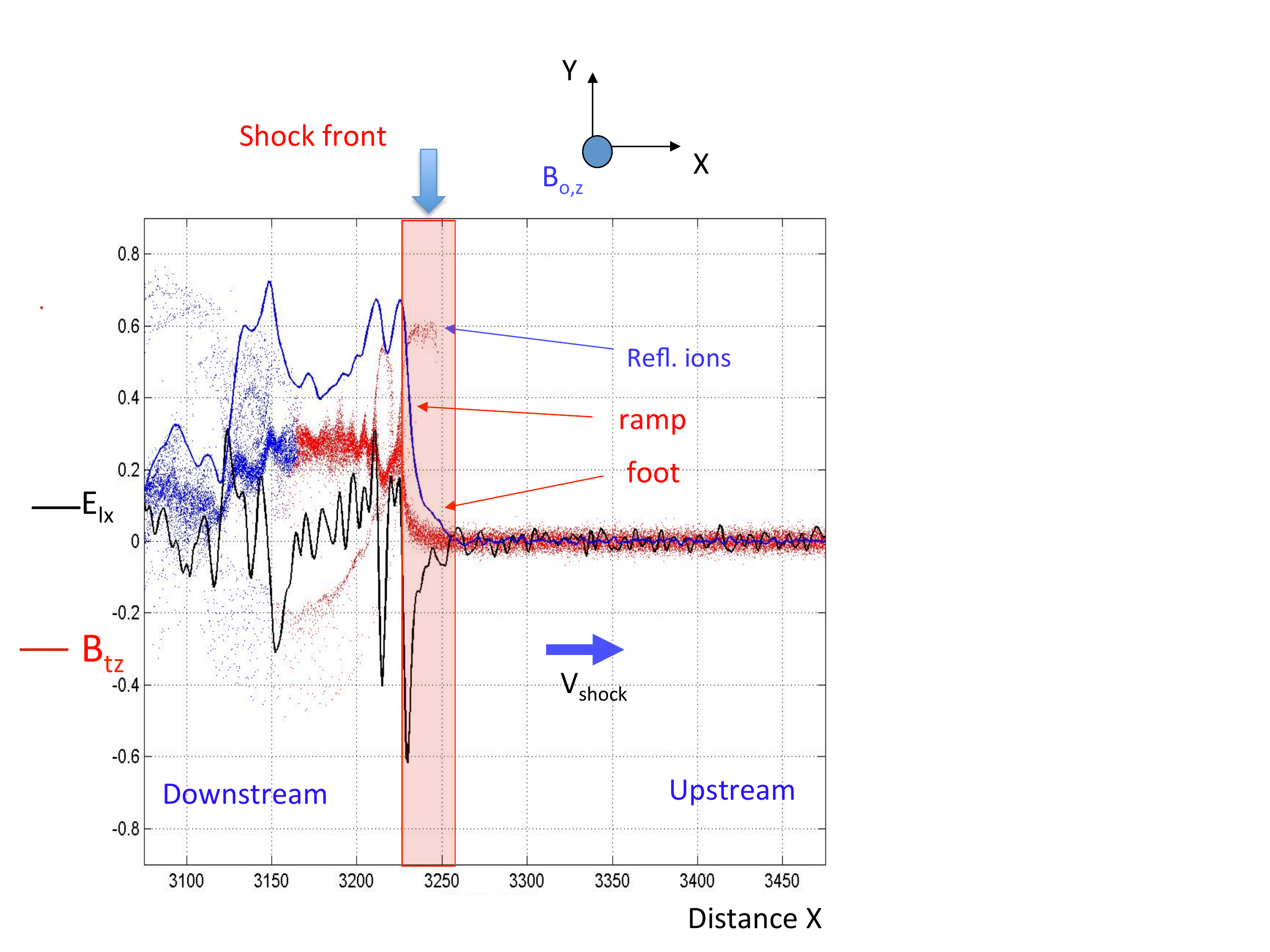}
\caption{Typical microstructures of the shock front for a strictly perpendicular shock obtained from a 1D-PIC simulation in a supercritical regime (characterized by a noticeable percentage of reflected ions as seen in the ions phase space). Dots, blue line and black line represent the ions phase space, the magnetic field and the electrostatic field profiles at a fixed time respectively. The shock front (indicated by the rectangle) includes the foot (due to the local accumulation of reflected ions) and the ramp (the steepest part of the front), and is limited by the overshoot (first maximum of the magnetic field). See \cite{Lembege10}}. \label{F:Msh}
\end{figure*}

Collisionless shocks are very common in space plasmas within our heliospheric system. One can consider three main groups of shocks : (i) the "obstacle-type" shocks when a plasma flow in supersonic regime (such as the solar wind emitted by the sun) meets an obstacle which can be a "magnetospheric-type" (for a magnetized planet as the Earth or Mercury) or a "ionospheric-type" (for an unmagnetized planet as Venus). Then, the regime of the flow suffers a transition from supersonic (upstream) to subsonic (downstream) through the shock front; (ii) the "CME-type" shock (coronal mass ejecta of gas and magnetic field or CME event) in solar physics as a huge quantity of hot and dense plasma is suddenly ejected from the solar corona into the solar wind. A shock front forms at the upstream edge of this ejected dense plasma, and propagates through the interplanetary space. Although it progressively dilutes during this propagation, it persists and becomes a self-sustained interplanetary shock (IPS). Eventually, such IPS can reach the terrestrial magnetosphere and collides with the Earth's bow shock; (iii) the "cometary-type" shock forms as a supersonic flow encounters a body emitting neutral matter as a comet; neutral atoms become ionized by different processes and form a secondary "pick-up" ion population in addition to the solar wind population; the common abrupt step-like profile of the shock front is replaced rather by a progressive transition layer through which the incident flow regime becomes subsonic, and where different ion populations are present.\\
In a simple approach, the electric field at the front of a {\it strictly perpendicular} shock has the appropriate sign to leave the electrons passing directly through the front and to reflect a part of incident ions. Most ions which have enough energy succeed to be directly transmitted (TI) too, and only a certain percentage is reflected by a large electromagnetic gradient, the ramp. Then, these reflected ions (RI) suffer a large gyromotion at the front due to the presence of the upstream static magnetic field (e.g. the interplanetary magnetic field or IMF) and accumulate locally. One key parameter is the percentage of RI which strongly depends on the Alfv\'enic Mach number $M_{A}$ of the incoming flow (ratio of the solar wind velocity over the local Alfv\'en velocity). For supercritical shock (where $M_{A}$ is typically larger than 2 roughly), this percentage is so high that accumulated ions are responsible for a bump in the magnetic field B and the pressure profiles upstream of the ramp, named the foot. RI describe only a single gyration (which corresponds to a ring in velocity space), and gain enough energy to penetrate the downstream region at later times. The signature of this ion ring may persist downstream of the front and is at the origin of an overshoot-undershoot pattern just behind the ramp, but becomes more dilute when penetrating further downstream (relaxation of the ion ring). In summary, the whole front of a supercritical shock includes three characteristic parts: the foot, the ramp and the overshoot-undershoot (see figure \ref{F:Msh}); the downstream region includes two ion populations: (i) the TI and (ii) the more energetic RI which have previously suffered one gyration at the ramp  \cite{Leroy1981, Leroy1983, Lembege1987a}. For subcritical shocks ($M_{A}$ is lower than 2), the percentage of RI is too weak to feed the foot formation, and neither foot nor overshoot-undershoot are present. \\
For oblique (quasi-perpendicular) shock, two features are relevant provided that the angular deviation from $90^\circ$ is large enough: (i) a finite velocity component parallel to the static magnetic field B allows both RI and electrons to return upstream along B; (ii) some dispersive waves can be emitted upstream from the ramp and form a so called "precursor" wave-train which competes with the foot formation. So, new incoming upstream particles do interact successively with the precursor and the foot before interacting with the ramp itself \footnote{The label "reflection" requires to be defined in order to avoid any confusion. Indeed, for quasi-perpendicular (supercritical) shocks, RI extend over a very limited region upstream from the ramp to form the foot (maximum width is of the order of convected ion gyroradius). In contrast, for quasi parallel (supercritical) shocks, ions are reflected along the static magnetic field to form field aligned beams (FAB) back-streaming into the solar wind, and extend "freely" over a much larger distance upstream from the ramp.}.\\
Let us stress out the following point: the fact that ions appear to play a key role in the structure of a collisionless shock does not mean that electrons do have a minor role and are only considered as light particles strongly energized by the shock itself. The impact of electrons is more important than expected. In addition to the large space charge field building up locally at the front (in particular at the ramp) where they strongly contribute, the electrons also contribute to the front dynamics itself i.e. to its non-stationary behavior (see \S \ref{S:M-NS}). Electrons permit the access to scales much smaller than ion scales, as the balance between nonlinear and dispersive/dissipative effects varies in time, i.e. this accessibility affects the steepness of the shock front profile itself; Electron or hybrid time scales via micro-instabilities can also have an impact on the shock profile and on its dynamics itself (see \S \ref{S:M-NS}). Using Particle-in-cell (PIC) simulations where both ions and electrons are fully included as a large assembly of individual particles, reveals to be quite helpful to analyse the shock dynamic versus time and its impact on particles' energization.

\subsubsection{Shock non-stationarity and instabilities}
\label{S:M-NS}
Non-stationarity of collisionless shocks has been clearly evidenced in early laboratory experiments \cite{Morse1972}, for the terrestrial bow shock  \cite{Walker1999} and has been analyzed more recently by using CLUSTER mission data (\cite{Horbury2002, Lobzin2008, Mazelle2010} and references  therein). Non-stationarity is not restricted to the terrestrial shock but appears to be a very common feature of shocks as mentioned also by \cite{Burlaga2008} to account for the multi-crossing of the terminal heliospheric shock by Voyager 2 in August 2007. Nowadays, the difficulty is to determine which of the different processes that have been proposed can be responsible for this effect. One way to clarify the situation is to concentrate on mechanisms which persist independently of some simplifying assumptions intrinsic to specific theoretical/numerical supports. In other words, these mechanisms should be retrieved in mono/multi- dimensional models and/or by using totally different types of numerical/simulation models. The label "non-stationary"  is often a source of confusion since it is commonly believed to be based on instabilities developing within the shock front, which is incorrect. For the purpose of clarity, we now separate non-stationarity processes excluding and including micro-instabilities. The latter is also discussed in the context of astrophysical shock waves. 
\begin{enumerate}
\item \underline{Non-stationarity group I}:  A large amount of works (mainly from simulations) have been already performed on the shock front non-stationarity excluding micro-instability processes. Among the different candidates, two main mechanisms generally emerge since these are recovered in different types of simulations and for different codes dimensionalities which is a good signature of their robustness \cite{Lembege2003}:

\begin{enumerate}
\item {\it Self-reformation due to the accumulation of RI} As the shock front propagates in supercritical regime, the number of RI increases locally at a foot distance from the ramp. As a consequence, the foot amplitude (at its upstream edge) increases until reaching a value comparable to the ramp. A new ramp is locally formed and starts to reflect a new set of upstream ions. Then, the process repeats cyclically with a period less than one upstream ion gyroperiod, and is at the origin of the so called "self-reformation" (SR) process. During the cyclic SR, the features of the shock front largely change \cite{Biskamp1972, Lembege1987a, Lembege1992, Hada2003, Chapman2005}:  (i) The amplitude of the shock front fluctuates in time where the variations of the overshoot versus the foot amplitude are anti-correlated (the overshoot decreases while that of the foot increases); (ii) The ramp thickness strongly varies from a large value (a fraction of the upstream ion inertial length) to a very narrow value (only a few electron inertial lengths); both features (i) and (ii) do have a strong impact on the particles' dynamics which depends on the front amplitude/ramp thickness as discussed in next subsections; (iii) The SR takes place as long as the RI keep a certain coherency during their gyromotion (narrow ion ring) as these accumulate far from the ramp, which is evidenced when the ratio of the shock velocity to the ion thermal velocity is quite large (several tens and above), or equivalently as the ion ratio  $\beta_{i}$ of the kinetic pressure over the magnetic pressure is weak (i.e. much less than 1) \cite{Hada2003, Scholer2004}; (iv) The SR process persists quite well for oblique propagating shock as long as the density of RI is high enough to feed the SR process itself; (v) Bursts of RI are cyclically emitted from the ramp with a period equal to that of the SR instead of a reflection at constant rate, similar bursts of reflected electrons emitted along the static magnetic field have been also evidenced \cite{Lembege2002} for quasi-perpendicular shock. One key feature is that the cyclic SR process reveals to be quite robust since observed in both PIC and hybrid simulations \cite{Hellinger2002, Hellinger2007} and references therein; it is basically a one-dimensional process but persists quite well in 2D and 3D simulations.

\item {\it Non-Linear Whistler Waves emitted by the shock front (so called "NLWW" process)} These waves do have the following features: (i) They are observed at least in 2D codes (i.e. are absent in 1D) and in both PIC and hybrid simulations which confirms their robustness; (ii) They do have a very large amplitude comparable to that at the ramp ($\Delta B /B  \approx. 1$); (iii) They propagate at an angle oblique to both the normal to the shock front and the static upstream magnetic field; (iv) Their wavelength along the shock front covers one or a few ion inertia lengths; (v) One key point is that no SR process is observed as these waves are emitted and inversely. This is explained by a loss of coherence of RI during their gyromotion as they interact with NLWW. This switches off the SR process; (vi) These waves have been observed originally in 2D simulations for a strictly perpendicular shock and as the magnetostatic field is lying within the simulation plane along the shock front. This configuration allows any waves to propagate between 0$^\circ$ and 90$^\circ$ to the magnetic field in the 2D plane \cite{Forslund1984}. This is in contrast with the other 2D perpendicular shock configuration where the static B field is perpendicular to the simulation plane; in this case, no NLWW are observed and the SR process is retrieved  \cite{Hellinger2007, Lembege2009}. At present, the mechanism responsible for the NLWW emission has not been clearly identified. The signature of this 2D non-stationarity (fluctuating large amplitude waves at the front) differs from that due to the cyclic SR process. So, this difference in the shock front dynamics observed for two different orientations of the magnetic field raises up the following questions: which process is dominant ? and in which configuration? A tentative answer has been given \cite{Shinohara2011} by performing 3D PIC simulations with realistic mass ratio and have shown that both processes can coexist but the SR process seems to be dominant. Indeed, the amplitude of NLWW appears to be smaller in 3D than in 2D simulations and consequently, the diffusion of RI by the waves is weaker.
\end{enumerate}
	
\item \underline{Non-stationarity group II}: The second category of non-stationary processes is based on micro-instabilities triggered within the shock front and are of prime interest for the case of astrophysical shocks treated at length in this review.
A large amount of instabilities excited in collisionless shocks have been already identified and may be found in some previous reviews \cite{Wu1984, Lembege1990}. Their mere nomenclature can be a source of confusion. For clarifying, they have been classified both in terms of their source mechanisms (see next), and in terms of the population involved, electrons or ions (see table \ref{tab:insta}), and the medium magnetization. When necessary, and even if the section is devoted to NR shocks, we will precise the NR and relativistic (R) shock type to these mainly apply, hence the discussion can also be transposed to the relativistic case. In short, four classes of micro-instabilities can be distinguished: (a) the "unmagnetized flow instabilities" (UFI) mainly analyzed for relativistic shocks; three other ones mainly analyzed for NR shocks: (b) the "cross-field current instabilities" (CFCI)  and the "field-aligned current instabilities" (FACI) which are responsible for micro-turbulence as detailed later on, (c) those based on "temperature anisotropy" (TAI) which builds up at the front/in the downstream region, and (d) those due to the formation of "out of equilibrium" distribution functions (both for electrons and ions) which form after upstream ions and electrons have interacted with the shock front (as f.i. the ring in the velocity space associated with RI, or electrons loss cone distribution). All these micro-instabilities are responsible for the turbulence which develops at the shock front and in the downstream region in electron, hybrid and ion ranges. Herein, we will focus mainly on the candidates of the above subgroups (a), (b) and (c):
\begin{enumerate}
\item {\it Unmagnetized flow instabilities or UFI} In the absence of a static magnetic field, the main source of anisotropy is defined with respect to the flow direction. In the context of astrophysical shocks, a special attention has been devoted to the {\bf Weibel instability} which simply occurs when the electronic distribution function is anisotropic in temperature. Weibel first uncovered that transverse waves with wave numbers $\mathbf{k}$ normal to the high temperature direction can grow exponentially \cite{Weibel59}. Kalman and co-workers proved later that while unstable waves grow for many orientations of $\mathbf{k}$, the Weibel modes are the fastest growing one \cite{Kalman1968}. Then, Davidson and co-workers assessed the non-linear regime \cite{Davidson1972}. Still within the range of unmagnetized, electronic instabilities, streaming instabilities have received an enormous amount of attention over the last 60 years. A typical setup consists in two counter-streaming electron beams initially compensating each other's current. As early as in 1949, Bohm \& Gross found that density perturbations along the flow can lead to exponentially growing waves in such systems \cite{BohmGross49}. This is the famous {\bf two-stream instability}, where electrostatic waves with $\mathbf{k}$ parallel to the flow grow exponentially. A few years later, Fried found perturbations with a $\mathbf{k}$ normal to the flow could result in exponentially growing transverse waves \cite{Fried1959}. These modes are now labeled {\bf filamentation} modes. Finally, some {\bf oblique} modes, with a $\mathbf{k}$ oriented obliquely to the flow, were also recognized as potentially unstable \cite{Watson60}. A global picture of the temperature dependent full unstable spectrum has emerged in recent years, evidencing the parameter domains where two-stream, oblique and filamentation instabilities govern the system \cite{Bret04, Bret10}. The filamentation instability is frequently called "Weibel'' in the literature, partly because they share some properties \footnote{A detailed discussion of this puzzling point has been made in \cite{Bret10}.}. All these unmagnetized instabilities have been discussed, among others, in the GRB context \cite{Medvedev1999, Nakar11} or in connection with primordial magnetic field generation \cite{Schlickeiser2003}.\\
\item {\it Cross field current instabilities or CFCI, and field aligned current instabilities or FACI} concern magnetized plasmas and has been analyzed in the context of NR shocks.\\
{\it b1)} For the first family two types of CFCI are commonly defined according to the currents direction: (i) When the relative drift is along the shock front (mainly at the ramp) and can trigger some instabilities (belonging to the family of lower hybrid drift instabilities or LHDI), responsible for small scale front rippling \cite{Lembege1992}. Indeed, let us remind that the ramp at the front (where the field gradient is the strongest) is supported by a strong cross-field current carried mainly by electrons. Moreover, note that other works have been dedicated on the front rippling both numerically \cite{Lowe2003} and experimentally with CLUSTER mission data \cite{Moullard2006}; (ii) When the relative drift establishes along the shock normal between three populations present within the foot region: the incident ions (II), the RI and the incident electrons (IE). The '"label" given to the instability varies according to the plasma parameters regime such as the strength of the relative drift (related to the Mach number regime $M_A$ of the shock front) and to the angle of the shock propagation. At present, works have been focussed on the {\bf electron cyclotron drift instability} (ECDI) defined for perpendicular shock, the {\bf modified two stream instability} (MTSI) defined for oblique (quasi-perpendicular) shocks. The MTSI is the two-stream instability with a magnetic field in the direction nearly normal to the flow \cite{McBride1972} (a flow aligned field leaves it unchanged). It is important to note that some instabilities which are basically electrostatic in linear regime may become electromagnetic in non-linear regime as shown recently for the ECDI \cite{Muschietti2013}. To our knowledge and within the context of R shocks, a nomenclature has not been systematically implemented, except for two cases: The {\bf Harris instability} is a kind (only a kind) of filamentation instability with a magnetic field normal to the flow \cite{Harris1959}. In contrast with the ECDI and MTSI defined for moderate drift, the {\bf Buneman instability} (defined for perpendicular shock) triggers for relatively high drift. Discovered on theoretical ground by \cite{Buneman59}, it amounts to a treatment of the two-stream instability including electrons/ions.\\
{\it b2)} The second family (FACI) has been mainly analyzed for oblique configurations where reflected particles (both ions and electrons) get enough energy during their reflection and are back-streaming along the ambient magnetic field into the incoming plasma flow. The resulting FACI are commonly invoked in foreshock region which is located upstream of the curved shock front and where two populations (incoming and back-streaming) co-exist. Both electron and ion foreshocks have been identified in 2D PIC simulations \cite{Savoini2001, Savoini2013} (and references therein), and have been clearly evidenced in experimental space missions \cite{Paschmann1980, Tsurutani1981}. The external edge of the electron foreshock is defined by the upstream magnetic field line tangent to the curved shock.\\
\item {\it Temperature anisotropy instabilities (or TAI)} TAI studies have been focused on ion population. The instabilities on ionic time scales, common in literature in connection with solar wind physics for example \cite{Maruca11,Schlickeiser11}, mainly concern the {\bf cyclotron}, {\bf mirror}, and {\bf firehose} instabilities. They all have to do with an anisotropic ionic component embedded in a background magnetic field $\mathbf{B}_0$. Denoting $R=T_\perp/T_\parallel$ ($\perp$ and $\parallel$ with respect to $\mathbf{B}_0$) and $\beta_\parallel=n k_BT_\parallel/(B_0^2/8\pi)$, the firehose instability can be triggered for $R < 1$ and $\beta_\parallel>1$. For $R > 1$, the mirror and cyclotron instabilities can grow and are responsible for the large scale front rippling \cite{Winske1988}. Still related to ion time scales, the {\bf Bell instability} \cite{Bell04} has been frequently evoked in connection with magnetic field amplification in SNR shocks. When considering a proton beam propagating along a guiding magnetic field into a background plasma, the so-called Bell modes are potentially unstable circularly polarized waves (see \S\ref{S:NRth}).
\end{enumerate}

\begin{center}
\centering
\begin{table}
\begin{tabular}{lcccc}
\textbf{Instability}       &\textbf{Conditions} &  \textbf{Stream}        &   $\mathbf{k}$               &   $\mathbf{B}_0$    \\
\hline\hline
Unmagnetized, electronic   &                   &                          &                              &                     \\
\emph{Weibel}              & $T_x > T_y$       &                          &    $\mathbf{k}\parallel y$   &                     \\
\emph{Two-stream}          &                   &  $\rightleftharpoons$    &     $\rightarrow$            &                     \\
\emph{Filamentation}       &                   &  $\rightleftharpoons$    &     $\uparrow$               &                     \\
\emph{Oblique}             &                   &  $\rightleftharpoons$    &     $\nearrow$               &                     \\
\hline
Magnetized, electronic     &                   &                          &                              &                     \\
\emph{Harris}              &                   &  $\rightleftharpoons$    &     $\uparrow$               &    $\rightarrow$    \\
\emph{Modified two-stream} &                   &  $\rightleftharpoons$    &     $\rightarrow$            &    $\uparrow$        \\
\emph{Electron cyclotron drift}  &                   &  $\rightleftharpoons$    &     $\rightarrow$           &    $\uparrow$    \\
\hline
Magnetized, ionic          &                   &                          &                              &                     \\
\emph{Bell}                &                   &  $\rightleftharpoons$    &     $\rightarrow$            &    $\rightarrow$    \\
\emph{Cyclotron}           &   $R>1$           &                          &     $\rightarrow$            &    $\rightarrow$    \\
\emph{Mirror}              &   $R>1$           &                          &     $\uparrow$               &    $\rightarrow$    \\
\emph{Firehose}            & $R<1$ \& $\beta_\parallel > 1$ &             &     $\rightarrow$           &    $\rightarrow$    \\
\hline\hline
\end{tabular}
\label{tab:insta}
\caption{Instabilities mentioned in \S \ref{S:NRS} with, when relevant, the conditions for their existence, the direction of the streaming motion, the orientation of the wave vector and of the static magnetic field. For magnetized ionic instabilities, $R=T_\perp/T_\parallel$ and $\beta_\parallel=nk_BT_\parallel/(B_0^2/8\pi)$, where $\perp$ and $\parallel$ refer to the magnetic field direction.}
\end{table}
\end{center}
What is the impact of Group II instabilities on the non-stationary behavior of the shock front identified in Group I, i.e. when no instability is involved ? Do these reinforce or inhibit the Group I non-stationary processes ? In summary, the ECDI leads to some ion scattering which remains too weak to have noticeable impact on the SR process  which persists quite well \cite{Muschietti2006}. However, the MTSI  which has a linear growth rate lower than that of the ECDI, plays a major role in the sense that an important ion scattering takes place, and a pressure gradient builds up locally at the edge of the diffusion region (within the foot). Then, a local new ramp starts reflecting a new set of incoming ions and initiates a new cyclic self reformation \cite{Scholer2004}. In this case, the MTSI (rather than  the accumulation of RI) is driving the SR process which takes place within a shorter time period.
\end{enumerate}

\subsubsection{Particle acceleration mechanisms and energization}
\label{S:M-PAME}
 Diffusive shock acceleration (DSA) \cite{Axford77, Bell78a, Bell78b, Krymsky77, Blandford78, Lee1983, Blandford1987, Webb1995} is a commonly accepted process for particle acceleration in quasi-parallel shocks, it will be discussed at length in this review. However, this process turns out not to work efficiently at low energies for (non-relativistic) quasi-perpendicular shocks, where the RI return to the shocks almost immediately due to their gyromotion in the upstream magnetic field. Therefore, shock drift acceleration (SDA) \cite{Hudson65, Webb83, Decker85, Decker88, Begelman90, Chalov01, Caprioli15} and shock surfing acceleration (SSA) \cite{Sagdeev66, Katsouleas83a, Zank96, Lee96, Lee1999, Shapiro2003} are considered to play important roles in ion acceleration or in pre-acceleration at quasi-perpendicular shocks. In the latter process, particles may repeat the process several times until they have acquired sufficient kinetic energy to overcome the macroscopic potential barrier at the shock front \cite{Zank96, Lee96, Lee1999, Lever01, Shapiro2003} and become transmitted. Simple models of shocks have shown that the SSA process is particularly efficient for a very narrow ramp. All these works have been based on stationary shock front. Recently, the impact of a non-stationary shock front has  been analyzed in details on the efficiency of SDA and SSA processes; this impact has been extended to hydrogen, heavy ions and to Maxwellian and pick-up (shell) ion distributions  \cite{Yang2009a, Yang2009b, Yang2011a, Yang2011b}. The SDA process appears to be largely dominant in most cases. However, these works have been restricted to an homogeneous shock front. In an improved and simple approach, Decker \cite{Decker1990} has considered the acceleration of ions within a rippled shock front by using a quasi-static surface corrugation described phenomenologically by a sinusoidal function, and have found that a few injected ions are trapped by the ripples, undergo many reflections within the front and are accelerated non-adiabatically. But, this work is based on a stationary shock front and the front rippling used is not consistent. More recent 2D test-particles simulations (where fields profiles are issued from self consistent 2D PIC runs) have analyzed the relative impact of both types of front rippling: one due to the emission of NLWW (excluding micro-instabilities) and the other due the front-aligned microinstabilities (CFCI). It clearly appears  that both SDA and SSA processes still persist and compete with each other \cite{Yang2012}, but the SDA mechanism appears to be still dominant in many cases even in the presence of front rippling. The electrons show a quite different dynamics. For strictly perpendicular shocks, they suffer an almost adiabatic heating in a first simple approach. In a more refined approach, differences sources of non-adiabaticity may be mentioned: (i) From the macroscopic fields at the shock front as the ramp thickness becomes very narrow as during a SR process, in this case, the electron only describes a very limited number of gyrations within the ramp before being transmitted downstream \cite{Savoini2005, Savoini2010}; (ii) From the microinstabilities triggered within the foot region as the ECDI \cite{Muschietti2006, Muschietti2013} where electrons suffer some preheating before reaching the ramp. Let us note that SSA mechanism which has been mainly proposed for ions energization has been also invoked as an efficient source of very energetic electrons in the context of relativistic shocks \cite{Hoshino2002}. For oblique (quasi-perpendicular) non-relativistic shocks, the electrons suffer different types of energization: (a) From the macroscopic fields at the shock front namely by specular reflection where electrons suffer a magnetic-mirror-type reflection by the magnetic field mainly (Fermi type 1) as in \cite{Leroy1984} and/or by the parallel component of the electrostatic field \cite{Lembege1987b}; (b) From the front rippling where the electrons can temporarily stay (trapping) \cite{Lembege2002}; (c) From the microinstabilities excited within the foot region such as the MTSI \cite{Scholer2004}. \\
One can wonder what is the impact of the different sources of shock front non-stationarity on electron dynamics ? The 1D and 2D PIC simulations of \cite{Lembege2002} have shown that, in absence of microinstabilities along the shock normal, the cyclic SR process (along the shock normal) leads to the formation of cyclic reflected electron bursts in time (with a period equal to  that of SR), while partial electron trapping takes place within the front rippling (due to CFCI) which results in the formation of electron packs in space. As a consequence, these results indicate that the electron reflection is not continuous in time neither homogeneous in space. However, a full understanding of the processes requires a 2D PIC simulation in conditions where ECDI / MTSI candidates are also fully included in order to check whether the electron packs/bursts persist (even partially) or are totally diffused by the local micro turbulence. Such works are under active investigation at present.

\subsection{Particle acceleration at astrophysical shock waves: observations}
As stated above DSA is a very promising mechanism for producing supra-thermal and \rel\ particles in a wide variety of objects ranging from the Earth bow shock \cite{Blandford1987, Ellison90, Jones91,Malkov01} to Mparsec (Mpc) scale size shocks in clusters of galaxies \cite{BykovDolag08}. This mechanism is believed to be efficient (see e.g. \cite{Helder09}) and capable of producing CRs of energies well above $10^{15}$\,eV in young SNRs \cite{Ptuskin10}, and even higher in active radio-galaxies such as Centaurus A \cite{Croston09}. By now SNRs are the most studied sites of DSA with high Mach number shocks. We will review multi-wavelength observations stating about supra-thermal particle acceleration in SNRs in the following sections.

\label{S:NRobs}
\subsubsection{Radio observations of young supernova remnants}
\label{S:NRobsrad}
This is a long time since SNRs are known as energetic particles sources and radio emitters \cite{Shklovskii53}. Radiation is produced in the MHz-GHz frequency band by synchrotron emission of non-thermal relativistic electrons with energies $E \sim 15 \ \rm{GeV} ((\nu/GHz)/(B/1\mu G))^{1/2}$, where $B$ is the local mean magnetic field measured in micro Gauss units. Radio observations provide informations about remnants morphology, about magnetic field strength and orientation and about particle acceleration processes \cite{Reynolds08, Dubner11}. Morphological studies are important to probe the explosion mechanism and the ambient medium, but also using self-similar hydrodynamical models (e.g. \cite{Chevalier82}) and X-ray observations, the position of the contact discontinuity with respect to the forward shock \cite{Decourchelle04}. The more compressible the fluid is, the closer the contact discontinuity and the forward shock are and the most efficient particle acceleration is. Efficient DSA where a substantial fraction, say more than 10\%, of the shock ram pressure is converted into CRs is most likely accompanied by the formation of strong magnetic turbulence in the shock vicinity. The first signatures of relativistic electron acceleration and magnetic field amplification were obtained from observations of synchrotron radio emission of SNRs (see for a review\cite{Ginzburg64}). Analyzing radio observations of Tycho's supernova remnant \cite{Chevalier77} revealed the presence of a collisionless shock wave undergoing turbulent magnetic field amplification by a factor of about 20. The authors pointed out that the amplification resembled some phenomena in heliospheric collisionless shocks. A few years later, a lower limit of about 80 $\mu$G (indicating again magnetic field amplification by a factor of 20 or higher) in the radio emitting shell of the supernova remnant Cassiopeia A was derived by \cite{Cowsik80}. To obtain this estimate the authors compared the observed upper limit of the gamma-ray flux to the expected bremsstrahlung gamma-ray flux derived from the detected synchrotron radio emission of GeV regime electrons. \cite{Cowsik80} speculated that this field strength must arise from magnetohydrodynamic instabilities in the expanding shell. In parallel, polarization studies provide information on the degree of order in the magnetic field as well as its global orientation. From the analysis of the early radio observations the non-adiabatic magnetic field amplification was expected in the shells of young SNRs, while the existing data of the extended old SNRs were consistent with just adiabatic compression of the interstellar magnetic field by the forward shock of supernova shell \cite{Vdlaan62}. The magnetic field polarization and orientation in young supernova remnants can be reproduced by invoking the development of a Rayleigh-Taylor instability at the interface between ejecta and shocked interstellar material \cite{Jun96}. Closer to the forward shock the Rayleigh-Taylor instability is unable to reproduce the magnetic field orientation unless, again, the contact discontinuity is closer to the forward shock \cite{Schure10}. But it is not clear wether another instability is able to produce this radio component \cite{Zirakashvili08}. 

Spectral studies show a mean radio spectral index of $\alpha = 0.5$ consistent with standard DSA theory (see section \ref{S:NRth}), although the indices are significantly dispersed around this value with a dispersion $\Delta \alpha \simeq 0.2$. The dispersion is possibly associated with either confusion by free-free emission (producing harder spectra) or non-linear particle acceleration (producing softer spectra). But a case by case explanation of the radio spectrum remains challenging. DSA relies on the ability for particles to get scattered by magnetic fluctuations. The upper limits on the scattering mean free paths of radio emitting electrons in front of supernova remnant shock waves have been established by \cite{Achterberg94} using high-resolution radio observations of four Galactic SNRs. The authors found that, for the sharpest synchrotron radio rims, the mean free path is typically less than one percent of the one derived for CRs of similar rigidity in the interstellar medium. The result suggested the presence of enhanced hydromagnetic wave intensity most likely generated by DSA. 

We finally mention the particular cases of very young SNRs (younger than 100 years old) \cite{Beswick06}. These objects are interesting as they can be monitored over time and hence provide information about the shock dynamics and the circum-interstellar medium properties. In some objects, the radio spectrum shows signatures of synchrotron-self absorption. The reconstruction of the turnover frequency (between the optically thin and the optically thick part of the spectrum) gives a direct estimate of the magnetic field strength. One of the most studied object SN 1993J \footnote{SN 1993J is a type IIb SN which blown off in M81. The early evolution showed a very fast shock with a velocity $v_{\rm sh} \simeq 0.1$c propagating in a dense red supergiant stellar wind.} where magnetic fields of the order of $200$ Gauss after the blow out have been inferred from radio emission modeling \cite{Bartel02}. These high values are likely quite in excess with respect to the equipartition magnetic field in the wind and point towards a strong amplification process possibly connected with proton and ions acceleration (see \S \ref{S:NRth}).

\subsubsection{Optical lines as a diagnostic tool for particle acceleration}
\label{S:NRobsopt}
The shocks of several young SNRs are often associated with very thin optical filaments dominated by Balmer emission. An important aspect of optical emission is the possibility of using the line shape and its spatial profile to check the efficiency of SNR shocks in accelerating CRs. 

The first detection of bright H$\alpha$ filaments around the remnants of Kepler, Tycho and the Cygnus Loop was reported by \cite{Minkowski56}. A peculiarity of this emission is the weakness of forbidden metal lines which implies an high temperature of the emitting region so that radiative cooling and recombination are unimportant. The interpretation of such optical emission remained a mystery up to the seminal works of \cite{Chevalier78,Chevalier80} who proposed that it can be produced by shocks propagating through a partially neutral gas. Their model was able to explain the intensity, spectrum and width of the filaments observed in Tycho's SNR, including the weakness of the forbidden metal lines. A peculiarity of Balmer dominated shocks, firstly reported by \cite{Chevalier80} for the Tycho's SNR, is that the H$\alpha$ line is formed by two distinct components, a narrow line with a FWHM of few tens km/s and a broad line with a FWHM of the order of the shock speed. Similar optical profiles are now observed from a bunch of young SNRs both in the Galaxy and in the Large Magellanic Cloud (for a review see \cite{Heng10}).
 
SNR shocks are collisionless and when they propagate in partially ionized medium, only ions are heated up and slowed down, while neutral atoms are unaffected to first approximation. However, when a velocity difference is established between ions and neutrals in the downstream of the shock, the processes of charge exchange (CE) and ionization are activated and explains the existence of two distinct lines: a narrow line emitted by direct excitation of neutral hydrogen after entering the shock front and a broad line that results from the excitation of hot hydrogen population produced by CE of cold hydrogen with hot shocked protons. As a consequence, optical lines are a direct probe of the conditions at the shock, in particular the width of the narrow and broad components reflect the temperature upstream and downstream of the shock, respectively. From the theoretical point of view, the main difficulty in describing the structure of a collisionless shock propagating in a partially ionized medium is that neutrals have no time to reach thermalization and cannot be treated as a fluid. Steps forward in relaxing the fluid assumption have been made by \cite{Heng07} and \cite{vanAdelsberg08}, even if these works neglect the modification induced by neutrals upstream of the shock. 

The first clue that Balmer emission could provide evidence for the presence of accelerated particles was put forward as a possible way to explain the anomalous width of narrow Balmer lines reported for the first time by \cite{Smith94} and \cite{Hester94}: FWHM ranging from 30 to 50 km/s was detected in four SNRs in the LMC and for the Cygnus Loop, implying a pre-shock temperature around 25,000-50,000 K. If this were the ISM equilibrium temperature there would be no atomic hydrogen, implying that the pre-shock hydrogen is heated by some form of shock precursor in a region that is sufficiently thin so that collisional ionization equilibrium cannot be established before the shock. Several explanations for this anomaly were proposed but only two of them were considered realistic: 1) the neutral-induced precursor and 2) the CR-induced precursor. 

{\it Neutral-induced precursor}:  When fast, cold neutrals undergo CE interactions with the slower hot ions downstream of the shock, some fraction of the resulting hot neutrals can cross the shock and move upstream. The relative velocity between these hot neutrals and the upstream ions triggers the onset of CE and ionization interactions that lead to the heating and slowing down of the ionized component of the upstream fluid. The system then tends to develop a neutral-induced shock precursor, in which the fluid velocity gradually decreases, and even more important, the temperature of ions increases as a result of the energy and momentum deposition of returning neutrals. A first attempt to investigate the broadening of the narrow line component induced by the neutral precursor was made by \cite{Lim96}, using a simplified Boltzmann equation for neutrals, but their calculation does not show any appreciable change of the narrow line width. This conclusion was confirmed by \cite{Blasi12, Morlino12}, using a fully kinetic approach able to describe the interaction between neutrals and ions in a more accurate way. The physical reason is that the ionization length-scale of returning hot neutrals in the upstream is always smaller than the CE length-scale of incoming cold neutrals. Interestingly enough, \cite{Morlino12} showed that the neutral precursor could produce a different signature, namely the presence of a third intermediate Balmer line due to hydrogen atoms that undergone charge exchange with warm protons in the neutral precursor.

{\it CR-induced precursor}: The anomalous width of narrow lines can be related to efficient particle acceleration where the ionized plasma is heated before crossing the shock. If the precursor is large enough, CE can occur upstream leading to a broader narrow Balmer line. The first attempt to model this scenario was done by \cite{Wagner08} using a two-fluid approach to treat ions and CRs but neglecting the dynamical role of neutrals.  A different model was proposed by \cite{Raymond11} where momentum and energy transfer between ions and neutrals is included, but the profile of the CR-precursor is assumed {\it a-priori} . Both works concluded that the observed width of 30-50 km s$^{-1}$ can be explained using a low CR acceleration efficiency. A more reliable interpretation of Balmer line profile requires an accurate description of the CR acceleration process where the mutual interplay between CRs, neutrals, ionized plasma and magnetic turbulence is simultaneously taken into account. Such an approach has been developed by \cite{Morlino13} using a semi-analytical technique. This work showed that the main physical effect able to broaden the narrow line is the damping of magnetic turbulence in the CR precursor while the adiabatic compression alone is ineffective. Hence the observed widths are compatible also with large acceleration efficiency provided the right level of magnetic damping. \\

\begin{figure}
\begin{center}
{\includegraphics[width=0.8\linewidth]{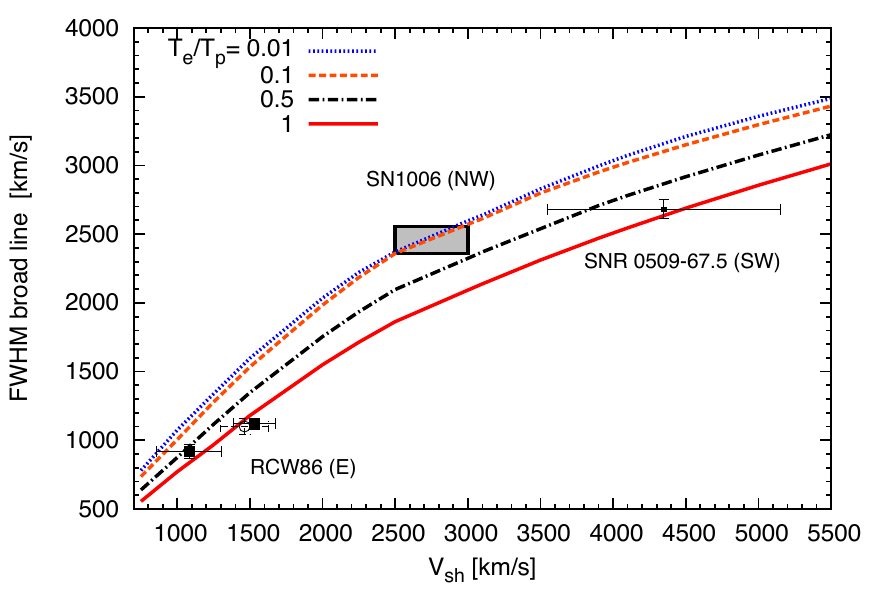}}
  \caption{Measured FWHM of the broad Balmer line as a function of the shock speed for three different remnant: RCW 86 (three different locations, one in the northeast (open circle) and two location in the southeast (filled square), data from \cite{Helder13}); northwest rim of SN 1006 \cite{Nikolic13}; southwest rim of SNR 0509-67.5 (the FWHM is taken from \cite{Helder10} while the uncertainty on $V_{\rm sh}$ is taken from the theoretical model in \cite{Morlino13b}). Lines show the theoretical prediction without CR acceleration for different values of electron to proton temperature ratio in the downstream medium \cite{Morlino13}.}
\label{fig:Balmer_broad}
\end{center}
\end{figure}

An efficient CR acceleration can also affect the width of broad lines. In fact, when a sizable fraction of the ram pressure is channeled into non-thermal particles, the plasma temperature behind the shock is expected to be lower, and this should reflect in a narrower width of the broad H$\alpha$ line. Remarkably, there are clues of this phenomenon in two different remnants, RCW86 \cite{Helder13, Morlino14}, and SNR 0509-67.5 in the LMC \cite{Helder10, Morlino13b}. In both cases the measured FWHM of the broad lines is compatible with theoretical predictions only assuming fast electron-proton equilibration downstream of the shock, a conclusion which seems to be at odds with both theoretical models and observations \cite{Rakowsky05}. In the left panel of Fig. \ref{fig:Balmer_broad} is compared the FWHM measured from these two remnants with those taken in the northwest rim of SN 1006 \cite{Nikolic13}, a region which does not show any signature of efficient CR acceleration and has a Balmer emission fully compatible with theoretical expectation assuming a low level of electron-proton equilibration. 

Finally, we mention that the presence of efficient CR acceleration could result in a third signature, namely the presence of Balmer emission ahead of the shock. This was claimed for the first time by \cite{Lee10}, where the authors measured a gradual increase of H$\alpha$ intensity just ahead of the shock front, in the so called {\it knot g} of the Tycho's SNR. This has been interpreted as emission from the thin shock precursor ($\sim1''$ which imply a thickness of $\sim3\times 10^{16}$ cm for a distance of 3 kpc) likely due to CRs, which, if confirmed, would represent the first direct proof of the existence of a CR precursor. On the other hand Balmer emission from the upstream can be also produced by the neutral-induced precursor, as showed in \cite{Morlino12}, and, in order to distinguish between these two possibilities, a careful modeling of the shock is required, able to handle the complex interaction between the CR and the neutral induced precursor. At the moment the most promising technique seems to be the kinetic theory developed in \cite{Morlino13}.

\subsubsection{X-ray structures in young supernova remnants}
\label{S:Xobs} 
\begin{figure*}[t]
\includegraphics[width=0.5\textwidth]{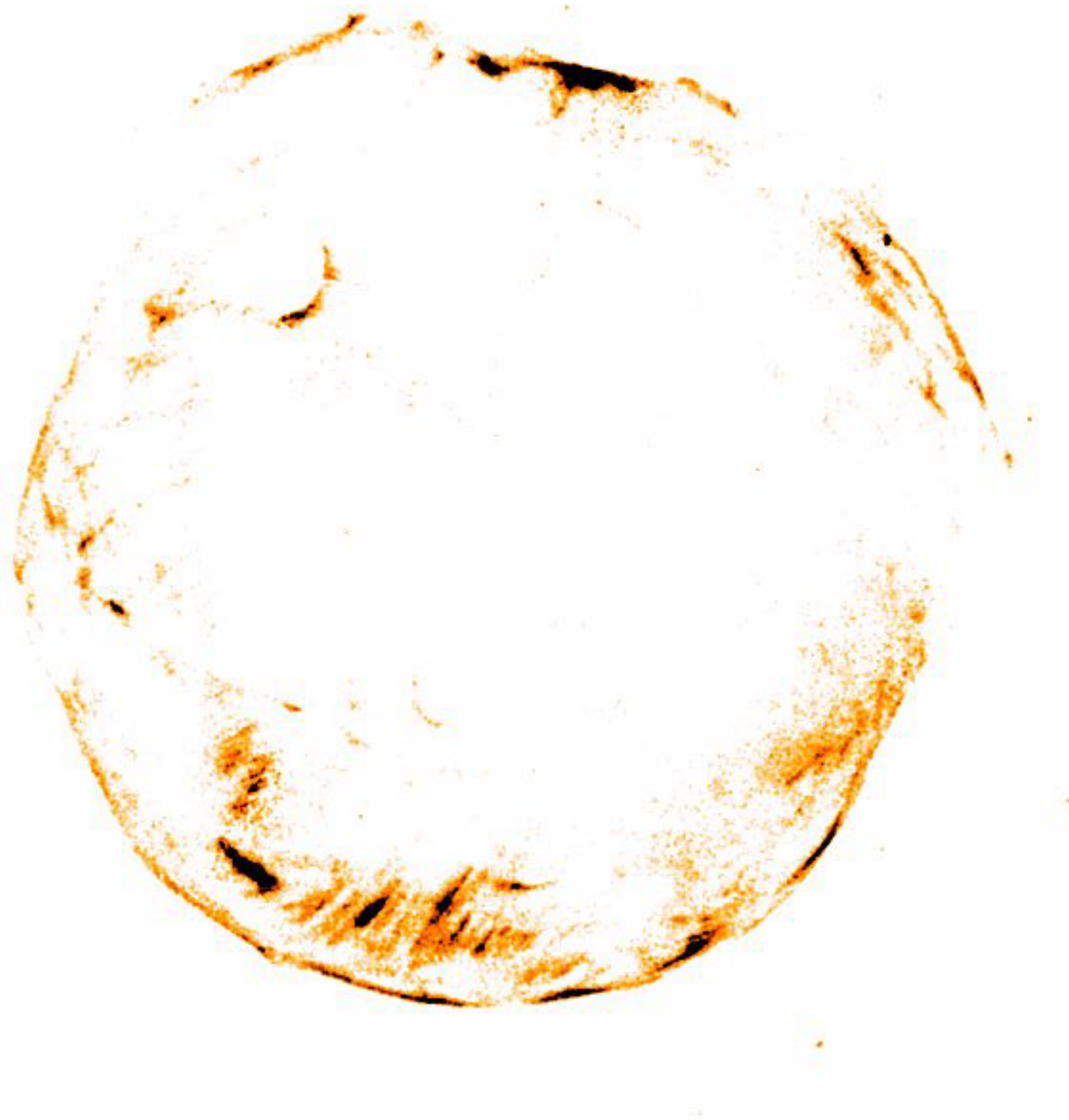}
\includegraphics[height=0.3\textheight, width=0.5\textwidth]{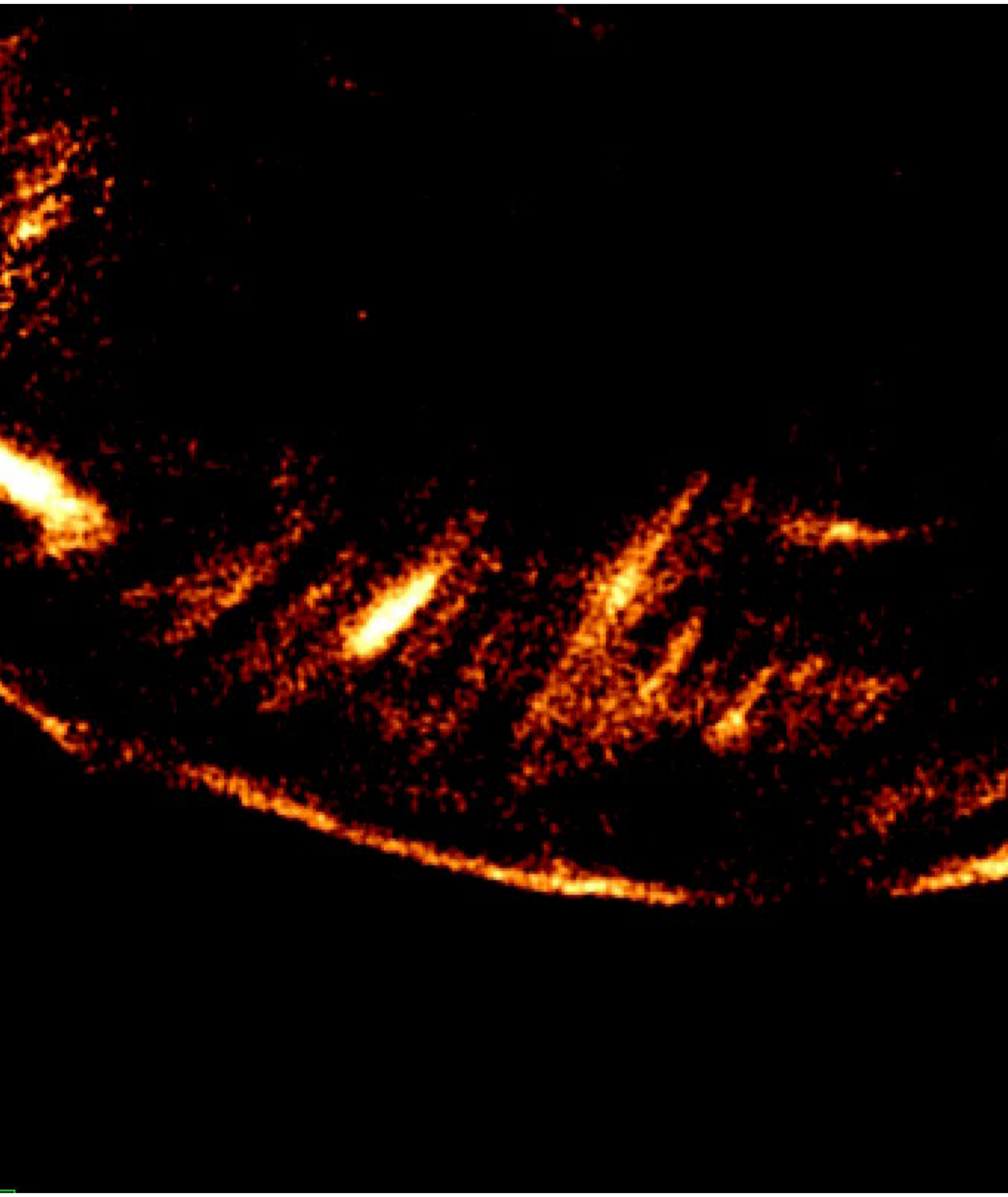}
\caption{Left: \chan\ X-ray image of Tycho's SNR in 4-6 keV photon energy regime. Right: Zoomed X-ray stripes in 4-6 keV \chan\ image of Tycho's SNR discovered by \cite{Eriksen11}.} \label{F:tycho}
\end{figure*}

\begin{figure*}[t]
\centering
\includegraphics[height=0.6\textheight,width=0.8\textwidth]{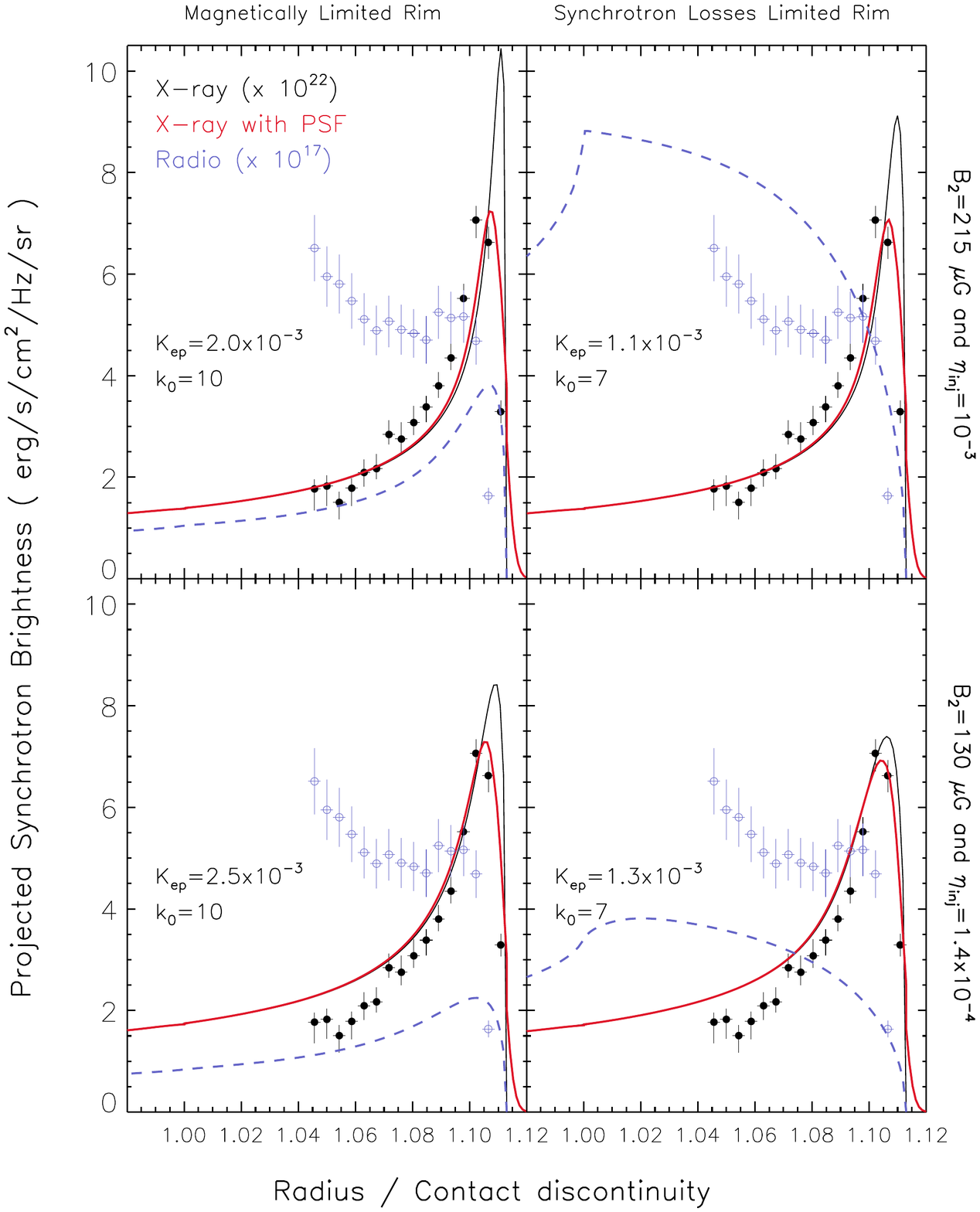}
\caption{The X-ray and radio profiles in the postshock flow in Tycho's SNR from \cite{Cassam07}. The red solid lines show the X-ray profiles convolved with a model of the \textit{Chandra} PSF. The radio \textit{VLA} data (marked with {\Large $\circ$}, in blue) and X-ray \textit{Chandra} data points (marked with{\Large $\bullet$}) show the profiles of the Western rim (see Fig.~\ref{F:tycho} left). The line-of sight projections of synchrotron brightness modeled by \cite{Cassam07} are shown with blue dotted lines in radio (1.4 GHz) and with black solid lines for X-rays (1 keV). The radio and X-ray profiles were multiplied by $10^{17}$ and $10^{22}$, respectively.} \label{Fig:cassam-chenai_proj}
\end{figure*}

X-ray observations can also be used to study DSA. \cite{Reynolds81} suggested that the featureless power-law X-ray spectrum of the SN 1006 remnant is the extension of the synchrotron radio emission. Observations of SN 1006 with {\sl ASCA} satellite by \cite{Koyama95} indicated that emission from the edges of the remnant shell is dominated by the synchrotron radiation from 100 TeV electrons accelerated by supernova shock. The Inverse Compton (IC) radiation of the TeV regime electrons is likely responsible for the TeV photons detected from SN 1006 with H.E.S.S. Tcherenkov telescope (see e.g.\cite{Acero10} and section \ref{S:GSNR}). 
The X-ray imaging of SNRs with the superb spatial resolution of \chan\ telescope have revealed the synchrotron emission structures in Cassiopeia A, Tycho's SNR, Kepler's SNR, SN1006, G347.3-0.5 (RX J1713.72-3946), and other SNRs \cite{Vink03, Bamba05, Uchiyama07, Reynolds08, Eriksen11, Vink12, Helder12}. The morphology of the extended, non-thermal, thin filaments observed at the SNR edges, and their X-ray brightness profiles, strongly support the interpretation that $\gsim 10$\, TeV electrons are accelerated at the forward shock of the expanding supernova shell and produce \syn\ radiation in an amplified magnetic field. We illustrate in the left panel of Fig.\ref{F:tycho} the filaments in Tycho's SNR X-ray image made with high resolution \chan\ telescope in 4-6 keV regime. There are no strong K-shell lines of astrophysical abundant elements in 4-6 keV band, therefore the X-ray emission is mostly continuum and in Tycho's SNR it is likely dominated by the \syn\ emission of TeV regime electrons (and possibly positrons). X-ray filaments associated with the forward shock are clearly seen at the edge of the image. \cite{Cassam07, Cassam08} studied the brightness profiles of Tycho's SNR both in radio and X-rays to distinguish between two possible models of the apparent synchrotron filaments. Both models assumed TeV regime electron acceleration and magnetic field amplification at the forward shock. However, in the first model magnetic field profile in the post-shock region was assumed to be flat and therefore the sharp X-ray profile was supposed to be due to the synchrotron losses of radiating electrons synchrotron losses limited rim in the right panel of Fig. \ref{Fig:cassam-chenai_proj}. The alternative model assumed magnetically limited rim due to possibly strong damping of the amplified magnetic field in the post-shock flow as it was proposed by \cite{Pohl05} and illustrated in the left panel in Fig. \ref{Fig:cassam-chenai_proj}. The observed radio profile is not consistent with the assumption on the fast decay of the amplified magnetic field. However, \cite{Cassam07} concluded that while the two models they used describe the X-ray data fairly well, they both fail to explain quantitatively the observed radio profile. Assuming that the observed thickness of the X-ray rims in young SNRs are limited by the synchrotron losses of the highest energy electrons in uniform and isotropic turbulence \cite{Berezhko03, Voelk05, Parizot06} derived constraints on the CR diffusion and acceleration parameters in these SNRs. Namely, \cite{Parizot06} concluded that the magnetic field in the shock downstream must be amplified up to values between 250 and 500 $\mu$G in the case of Cassioppeia A (CasA), Kepler, and Tycho, and to about 100 $\mu$G in the case of SN 1006 and G347.3-0.5.

The angular resolution of \chan\ telescope is about 1$\arcsec$ corresponding to a spatial scale of about 7 $\times$ 10$^{15}$ cm for an SNR at 1 kpc distance. The resolution scale roughly corresponds to the gyroradii of a CR proton of energy 2 $\times B_{\mu G}$ TeV. Since the typical amplified magnetic fields in young SNRs was estimated to be above 50 $\mu G$ see e.g.\cite{Parizot06} the resolution scale corresponds to about 100 TeV CR proton gyroradii. The energy containing scale of simulated spectra of CR-driven turbulence in the case of efficient CR acceleration is comparable with or larger than the gyroradius of the maximal energy proton \cite{Vladimirov06, Pelletier06, Zirakashvili08, Bykov12, Schure12, Bykov13}. This imply a possibility to study synchrotron structures associated with turbulent magnetic fields amplified by CR driven instabilities.

Apart from the extended filaments other types of X-ray synchrotron structures were discovered with \chan\ telescope. Amazing structures consisting of ordered sets of bright, non-thermal stripes with the apparent distance between the stripes about 8$\arcsec$ were discovered by \cite{Eriksen11} in Tycho's SNR with a deep \chan\ exposure (see the right panel in Figure \ref{F:tycho}). \cite{Eriksen11} pointed out that if one associates the apparent distance between the strips with two times of the proton gyroradius then the maximal CR proton energy should be about 10$^2$-10$^3$ TeV  for the distance to Tycho's SNR estimated as about 4 kpc. Interpretation of these structures presents a formidable challenge for current models of X-ray \syn\ images of young SNRs because of its non-trivial quasi-regular structure. An explanation connected with cosmic-ray generated magnetic turbulence at the SNR blast wave is addressed in \S \ref{S:NRth}.

Small scale variable X-ray structures which are likely of synchrotron origin were discovered with \chan telescope in the shells of SNR G347.3-0.5 (RXJ1713.72-3946) \cite{Uchiyama07} and CasA \cite{Patnaude09}. Namely, \cite{Uchiyama07} reported a year timescale variability of a few X-ray brightness enhancements in the shell of SNR G347.3-0.5. The authors attributed this X-ray variability to synchrotron radiation losses and, therefore, suggested the ongoing shock-acceleration of electrons in real time. They concluded that the magnetic field in the shell of G347.3-0.5 would need to be about mG (i.e. to be amplified by a factor of more than 100) to provide the very rapid radiation losses of the emitting electrons. However, the multi-wavelength data analysis by \cite{Butt08} concluded that the mG-scale magnetic fields estimated by \cite{Uchiyama07} cannot fill in the whole non-thermal SNR shell, and if small regions of enhanced magnetic field do exist in SNR G347.3-0.5, it is likely that they are embedded in a much weaker extended field. An alternative interpretation to the observed fast variability of the X-ray clumps in SNR G347.3-0.5 attributed the effect to quasi-steady distribution of X-ray emitting electrons radiating in turbulent magnetic field was proposed by \cite{Bykov08}. The model allowed modest magnetic field amplification. The lifetime of X-ray clumps can be short enough to be consistent with that observed even in the case of a steady particle distribution.

\subsubsection{Gamma-rays: hadronic or leptonic scenarii} 
\label{S:GSNR}
Gamma-ray radiation can be produced in three different ways in SNR shocks. Electrons (leptons) can produce IC radiation scattering off photons from the cosmic microwave or infra-red backgrounds. They can also produce Bremsstrahlung photons if the medium is dense enough (this may be the case for SNR shocks in interaction with molecular clouds). Protons (hadrons) can radiate gamma-rays through neutral pion production induced in p-p interaction. Gamma-rays are interesting because they probe the highest particle energies and in case of pion production as they probe the hadrons accelerated at SNR shocks.\\
We first consider the case of shell-like SNR. At the time of this review, there are seven historical SNR detected at TeV energies by the current Tcherenkov telescopes CasA, SN 1006, Tycho, RCW86, HESS 1731-347, RX J1713-3946.5, Vela Jr, G0.9+0.1) whereas no TeV gamma-rays have been detected from Kepler SNR yet. Among them only four (CasA, Tycho, RX J1713-3946.5, Vela Jr) have been detected at GeV energies by the {\it Fermi} telescope (see e.g. \cite{Aharonian13} for a review of combined GeV and TeV observations). Often the gamma-ray spectrum appears to be soft with differential energy indices larger than 2 (CasA, Tycho) as may be expected from DSA in the test-particle limit. Objects like RXJ 1713-3946.5 and Vela Junior on contrary do show a spectrum harder than 2 in the Fermi domain that is usually difficult to reconcile with pure {\it one-zone} (see however \cite{Gabici14}) hadronic scenarios even considering non-linear back-reaction effects (see \cite{Aharonian13}). The maximum gamma-ray energy is often limited to the TeV range which in the hadronic scenario corresponds to highest CR energies in the range 10-100 TeV, still under the CR knee. These points are among the issues that question SNR as being the origin of galactic CRs (see \S \ref{S:NRcr}). Due to the modest angular resolution of the instruments the origin of the gamma-ray radiation can not firmly be associated with the forward shock only, some contribution from the reverse shock is possible especially in CasA \cite{Abdo10}.\\
Gamma-ray filaments have been proposed to help in discriminating the dominant accelerated population and to probe the magnetic field structure. A first work by \cite{Marcowith10} did considered leptonic acceleration only but obtained X-ray and gamma-ray filaments produced by different particle populations in Kepler and RX J 1713-3946.5 SNRs. It appears especially that gamma-ray filaments can have an important component produced by electrons scattering soft photons upstream the shock. Another work by \cite{Lee13} did propose synthetic gamma-ray profiles of Vela Junior SNR filaments produced by electrons or hadrons in a leptonic or hadronic scenario respectively. The authors have shown that electron or hadron induced gamma-ray filaments do not strongly differ as hadron are less sensitive to radiative losses whereas electrons suffer from less losses in a leptonic scenario due to lower magnetic field values. In each cases, it appears difficult to resolve the gamma-ray filaments by the on-going gamma-ray instruments. However, improved angular resolution of the next generation gamma-ray Tcherenkov instrument C.T.A. (Cerenkov Telescope Array) \footnote{see https://portal.cta-observatory.org/} may resolve the gamma-ray filaments of large SNR like Vela Junior or RX J 1713-3646.5 \cite{Acero13}.\\
Most of the gamma-ray SNR sources detected by the {\it Fermi} telescope are associated with a system of a shock in interaction with molecular clouds (see \cite{Uchiyama11}). The spectra of these objects do show a convex $\nu F_{\nu}$ spectrum with a spectral break in the GeV domain. The origin of the emission is very likely hadronic as a leptonic scenario would require unrealistic low densities and magnetic fields and would not fit radio data in a satisfactory way. The origin of gamma emission and the GeV break is still debated. Two scenarios tend to emerge actually: the gamma-ray may be produced by CR escaping from the SNR shock and interacting with the cloud material \cite{Ohira11} or may result from a pre-existing population of CR compressed by the SNR shock \cite{Uchiyama10}.

\subsection{Theory: Magnetic field amplification by Cosmic Ray-driven instabilities}
\label{S:NRth} 
Fast and efficient particle acceleration by Fermi mechanism at astrophysical shocks assumes that particles are multiply scattered by magnetic fluctuations in the shock vicinity (see \S \ref{S:Intro} for references). The amplitude of the magnetic turbulence is substantially higher than the ambient field fluctuations forcing a bootstrap scenario where the accelerated particles amplified the turbulence required for their acceleration \cite{Blandford1987, Bell01, Malkov01}. We discuss also combined CR - fluid mechanisms of magnetic field amplification in the shock vicinity. These include a dynamo like process \cite{Beresnyak09} or a modulation instability of CR excited Alfv\'en waves scattering off ambient density perturbations \cite{Diamond07} as well as pure fluid MFA models \cite{Giacalone07, Fraschetti13}. The two next sections provide a detailed analysis of the CR induced streaming instabilities in the conditions that prevail in the shocks of SNR. We first detail the streaming instabilities and then provide a more general framework of instability analysis. The numerical studies of CR induced streaming instabilities are addressed in \S \ref{S:MHDNR}.\\

\subsubsection{Streaming instabilities}
The streaming instabilities are known for decades to generate Alfv\'en waves which interact resonantly with energetic particles in the interstellar plasma (see for a review \cite{Amato11} and references therein). The galactic CR propagation models rely on the resonant interaction of CRs with Alfven waves \cite{Berezinskii90, Schlickeiser02}. Numerical models demonstrated that strong self-excited turbulence may reduce the CR diffusion coefficient close to the Bohm limit \footnote{equal to 1/3 $R_g v$.} with important implications for cosmic-ray transport \cite{Casse02, Candia04, Reville08}. In recent years the CR streaming instability at shocks has been the subject of a number of theoretical developments. The {\sl resonant} CR streaming instability is a kinetic instability which involves the production of modes in resonance with particle gyromotion at wavenumbers $k$ such that $k R_g > 1$. This instability was shown to be able to amplify the magnetic field fluctuations above the mean field level in the vicinity of a strong shock accelerating CRs by first order Fermi mechanism \cite{McKenzie82, Blandford1987, Malkov01}. The CR pressure gradient in the shock upstream can induce magnetic turbulence upstream of the supernova blast wave \cite{Dorfi85, Drury86, Drury12}.\\
\cite{Bell04} showed that the presence of a strong CR current (that is expected in DSA scenario) should result in a {\sl non-resonant} instability amplifying fluctuations of scales shorter than CR particle gyroradius (see \S \ref{S:M-NS} for the properties of the modes). We describe here this {\sl non-resonant} CR-current driven instability in more details. \\
In the simplest theory of DSA, where supra-thermal particles satisfy the standard advection-diffusion equation, the isotropic steady-state particle distribution as measured in the upstream plasma rest frame, is: 
\begin{equation}
 f_0(x,p) ={n_{CR} N(x,p) \over 4\pi}=  f_0(0,p)\exp\left(-\int \frac{u}{\kappa} dx\right) \ ,
\end{equation}
where  $n_{\rm CR}$ is the number density of CRs and $x$ is the distance from the shock front. The quantities $u$ and $\kappa$ represents the upstream fluid velocity and CR spatial diffusion coefficients respectively \cite{Drury83}. 
The net CR current in the upstream (background) plasma is therefore
\begin{equation}
 \bf{j} = e\int \bf{v} f_0(x,p)\left(1+3\frac{u_{\rm sh}}{c}\cos\theta\right) d^3 p 
\approx en_{\rm CR} u_{\rm sh}\bf{\hat{x}} \ ,
\end{equation}
where $\theta$  is the pitch-angle of a CR particle, we will note hereafter $\mu=\cos\theta$.\\
The discussion focuses on the parallel shock configuration where the shock velocity and the background magnetic field are oriented along the $\hat{x}$ axis. \cite{Bell05} generalized these calculations to other shock configurations. It is this current that ultimately does the work on the ambient plasma. In any numerical investigation of CR acceleration or magnetic field amplification, how this current is calculated or determined is central to the problem (see \S \ref{S:NRsim}). With regard to the acceleration of relativistic protons, the time-scales of interest are of the order of the proton Larmor period $R_g/c \sim 100 \gamma B_{\rm \mu G}$s. (Here $\gamma$ is the Lorentz factor of the proton, and $B_{\rm \mu G}$ the magnetic field measured in microGauss.) Clearly, the relevant times are orders of magnitude larger than those associated with the kinetic timescales of the background thermal plasma, suggesting that a fluid treatment is sufficient for the study of particle acceleration in non-relativistic plasmas, ie. the background plasma can be shown to satisfy the momentum equation
\begin{eqnarray*}
\rho\frac{d\bf{u}}{d t} = - \bf{\nabla} P_{\rm gas}  
+\frac{{\bf{j}_{\rm th}} \times \bf{B}}{c} +e(n_{\rm p}-n_{\rm e})\bf{E}
\end{eqnarray*}
where it is safe to assume $\bf{E} = -\bf{u}/c\times\bf{B}$.
Using Ampere's Law
\begin{eqnarray*}
{\bf j}_{\rm th}=\frac{c}{4\pi}\bf{\nabla}\times {\bf B}-{\bf j}_{\rm cr}
\end{eqnarray*}
this becomes  
\begin{eqnarray}
\label{MHD_cr}
\rho\frac{d\bf{u}}{d t}+ \bf{\nabla} P_{\rm gas}  
+\frac{1}{4\pi}\bf{B}\times(\bf{\nabla}\times \bf{B}) 
&=&-n_{\rm cr}e{\bf E}-\frac{{\bf{j}_{\rm cr}} \times \bf{B}}{c} \ .
\end{eqnarray}
One approximation that is frequently used at this point is to take a fluid approximation for the cosmic-ray momentum conservation. Neglecting the inertia of this fluid, it follows that 
\begin{eqnarray}
n_{\rm cr} e\bf{E} + \frac{\bf{j}_{\rm cr}\times \bf{B}}{c} = 
\bf{\nabla} P_{\rm cr} 
\end{eqnarray}
which on substituting into equation (\ref{MHD_cr}) gives
\begin{eqnarray}
\rho\frac{d\bf{u}}{d t}+ \bf{\nabla} (P_{\rm gas} + P_{\rm cr})  
+\frac{1}{4\pi}{ \bf B}\times(\bf{\nabla}\times {\bf B}) =0 \ .
\end{eqnarray}
This is the typical starting point for non-linear diffusive shock acceleration models \cite{McKenzie82, Malkov97}. This  approximation is useful for studying effects on large length-scales $\ell\sim f/|df/dx|$, such as dynamics and feedback of magnetized CRs in the precursor. However, if strong non-linear magnetic field amplification is occurring, this approximation is no longer ideal, and Eq.(\ref{MHD_cr}) is the preferable approach and was adopted in \cite{Bell04}. Eq.(\ref{MHD_cr}) can be tested against linear perturbations $\propto\exp\left(ikx- i\omega t\right)$. The dispersion relation in the MHD limit reads:
\begin{equation}
\omega^2 - k^2 v_{A,0}^2 \pm {{\xi} \over 2} {k \over r_{*}} = 0\ ,
\end{equation}
for particles of gyro-radius $r_*$ in the background magnetic field. The different signs correspond to the case of purely growing or decaying modes. The parameter $\xi$ is connected to the maximum CR momenta $p_{max}$ and the energy imparted into CR $U_{CR}$ through 
\[
\xi = 1/\ln(p_{max}/mc) \times (U_{CR}/\rho_u v_{sh}^2) \times \beta_{sh} \ .
\]
The maximum growth rate $\gamma_{max} = k_{max} V_{a,0}$ is obtained at a wave number $k_{max} r_* ={\xi}/2 M_{a,0}^2$ usually $\gg 1$ (small scale perturbations). The instability grows much faster than the resonant streaming instability as demonstrated in a series of analytical \cite{Bell04, Bell05, Pelletier06, Marcowith06, Amato09, Riquelme09, Bykov11, Bykov12} and numerical studies \cite{Bell01,Zirakashvili08, Niemiec08, Vladimirov09, Gargate10, Riquelme10, Rogachevskii12, Reville12, Schure13, Bai15} (see \S \ref{S:NRsim}). This process of fast amplification of short-scale modes proposed by \cite{Bell04} can be accompanied with amplification of the long-wavelength fluctuations that would allow the effective confinement and acceleration of higher energy particles \cite{Bykov11,Reville12, Schure12, Bykov13}. The latter are discussed now using the general description CR induced analysis that follows.

\subsubsection{A general linear analysis of CR driven instabilities}
\label{S:LWNR}
In efficient DSA, wave-particle interaction can be strongly nonlinear where CRs modify the plasma flow and affect the specific mechanisms of magnetic field amplification \cite{Vladimirov08,Vladimirov09, Reville09}. The basic results of the linear analysis of the CR-driven instabilities were reviewed recently in \cite{Amato11,Schure12, Bell13, Bykov13} and here we exemplify some of the results. We will limit ourself to a brief review of instabilities due to anisotropic distributions of CRs and discuss incompressible modes propagating along the mean homogeneous magnetic field $\mathbf{B}_{0}$ in the rest frame of the background plasma. The situation is typical for the upstream flow of a collisionless shock moving with non-relativistic speed $\displaystyle \beta_{\rm sh} = u_{\rm sh}/c \ll$1, where the CR distribution is nearly isotropic in the rest frame of the shock.
We consider now the unperturbed anisotropic CR distribution $F^{cr}_{0}$, that is the source of the instability free energy and can be parameterized with account for two spherical harmonics as \cite{Drury83}
\begin{equation}\label{Eq:distrF0}
F^{cr}_{0}=\frac{n_{cr}N\left(x, p\right)}{4\pi}\left[1+3\beta_{\rm sh}\mu+\frac{\chi}{2}\left(3\mu^{2}-1\right)\right] \ .
\end{equation}
The multipole moments of the CR angular distribution are represented by $\beta_{\rm sh}<$1 (the dipole) and $\chi<$ 1 (the quadrupole). The unperturbed state can be a steady state of a system with CRs where both the anisotropy and the spectral distribution $N(p)$ are determined by the energy source and sink as well as the magnetic field geometry. The unperturbed state can be derived from the kinetic equation with some appropriate boundary conditions. The most interesting application of the formalism is related to the diffusive shock acceleration model \cite{Blandford1987, Malkov01, Bykov12, Schure12, Bell13}. In this case the normalized spectrum of the shock accelerated particles at the shock front position can be presented as:
\begin{equation}\label{Eq:spektrNp}
N\left(p\right)=\frac{\left(\alpha-3\right)p_{0}^{\left(\alpha-3\right)}}{\left[1-\displaystyle{\left(\frac{p_{0}}{p_{m}}\right)}^{\alpha-3}\right]p^{\alpha}},\,\,
p_{0}\leq p\leq p_{max},
\end{equation}
where  $\alpha$  is the spectral index,  $p_{0}$ and $p_{max}$   are the minimal and maximal CR momenta, respectively. In the test particle DSA model $\alpha =$4, while in the case of the efficient CR acceleration with nonlinear back reaction of the CR pressure on the shock flow the spectrum shape depart from a simple power law \cite{Jones91,Malkov01,Vladimirov08}.\\

The linear dispersion relations for CR-driven modes can be obtained by the standard perturbation analysis of the kinetic equation for CRs. The CR interactions with the magnetic fluctuations ensemble can be accounted for using the relaxation time approximation for the CR collision operator (the Coulomb collisions of CRs are almost negligible) \cite{Bykov13}. The relaxation time $\tau_s$ in the collision operator is parameterized by dimensionless value $\displaystyle a =ceB_{0}\tau_s/E=\tau_s/\tau_L$, the ratio of the relaxation time to the gyration time of the CR, where $E$ is the CR particle energy. The background plasma is treated using MHD equations. Then assuming the perturbations of the magnetic field $\delta \mathbf{B}$, plasma bulk velocity $\mathbf{u}$ and the CR distribution $F^{cr} \propto\exp\left(ikx- i\omega t\right)$, one can present the dispersion equation in the form
\begin{equation}\label{dispers1}
\frac{\omega^{2}}{V_{a}^{2}k^{2}}= 1 + \Phi_{\pm}(\omega, k, k_{0},x_{0},x_{m},\beta_{\rm sh},\chi),
\end{equation}
where the  $\pm$ signs correspond to the two possible circular polarizations defined by $\mathbf{b}= b\,(\mathbf{e}_{y}\pm i\mathbf{e}_{z})$, with the $x$-axis along the mean field $\mathbf{B}_{0}$, $k_{0}=\displaystyle\frac{4\pi}{c}\frac{j_{0}^{cr}}{B_{0}}$, $j_{0}^{cr}=en_{CR}u_{\rm sh}$, $\displaystyle x_{0}=\frac{kcp_{0}}{eB_{0}}$ , $\displaystyle x_{m}=\frac{kcp_{m}}{eB_{0}}$.\\

\begin{figure*}[t]
\centering
\includegraphics[width=0.75\textwidth, angle=90]{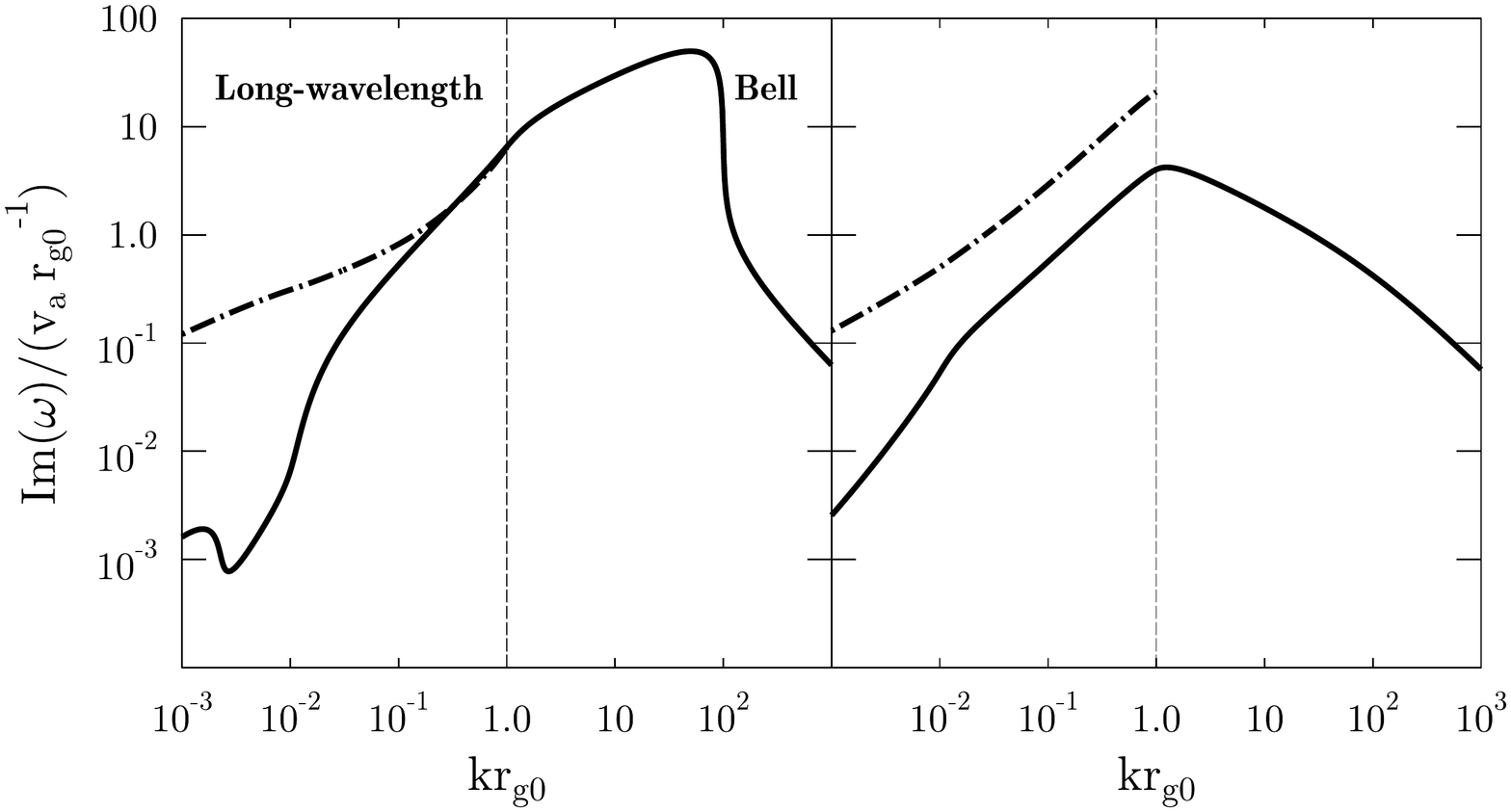}
\caption{The growth rates of the CR driven modes of two circular polarizations as function of the wavenumber for the anisotropic CR distribution given by Eqs. (\ref{Eq:distrF0}), (\ref{Eq:spektrNp}) for $p_{max}/p_{0} = 100$ and $\beta_{\rm sh}$ = 0.01. The model assumed the moderate DSA efficiency with the pressure of accelerated CRs to be about 10\% of the shock ram pressure. The right hand polarized mode (the left panel) and the left hand mode (the right panel) are propagating along the mean magnetic field. The CR-driven modes are derived from the kinetic model with account for particle scattering by waves (with the collision parameter $a$=0.1). The nonresonant instability by \cite{Bell04} with account for the firehose instability $\chi = 6 \beta_{\rm sh}^2$ are shown by solid lines. The dot-dashed line curves illustrate the growth rates of the long wavelength instability by \cite{Bykov11} for the dimensionless {\it r.m.s.} amplitude of Bell's turbulence $N_{B}=1$ (the modest case), and the mixing parameter $\xi=3$ [see for details]\cite{Bykov13}.} \label{figGrowth}
\end{figure*}

The growth rate of an unstable mode is determined by the imaginary part of the mode frequency $\gamma(k) = \Im(\omega(k))$. In the wavenumber range $k_{0}R_{g0} >kR_{g0}>1$, we recover the fast non-resonant instability discovered by \cite{Bell04} (and detailed above) where the right hand polarized mode has the growth rate
\begin{equation}\label{Bell}
\gamma_{b} \approx V_{a}\sqrt{k_{0}k-k^{2}}.
\end{equation}
The growth of the right hand polarized mode (left panel in Figure \ref{figGrowth}) is much faster than the left hand mode (the right panel in Figure~\ref{figGrowth}). The preferential growth of the right hand polarized Bell's mode may result in helicity production. On the other hand, in the weak collisions regime $a < kR_{g0}< 1$ the left hand polarized dynamo-type mode of the long-wavelength instability is growing faster than the right hand mode (see Fig.~\ref{figGrowth}). This may reduce the gross helicity production. The mode growth rate in this regime can be approximated by
\begin{equation}\label{gam_alpha}
\gamma_{lw}^d \approx 4\pi\sqrt{\xi}N_{B}V_{a}k.
\end{equation}
In the the hydrodynamical (collision dominated) regime with $kR_{g0} < a$ both circular polarizations grow with the same rate given by
\begin{equation}\label{gam1}
\gamma_{lw}^h\approx
\sqrt{\frac{\pi a \BellAmp}{2}}\sqrt{kk_{0}}V_{a},
\end{equation}
where $\BellAmp= \delta B/B_0$ is the dimensionless saturation level of Bell's short scale turbulence.

The non-monotonic behavior of the growth rate at the long-wavelength regime in Fig.~\ref{figGrowth} is due to the transition from the collisional regime of Bell's turbulence where $\gamma \propto k^{1/2}$ to the firehose instability where the growth rate scales $\gamma_{\rm fh} \propto \chi^{1/2}k$. Note that for the collisionless regime (where $a$=0) there is a dip between Bell's and the firehose branches. The dependence of the firehose growth rate on the collision parameter $a$ is discussed in \cite{Bykov13}. In Fig.~\ref{figGrowth} the growth rate of firehose instability is shown for $\chi = 6 \beta_{\rm sh}^2$. The contribution of the firehose instability to the long-wavelength fluctuations growth may be comparable to that of the current-driven if $\chi \geq \beta_{\rm sh}$. The growth rates illustrated in Fig.~\ref{figGrowth} are fast enough to highly amplify magnetic fields in a thousand years old SNR with the forward shock velocity about 3,000 $\kms$.

\subsection{Numerical simulations}
\label{S:NRsim} 
Numerical simulations play a central role in the development of modern studies of astrophysical shocks. The rapid growth in computing power over the last decade or more have made it possible to perform massive, multi-dimensional kinetic simulations of shock formation, particle acceleration, magnetic field amplification, and other phenomena. However, in the context of particle acceleration at astrophysical shocks, one is often dealing with a considerable separation of length, time and energy scales, and choosing the correct simulation for the problem at hand is important. While a full description of Maxwell's equations including particle kinetics (e.g. particle-in-cell simulations or PIC) is essential to understand the self-consistent formation of shocks and the injection of particles into the non-thermal acceleration process (see \S \ref{S:PICNR}), a fully kinetic treatment of particles ranging from eV to TeV, or more, is not currently possible. However, the large separation in energy/length scales between thermal and non-thermal particle can be taken advantage of. Discarding the kinetic description of the background plasma, i.e. adopting a fluid treatment, one can focus on the low frequency plasma modes that dominate the high-energy particles' interaction with the background fields. The high energy particles are still treated kinetically, this is the so-called MHD-PIC approach (see \S \ref{S:MHDNR}). Finally, one can relax the detailed multidimensional description of the problem to concentrate on the non-linear impact of energetic particles over the conservation laws of the flow. The kinetic treatment of energetic particles can be modeled either using a semi-analytical calculation or using Monte-Carlo techniques (see \S \ref{S:DCNR}). The two last techniques allow to discuss the maximum CR energies reached at non-relativistic SNR shocks versus shock dynamics. This information is essential to understand the way the whole cosmic-ray spectrum is built up. In this section, we describe the development of these different numerical techniques and treat various aspects of magnetic instabilities and particle acceleration occurring at non-relativistic shocks.  

\subsubsection{Injection at non-relativistic astrophysical shocks: particle-in-cell and hydrid simulations} 
\label{S:PICNR}
Particle-In-Cell (PIC) codes can model astrophysical plasmas in the most fundamental way. This is, as a collection of charged macro-particles moved by the Lorentz force. The currents deposited by the macro-particles on the computational grid are then used to compute the electromagnetic fields via Maxwell's equations. The loop is closed self-consistently by extrapolating the fields to the macro-particle locations, where the Lorentz force is computed (see \cite{Birdsall91} for a review). \\

In particular PIC and hybrid simulations are well adapted to investigate the injection of energetic particles in the DSA process, i.e. how efficiently (in number and energy) do particles participate in this process for a given set of shock conditions. Solving this problem involves understanding a series of plasma instabilities and wave phenomena occurring near the shock transition region (see \S \ref{S:M-NS} and table \ref{tab:insta}), which play a crucial role both in the generation of the shock and in the heating and acceleration of particles. Here we describe recent progresses made in the study of injection of electrons and ions in astrophysical, collisionless shocks, making used of kinetic simulations of the shocks.\\

\begin{figure}
\centering
\includegraphics[width=\textwidth]{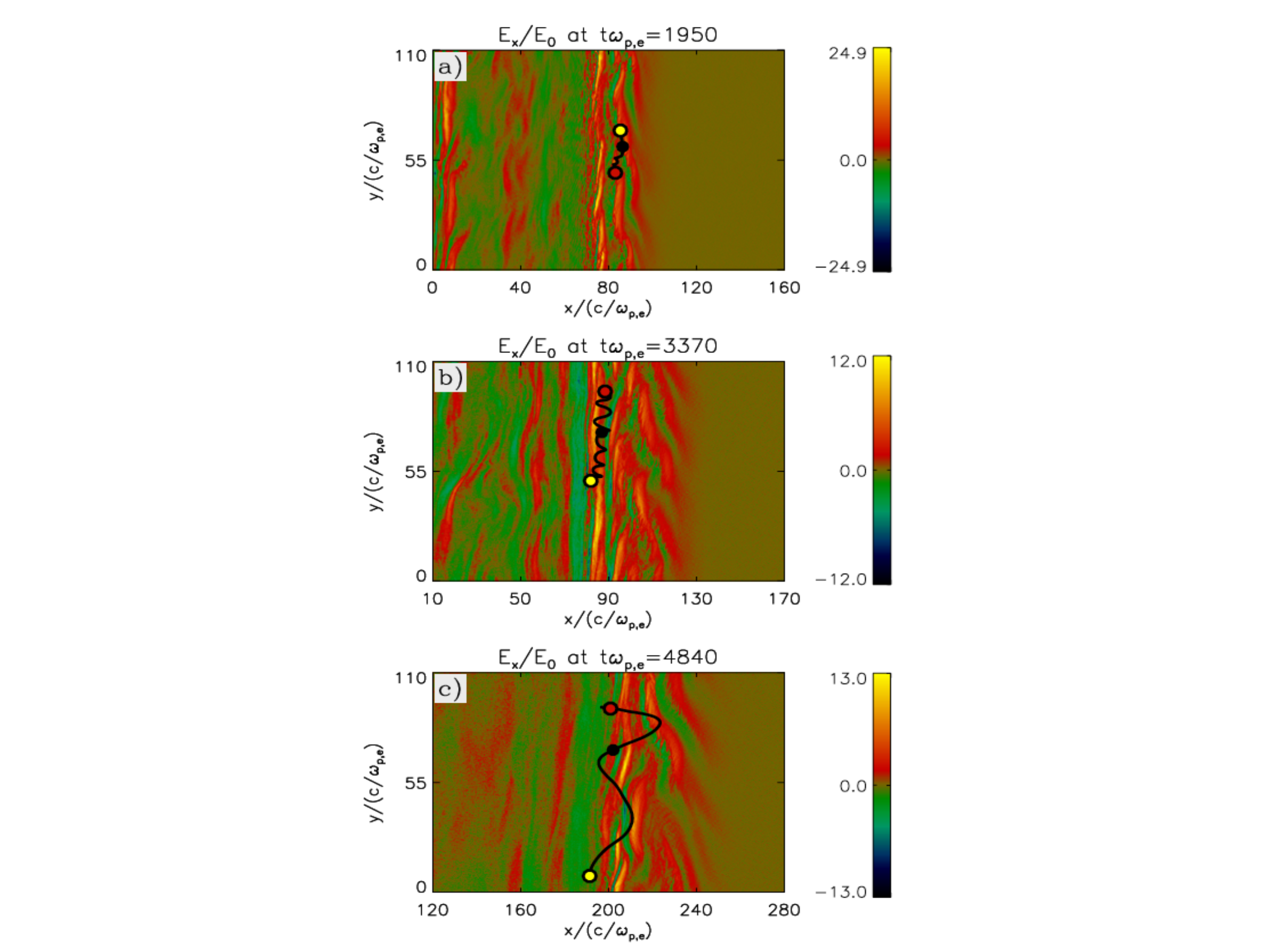}
\caption{The shock transition region at three different times for a simulation where significant electron injection occurs \cite{Riquelme11}. The color bars show the $x-$component of the electric field, $E_x$, normalized in terms of the total upstream electric field. $E_x$ is maximum at the shock overshoot. In front of the overshoot there are oblique waves that travel in front of the shock. This modes, which \cite{Riquelme11} identified as whistler modes, are key in giving suprathermal energies to a small fraction of the electrons. The black line (ended by red and yellow dots) show partial trajectories for one selected electron (the black dot shows the electron position at the time of each snapshot). The amplitude of the electron's orbit around the ambient magnetic field has increased significantly by $t\omega_{p,e}=4840$ (panel $c)$, denoting the suprathermal acceleration.}\label{fig:einject}
\end{figure}

\begin{enumerate}
\item \underline{Electron injection}: The non-thermal acceleration of electrons is ubiquitous in non-relativistic collisionless shocks.  For instance, electrons accelerated to $\sim$ keV are usually observed in interplanetary shocks in the solar neighborhood (see, e.g., \cite{Oka06}). The forward shocks of young SNRs are another site for efficient electron acceleration, as revealed by radio and X-ray observations of synchrotron-emitting, ultra-relativistic electrons (e.g. see \cite{Koyama95}). Although there seem to be agreement that the DSA is the most likely mechanism for this electron acceleration, the injection process, in particular its efficiency dependence on the shock conditions, is still an open question. 


Significant effort has been put into using PIC simulations to study electron injection in different shock regimes (see \cite{Riquelme11}, \cite{Matsumoto12}, \cite{GuoSN14a}) and a few injection mechanisms have been proposed. In the low $M_A$ ($\sim 3$), high $\beta_e$ (larger than $\sim 20$) regime, relevant to shocks in galaxy clusters and solar flares, it has been shown that significant non-thermal electron acceleration can be obtained via shock drift acceleration (SDA; \cite{Park13}, \cite{GuoSN14a}, \cite{GuoSN14b}). In quasi-perpendicular shocks, the electrons can gain energy from the motional electric field of the upstream medium (perpendicular to the upstream magnetic field). This is because the electrons tend to drift parallel to the motional electric field, given their different Larmor radii in the upstream and downstream magnetic fields. This process have been able to reasonably account for part of the X-ray spectra observed from solar flares, and the bright radio synchrotron emission observed from the outskirts of galaxy clusters. 

In the case of high $M_A$ shocks, a successful shock injection mechanism must be able to show a transition into the DSA for a significant fraction of the electrons. This implies that a sizable fraction of the electrons should, at least, reach Larmor radii ($R_{g,e}$) comparable to the one of the ions ($R_{g,i}$), since this parameter controls the width of the shock transition region. \cite{Matsumoto12} proposed that the shock surfing acceleration (SSA) would be an effective mechanism to inject electrons in high $M_A$ ($> (m_i/m_e)^{2/3}\approx$150), quasi-perpendicular shocks. In the SSA the non-thermal electrons gain energy from the motional electric field of the upstream, being helped by efficient scattering provided by the Buneman instability. In this case, the Buneman modes grow in the shock foot and are driven by counter-streaming ions that, instead of being thermalized at the shock, bounce and propagate into the upstream medium (by a distance close to the ion Larmor radii). Although this mechanism can produce non-thermal electrons, it has not shown to produce electrons with $R_{g,e}\sim R_{g,i}$.

An alternative mechanism, based on the excitation of whistler waves in quasi-perpendicular shocks, has shown to accelerate non-thermal electrons up to $R_{g,e} \sim R_{g,i}$ \cite{Riquelme11}. Similar to the Buneman modes in the case of the SSA, whistler waves can be excited by the modified two-stream instability (MTSI; \cite{Wu83}, \cite{Matsukiyo03}, \cite{Matsukiyo06}) driven by counter-streaming ions in the shock foot (see Figure \ref{fig:einject}). This mechanism has been able to produce a non-thermal, energy spectrum with a power-law tail of index $\alpha \approx -3$. The acceleration is most efficient in the case of $M_A$ smaller than $\sim 20$, which is consistent with the growth condition of the MTSI (where $M_A$ needs to be smaller than $(m_i/m_e)^{1/2}$). 

Both in-situ measurements of interplanetary shocks and the PIC simulations results presented by \cite{Riquelme11} suggest that quasi-perpendicular shocks constitute a suitable environment for the acceleration of electrons in non-relativistic shocks only for moderately low $M_A$ (less than $\sim 20$). This is opposed to the case of ion injection, which would happen most efficiently in high $M_A$, quasi-parallel shocks (see next point and \cite{Caprioli14a}). It is important to notice, however, that the acceleration of ions in these environments can render the shock suitable for electron injection. Indeed, ion acceleration can strongly amplify the shock magnetic field and change its direction, possibly transforming it into an {\it effectively} low Mach number, quasi-perpendicular shock in some regions \cite{Caprioli14b}. This way, electron injection by whistler waves could be at work even in {\it globally} high $M_A$, quasi-parallel shocks, making these shocks efficient accelerators of both ions and electrons. Finding out whether this scenario is correct will require further study of the electron injection problem. In the case of injection due to whistler waves, additional work is still required to show the transition into the DSA, where the electrons move diffusively in the shock vicinity and a $f_e(p) \propto p^{-4}$ distribution function dominates. Also, further study is needed to determine whether high $M_A$, quasi-parallel shocks can also act as efficient electron accelerators (as observationally suggested by \cite{Masters13} and theoretically proposed by \cite{Levinson92} and \cite{Levinson94}). 

\item \underline{Ion injection}:
Significant attempts to capture the physics of ion injection have been made using full particle-in-cell (PIC) simulations (e.g., \cite{Niemiec12}). However, the need to simultaneously resolve the electron- and ion-scale physics in PIC simulations makes the modeling of ion injection numerically challenging if moderately realistic ion to electron mass ratios $m_i/m_e$ are used. In order to overcome this difficulty, significant effort has been made in using kinetic hybrid simulations, where ions are modeled as particles while electrons are treated as a massless fluid (see, e.g., \cite{Gargate12}, \cite{Guo13}, \cite{Caprioli14a}, \cite{Caprioli14b}). Since in this case the electron-scale physics does not need to be resolved, hybrid simulations are computationally more efficient than PIC simulations, while being able to capture essentially the same ion acceleration physics in most cases.\newline

\noindent The hybrid studies have paid especial attention to the dependence on the upstream $\vec{B}$ field angle $\theta$ (with respect to the shock propagation velocity, $\vec{V}_{sh}$) and the shock Alfv\'enic Mach number $M_A$ ($\equiv V_{sh}/V_A$, where $V_A$ is the upstream Alfv\'en velocity). Especially encouraging has been the recent results presented by \cite{Caprioli14a} who have obtained efficient injection of non-thermal ions (with a maximum of $\sim 15\%$ in energy). The obtained injection is most efficient for quasi-parallel ($\theta < 45^{o}$), high $M_A$ shocks (the maximum $M_A$ studied was 100). Remarkably, a power-law tail consistent with the DSA theory, $f_i(p) \propto p^{-4}$, where $p$ is the non-thermal particle momentum, was obtained. Also, the authors found evidence for shock modification due to the dynamically important non-thermal particle pressure. These modifications include  an increase in the shock compression ratio $r$ to values beyond the standard Rankine-Hugoniot conditions ($r > 4$), as well as the formation of a shock precursor with significant magnetic field amplification due to instabilities driven by the streaming non-thermal particles (with amplification factor proportional to $M_A^{1/2}$; \cite{Caprioli14b}).\newline

\noindent Although these results constitute a significant advance in the study of ion injection in non-relativistic astrophysical shocks, the actual mechanism by which a small fraction of the ions are injected into the DSA is still unclear (although there seems to be evidence that the ions are injected from the shock transition region - instead of "leaking" from the downstream region, as also discussed by \cite{Guo13}).
\end{enumerate}

Note that recent long-term and large-scale 1D PIC simulations have tried to capture the main process controlling particle injection at parallel shocks. The injection of protons is related to phase-trapping by finite amplitude waves upstream \cite{Kato15} or associated with SDA \cite{Park15}. Both latter works find Fermi-like particle acceleration in the non-thermal regime for protons and electrons. Non-resonant hybrid (Bell) modes are observed in the shock CR precursor by \cite{Park15}. They can trap electrons and contribute to inject them in the relativistic regime.

\subsubsection{Coupled kinetic and magnetohydrodynamic simulations: large scale driven instabilities and maximum CR energies}
\label{S:MHDNR}
In order to access the full range of scales, a full kinetic description of the CR distribution, and its evolution, is required. So far, this has been achieved by a number of methods. One such approach is to use the techniques of particle-in-cell codes to treat the CRs. This determines the cosmic-ray current at each time step, which is then used to update the MHD fluid equations. \cite{Zachary86, Zachary89} used a one-dimensional hybrid MHD-PIC code and carried out detailed investigations on the growth of beam plasma instabilities, gyro-resonant particle interactions, trapping, and the break-down of quasi-linear theory. In \cite{Zachary87}, it was demonstrated that Alfv\'en waves were indeed amplified, consistently with the estimates from linear theory, but in all cases, the total field density was found to saturate at a rather modest level $\left\langle \delta B_\bot/B_0\right\rangle_{\rm rms} \lesssim0.5$. It was not until more than a decade later, that the technique was revisited by \cite{Lucek00}, who demonstrated with multidimensional simulations, that magnetic fields could in fact be amplified to values exceeding the initial seed field, if enough free energy was available in the CRs. Narrow X-ray filaments along the outer shocks of several young SNRs were discovered around the same time (section \S \ref{S:Xobs}) providing direct evidence of nonlinear ($\delta B \gg \langle B_{\rm ISM}\rangle$) magnetic field amplification at shocks. As magnetic field strength had long been seen as a limiting factor for acceleration beyond the knee in the cosmic-ray spectrum, this realization was of enormous significance to the cosmic-ray/particle acceleration community.

However, a major short-coming of these hybrid MHD-PIC simulations remained the finite free energy available in the systems. Since the numerical constraints at the time limited the computational domain to modest sized periodic boxes, full shock simulations were not possible, and free-energy in the streaming particles could not be replenished self consistently. However, soon after, \cite{Bell04} identified a non-resonant instability, growing on short (sub-Larmor) wavelengths (see \S \ref{S:NRth}). In this case, since the fields are evolving on a scale  $\ell \ll  f/|df/dx|$, the cosmic-ray current can be held fixed and uniform. This method avoids the problems associated with finite free-energy, and has the advantage of being easily implemented into any standard MHD code. This approach has been used extensively in the last decade for both MHD models \cite{Bell04, Bell05, Reville08, Zirakashvili08, Rogachevskii12, Bai15} and also PIC and hybrid models \cite{Ohira09, Riquelme09, Gargate10, Caprioli14b, Kato15, Park15}. These simulations have played a central role in demonstrating the possibility of strong non-linear amplification of magnetic fields in the precursors of supernova remnants. However, while the growth of this instability is rapid and indeed capable of significantly amplifying the magnetic fields, due to the short wavelength of the fastest growing linear mode, it is not immediately obvious that this mechanism has a significant effect on the acceleration of particles to higher energies. It is well-known that energetic particles will interact most effectively with magnetic structures that have a size comparable to their gyroradius. This places an interesting constraint on the maximum energy particles can be accelerated to if they resonantly excite their own scattering waves.

In the precursor of a young SNR, the background fluid is well approximated as an infinitely conducting plasma, and as such the magnetic field is completely tied to the background fluid. Following the argument in \cite{Bell13b}, in order to scatter resonantly a particle of energy $E$, a plasma fluid element or magnetic field line must be displaced by a distance comparable to the gyroradius of this particle. It is reasonable to assume that magnetic tension, and thermal pressure gradients are unimportant, and it follows that the only force acting on the background, is the CR Lorentz force 
\begin{equation}
\rho\frac{d\bf{u}}{d t}=-\frac{{\bf{j}'\bf_{\rm cr}} \times \bf{B}}{c} \ .
\end{equation}
Assuming all quantities are slowly varying, the maximum displacement in a time $t$ is
\begin{equation}
s_{\rm max}\sim \frac{{j'}_{\rm cr}B_0t^2}{2\rho c}.
\end{equation}
Defining an energy conversion efficiency parameter $j_{\rm cr} E/e = \eta \rho u_{\rm sh}^3$, and equating this displacement to the gyroradius of a particle of energy $E$, it follows that the maximum energy is  
\begin{equation}
E \sim 150 (\eta u_9^3)^{1/2}B_{\rm \mu G} t_{100}~{\rm TeV}
\end{equation}
where $u_9$ is the shock velocity in units $10^9$ cm/s, $t_{\rm 100}$ time in units of 100 years, and $\eta$ is generally thought to be at most on the order of a few percent. This falls very short of the energies required to explain the galactic cosmic-ray spectrum. Fortunatly there are a number of ways around this problem.

One approach is to rely on non-local field amplification, such as the cosmic-ray filamentation instability investigated in \cite{Reville12}. This instability couples the growth of magnetic structures on small scales to those on large scales, and appears to be a natural consequence of the non-resonant streaming instability. As localized regions of plasma expand under the action of the $\bf{j}_{\rm cr}\times\bf{B}$ force, they push on neighbouring expanding regions. The net effect is that, any loop enclosing these two, or indeed multiple regions, must also expand resulting in field growth on large scales. This instability has been studied using 2D MHD-PIC simulations, and appears also to occur in Hybrid simulations \cite{Caprioli13}. This type of instability is in some sense similar to the negative effective-pressure instabilities often discussed in mean-field dynamo theory \cite{Kleeorin93, Kemel12}.

Another possibility, is that the non-resonant streaming instability, in its non-linear evolution, naturally grows to larger length-scales. This was already evident in the first simulations of the instability \cite{Bell04}, where in the early stages of the non-linear evolution, the magnetic field structures grew to the scale of the simulation box. However, investigating this multi-scale problem, in three dimensions is numerically challenging. An approach is to solve numerically the Vlasov-Fokker-Planck (VFP) equation. While, in principle, this involves solving an equation in 6+1 dimensions, spectral methods can be used to reduce the number of dimensions required in momentum space, as is commonly done in the laser-plasma community \cite{Bell06}. Since most problems of interest in CR physics involve small departures from isotropy, it is desirable to select the spectral-basis functions such that this geometry is handled most efficiently. A particularly effective method, is to use spherical harmonics to expand  the momentum space distribution, 
\begin{equation}
f(\bf{x},\bf{p},t) = \sum_{\ell=0}^\infty\sum_{m=-\ell}^\ell f_\ell^m(\bf{x},p,t) 
P_\ell^{\vert m\vert}(\cos\theta) e^{im\varphi}
\end{equation}
since these functions form a natural basis for quasi-isotropic distributions. Using the orthogonality relations of the spherical harmonics, the VFP equation is reduced to a system of coupled differential equations, that can be truncated after a finite number of terms, depending on the problem at hand \cite{Reville13}. The transport equation, is in fact a simplification of the zeroth and first order equations in this expansion. Using this technique, it is possible to perform accurate 3D simulations that resolve both the gyroradius of the particles, and the fastest growing modes, using relatively modest computing power. This has resulted in the first genuine demonstration of sub-Bohm diffusion, with respect to the ambient magnetic field, of high energy particles in self-generated fields \cite{Reville13} \footnote{Sub-Bohm diffusion of lower energy particles, resonating with the short-wavelength modes of the non-resonant streaming instability has previously been shown \cite{Reville08}, but this does not help accelerate beyond the CR knee.}. The growth of fields, initially on scales small compared with the gyroradius of the driving particles, accumulate and concentrate in localized volumes with sufficient amplitude to scatter particles through large angles. The time required to scatter through $90^\circ$ is thus considerably reduced, and acceleration to higher energies is possible. However, full shock simulations are necessary to determine the global effect.

To this end, \cite{Bell13b} have performed hybrid MHD-VFP simulations using a novel expansion closure technique, that truncates after 2 terms. The reduced formalism was necessary in order to minimize the memory requirements, while capturing all the essential physics. The simulations track both the acceleration of particles at a strong planar shock, and the self-generation of magnetic fluctuations. The upstream plasma is initialized with small magnetic perturbations that are insufficient to confine the cosmic-rays, and as such must be amplified by the escaping current. The current is strong enough to trigger the growth of the non-resonant streaming instability. The authors argue that approximately 5-10 e-folding times of the non-resonant streaming instability are required to confine the particles:
\begin{equation}
\label{eqn:escapecond}
 \int \gamma_{\rm NRI} dt = \sqrt{\frac{\pi}{\rho c^2}}\int j_{\rm cr} dt \sim 5 
\end{equation}
i.e. the requirement for particle trapping depends solely on the total areal charge that has traversed a given fluid element over the lifetime of the shock. Hence, the escape of cosmic rays from supernova remnants is essential for the acceleration of yet higher energy particles. The escaping flux follows directly from the acceleration theory, where, assuming isotropy near the shock, the flux upward in momentum associated with the shock crossings is 
\begin{equation}
 \phi(p)  \approx \frac{4\pi p^3}{3}f(p)(u_u - u_d) 
\end{equation}
If particles at, or close to, the maximum energy are not confined, the flux can be associated with the escape of cosmic-ray protons {\it upstream} of the shock, which for a strong shock with compression ratio 4 gives 
\begin{equation}
j_{\rm cr} = e \pi p_{\rm max}^3 f(p_{\rm max})u_{\rm sh} \enspace.
\end{equation}
Inserting into Eq.(\ref{eqn:escapecond}), it has been shown that the resulting value for total areal charge traversed by a fluid element agrees remarkably well with the results from simulations \cite{Bell13b}.

Combining these terms, it is possible to evaluate an expression for the maximum energy particle based on the above confinement constraint. For a power-law cosmic-ray spectrum $f(0,p) \propto p^{-4}$, Eq.(\ref{eqn:escapecond}) can be reformulated to give an expression for the maximum energy
\begin{equation}
\label{Eq:escapethird}
E_{\rm max} = p_{\rm max}c \sim  
\frac{3\sqrt{\pi}}{4}\frac{e}{5\sqrt{\rho} c}\frac{P_{\rm cr}}{\rho u_{\rm sh}^2} 
\frac{\rho u_{\rm sh}^3}{\ln (p_{\rm max}/mc)} t
\end{equation}
where $P_{\rm cr}$ is the cosmic ray pressure at the shock. This form for the maximum energy, in contrast to both the Hillas (geometry) condition, and Lagage-Cesarsky (time) limit, is now completely independent of the strength of the magnetic field, and relies purely on the ability of the escaping cosmic-ray flux at maximum energies to self-confine. Hence, the acceleration of the highest-energy cosmic rays is found to proceed in a self-similar fashion, where the highest-energy particles escape into the upstream, ultimately self-confining and facilitating acceleration to ever higher energies.

On the topic of faster shocks, it should not be forgotten that the standard non-relativistic theory is accurate only to order $u/c$. As the shock velocity becomes a non-negligible fraction of the speed of light, higher order terms can become important, particularly at oblique shocks, where particle trajectories can change abruptly across the shock surface. VFP simulations have also been used to study this effect \cite{Bell11}. Shock acceleration simulations were performed in one dimension, where the scattering rate, shock velocity and magnetic field angle were varied as free parameters. The resulting asymptotically steady-state solutions confirmed that even in the test particle regime, large deviations from the standard theory could be achieved. For superluminal shocks, it was found that there was a general trend of spectral steepening with shock velocity, as particles are carried downstream on the field lines. At intermediate obliquities, the acceleration is more efficient due to the combined actions of shock drift and shock acceleration. The spectrum can in fact flatten in this case, depending on the scattering rate. The primary reason for the departure from the standard diffusion theory, is the inability to match the oblique drifts produced at non-parallel shocks, across the fluid discontinuity in the diffusion approximation. These drifts must therefore persist in the immediate downstream of the shock, gradually damp towards the asymptotic state. This damping can lead to a decrease or increase in the isotropic part of the distribution. As pointed out in \cite{Kirk96}, the difference between the distribution at the shock and downstream infinity modifies the power-law spectrum 
\begin{equation}
\frac{\partial \ln f}{\partial \ln p} = -3\left[1+\frac{f(\infty)}{f(0)(r_{\rm c}-1)}\right]
\end{equation}
where $r_{\rm c}$ is the shock compression ratio. The difference between $f(\infty)$ and $f(0)$ involves some, as yet unknown, dependence on shock velocity, obliquity and scattering rate. Further hybrid simulations are required to investigate this behaviour in a more self-consistent manner. Preliminary work has already demonstrated that the oblique drifts can damp in the upstream in the presence of strong magnetic field amplification \cite{Reville13}, suggesting that the drifts might not be important. However, particle trapping in magnetic bottlenecks can steepen the spectrum quite considerably \cite{Kirk96, Bell13b}. Full shock simulations are required to make genuine quantitative predictions for observational signatures.

\subsubsection{Semi-analytical calculations and Monte-Carlo approach: magnetic field amplification and non-linear DSA}
\label{S:DCNR}
Although the problem of DSA including efficient CR back reaction is non-linear some semi-analytical calculations are possible in 1D in space (and 1D in momentum) \cite{Eichler84, Blasi02}. The model involves the conservation laws of hydrodynamics including the CR pressure coupled to diffusion-convection equation describing the evolution of CR in the phase space. The solution of the fluid velocity profile upstream can be found after several iterations adapting the conservation laws and the solution of the kinetic equation. Recent developments involved the inclusion of an equation for the resonant waves \cite{Caprioli09a} as well as a far escape boundary upstream that mimics the energy losses produced by the escape of the highest CR from the shock precursor \cite{Caprioli09b}. In the same spirit, Monte-Carlo simulations have been developed that calculate the solutions of the particle distribution \cite{Ellison90}. Here again recent developments included an equation for the waves that can be resonant or not \cite{Vladimirov09, Bykov14} and the investigation of the transition towards relativistic shock waves \cite{Ellison13}.

\begin{figure*}[t]
\centering
\includegraphics[height=0.4\textheight, width=0.8\textwidth]{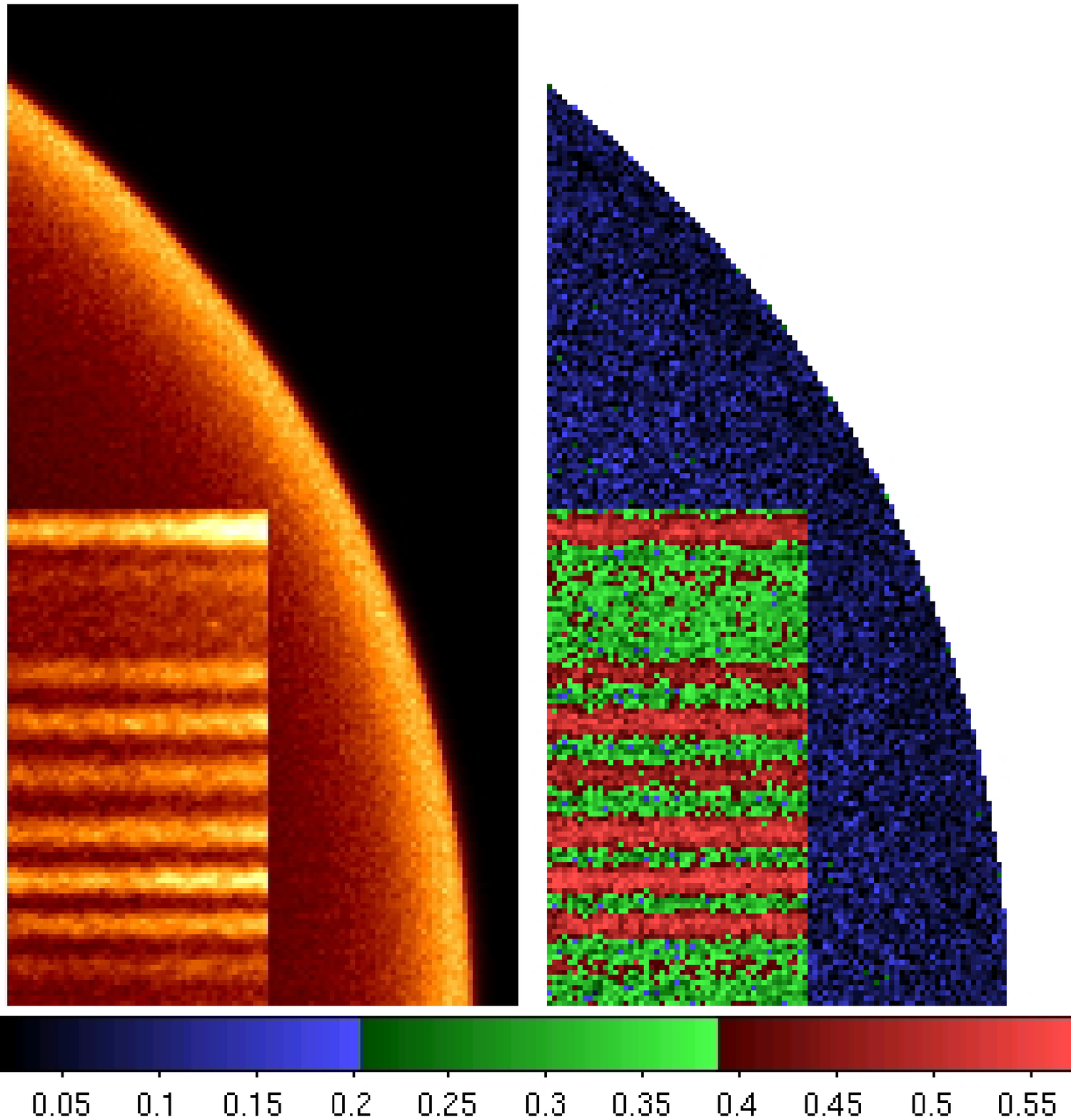}
\caption{Tycho SNR polarized synchrotron X-ray intensity at 5 keV see \cite{Bykovlett11}. The degree of polarization of the X-ray emission is shown in the color bar. The relatively high polarization fraction is mainly due to the peaked structure of the magnetic fluctuation spectrum and the steepness of the distribution of \syn\ emitting electrons.} \label{F:stripes-th}
\end{figure*}

In particular, the model can reproduce synchrotron structures consistent with the stripes observed by \chan\ in Tycho's SNR (see figure 1 in \cite{Bykovlett11} for an explanation of the stripes geometry). These stripes are the most likely results of CR generated magnetic turbulence at the SNR blast wave (see \S\ref{S:NRth}). The amazingly regular pattern of these stripes that appear in a number of shock-plasma phenomena must be in action simultaneously. The coherent appearance of the X-ray stripes suggests that the underlying magnetic turbulence is strongly anisotropic. (Isotropic turbulence would not produce extended coherent structures with thin stripes.) Both the Bell short-wavelength instability and the long-wavelength instability (see \S \ref{S:LWNR}), produce anisotropic turbulence with a prominent growth-rate maximum along the mean ambient magnetic field direction. The local ambient mean magnetic field geometry determines the orientation of the stripes and therefore it can be reconstructed with the high resolution X-ray imaging (see figure \ref{F:stripes-th}). The turbulent energy cascading spreads out the peaks in turbulence power and eliminate the key feature needed to produce an ordered pattern of stripes. Therefore, such stripes can form if the turbulent cascading along the mean magnetic field is quenched. Stripe-like structures should form in a section where the local field lies along the shock surface and where the turbulence cascading is suppressed. The stripe structure in synchrotron images requires narrow peaks in the magnetic turbulence in a perpendicular shock that can be understood in the frame of non-linear DSA model. The estimated maximum energy of the CR protons responsible for the strips is $\sim 10^{15}$\,eV. The model by \cite{Bykovlett11} (see also a recent work by \cite{Laming15}) also predicts a specific X-ray polarization pattern of the strips, with a polarized fraction $\sim$ 50\%, which can be tested with future X-ray polarimeter missions. \\

Turbulent magnetic fields, with energy densities approaching a substantial fraction of the shock ram pressure, are a generic ingredient in the efficient DSA mechanism. The CR-driven instabilities discussed in \S\ref{S:NRth} producing strong fluctuating magnetic fields in a broad dynamical range are a promising way to provide the required magnetic field amplification mechanism. The question arose how the strong turbulent fluctuations affect the observed X-ray images and spectra? The effect of magnetic turbulence on synchrotron emission {\sl images} and spectra was addressed in \cite{Bykov08, Bykov09, Stroman09}. Variable localized structures (dots, clumps and filaments), in which the magnetic field reaches local high values arise in the random field. The magnetic field concentrations dominate the synchrotron emission (integrated along the line of sight) from the highest energy electrons in the cut-off regime of the electron distribution. The resulting image has an evolving clumpy appearance. The simulated structures resemble those observed in X-ray images of some young SNRs. The spectral shape of the synchrotron radiation from the cut-off regime in the electron spectrum is strongly modified by the redistribution of photons towards the highest energies in a turbulent field compared to emission from a uniform field of the same magnitude \cite{Bykov08, BykovPavlov12}. This effect should be accounted for estimating of the maximal energies of accelerated electrons.

\subsection{Supernova remnants and the origin of galactic cosmic rays}
\label{S:NRcr}
SNRs are known as possible sources of CRs since long (the main arguments can be found in the reviews by \cite{Ginzburg64, Drury01}). Recent observations prove that electrons are accelerated at the forward shock of the SNR blast waves up to TeV energies (see \S \ref{S:Xobs}). Gamma-ray detection (section \S \ref{S:GSNR}) either support this conclusion or point towards the acceleration of a hadronic component that can contribute to the CR spectrum observed on Earth. But there are not definite observational proof yet that CRs are accelerated in SNRs (see \cite{Morlino12b} though).

The theoretical progresses made on the understanding of MFA have led to the emergence of a scenario of particle acceleration in SNR. The possibility to reach magnetic strength two orders of magnitude above the mean ISM values should produce higher energies and help to reach the CR knee at $3\times 10^{15}$ eV \cite{Ptuskin03}. However the efficiency of DSA including MFA in the non-linear regime involves complex plasma physics that require multiple scales and multiple numerical techniques approaches as exposed in \S \ref{S:NRsim}. In particular, the properties of the self-generated turbulence are important to constraint the maximum energies \cite{Bykov14}. If the main driving process of the MFA is the non-resonant streaming instability hence the Eq.\ref{Eq:escapethird} can be used inserting numerical parameters typical for nearby Galactic SNRs where measurements of MFA have been made:
\begin{equation}
E_{\rm max} = 10^{13} \left(\frac{P_{\rm cr}}{\rho u_{\rm sh}^2}\right) \times \frac{\sqrt{n}~u_{sh,8}^3~ t_{100}}{\ln (p_{\rm max}/mc)} ~{\rm eV} \ .
\end{equation}
With typical shock velocities on the order of a few thousand km/s (or $10^8$ cm/s), and age of a few hundred years, we see that even for acceleration efficiencies as large as $P_{\rm cr}/\rho u_{\rm sh}^2 = 0.5$,  the above estimates would imply that these remnants are not currently accelerating cosmic rays to PeV energies. However, the strong scaling with shock velocity suggests that younger remnants, particularly those expanding in a dense wind may in fact be the primary source of cosmic rays above the knee \cite{Bell04, Schure13, Marcowith14}. The models predict a charge dependent CR knee. A trend that seem to be compatible with the very last experiments results \cite{Fuhrmann13}. If most of the multi-PeV CRs are accelerated in the very early timescales of the SNR history then their detection at multi-hundred of TeV in gamma-rays may be difficult even with the next sensitivity improved instruments like the Cerenkov Telescope Array \cite{Cristofari13}.

Even if one could be more optimistic about the performances of SNRs to produce high-energy CR up to the knee it remains quite difficult to explain the component extending up to energies of $10^{17-18}$ eV where the extragalactic contribution takes over (see \cite{Blasi14}). One possibility is that some particular extreme events involving mildly relativistic flows could produce such highly energetic particles (see \S \ref{S:Ruhecr}).

\section{Mildly to ultra relativistic shock waves}
\label{S:RS}

\subsection{Observational clues on the magnetic field up- and downstream of relativistic shock waves}
\label{S:Robs}
The afterglow spectra radiated by a population of electrons accelerated at the external shock of a gamma-ray burst outflow provide one of the best observational probes of acceleration physics in the ultra-relativistic regime. In the standard scenario \cite{Rees92, Paczynski93, Katz94, Meszaros97, Waxman97, Sari97, Vietri97}-- see also the review \cite{Piran04} and detailed analytical estimates in \cite{Panaitescu00}-- the collisionless relativistic blast wave is formed as the outflow with bulk Lorentz factor impinges on the circumburst medium, generally considered to be either the interstellar medium of the host galaxy or the wind of the progenitor star. Due to the large density contrast between the blast wave and the circumburst medium, the forward shock propagates with Lorentz factor $\gamma_{\rm sh}\simeq \sqrt{2}\gamma_{\rm ej}$, while the reverse shock propagates back into the outflow at non- or mildly relativistic velocities. The blast wave picks up and shock-accelerates the circumburst medium electrons, leading to the appearance of synchrotron (and possible inverse Compton) spectrum, giving rise to the so-called ``afterglow''.

\subsubsection{The upstream field}
Li \& Waxman have noted in \cite{LiWaxman06} that X-ray afterglow observations offer the possibility to probe the magnetic field upstream of the blast through the time at which the characteristic frequency associated to the electrons of the maximum Lorentz factor exits the X-ray domain. The maximum Lorentz factor is determined by the competition between acceleration and losses and the detailed analysis of \cite{LiWaxman06} indicates that for an upstream magnetic field $B\sim1\,\mu$G, a density $n\sim1\,$cm$^{-3}$, the maximal frequency falls short of the X-ray domain on a day timescale. The detection of X-ray afterglows on longer timescales should thus point to amplification of the magnetic field beyond the interstellar value. In a few cases, their analysis leads to:
\begin{equation}
B\,>\, 200\,\mu{\rm G}\, n_0^{5/8}\ ,\label{eq:LW1}
\end{equation}
which can be rewritten in terms of the upstream magnetization as
\begin{equation}
\sigma_{\rm u}\,>\,10^{-6}n_0^{1/4}\ .
\end{equation}
This value lies well below the expected value for Weibel turbulence, $\sigma_{\rm u}\sim 10^{-3}-10^{-2}$, but well above the typical interstellar magnetization $\sigma_{\rm ISM}\,\sim\,10^{-9}$.

More recently, \cite{LiZhao11} has revisited this constraint by considering the extended GeV emission seen in a fraction of gamma-ray bursts by the Fermi-LAT instrument. This extended GeV emission is detected up to $10^3\,$s and its main characteristics (spectral index, flux level and slope of the light curve) strongly point towards synchrotron emission from the forward shock. Requiring that the afterglow can contribute to energies $>100\,$MeV as late as $10^3\,$s then leads, following an argument similar to the above, to a bound that is slightly stronger, by a factor of a few to ten in $B_{\rm u}$, depending on the value of the downstream magnetic field. Other groups have conducted similar analyses and find values in rough agreement with the above, up to some differences in the choice of the parameters \cite{Barniol11, Sagi12}: \cite{Barniol11} finds that $B\sim1\,\mu$G cannot be excluded for GeV afterglows, however the corresponding density is also much smaller than unity, leading to a magnetization well in excess of the interstellar value, while \cite{Sagi12} finds no strong evidence for amplification, but using a somewhat unrealistic Bohm assumption for the acceleration timescale.

In fine these results indicate a magnetic field significantly stronger than the interstellar value, upstream of a relativistic blast wave. Of course, it is tempting to attribute this apparent high magnetization to streaming instabilities triggered by the accelerated particle population penetrating the ambient plasma. However, at the present time, one cannot exclude that the circumburst medium is strongly magnetized to the above level. This might happen if, for instance, the gamma-ray burst explodes in a wind with a sub-equipartition magnetic field: e.g. a magnetic field $B\sim 10^3\,{\rm G}\left(r/10^{12}\,{\rm cm}\right)^{-1}\,$ remains dynamically unimportant in a wind with standard mass loss $\dot M_{\rm w}\sim 10^{-5}\,M_\odot/{\rm yr}$, $v_{\rm w}\,\sim\,10^3\,$km/s~\cite{Ignace98}, but leads to a magnetization $\sigma_{\rm u}\sim 10^{-4}$; such a scenario would produce radical signatures in the light curves, as discussed in \cite{Lemoine11}, because Fermi acceleration becomes inefficient at such high magnetization and large Lorentz factors, see \S \ref{S:PARMS} of this report.

\subsubsection{The downstream field}
The results of \cite{LiZhao11} depend somewhat on the value of the downstream magnetic field, which has become more uncertain with recent data. While early determinations have led to values $\epsilon_B \sim 10^{-2}$ (with a large uncertainty) ($\epsilon_B = \delta B^2/\left(16\pi\gamma_{\rm sh}^2 nm_pc^2\right)$ gives the equipartition fraction of energy in the magnetic field, downstream of the shock), e.g. \cite{Waxman97, Wijers99, Panaitescu01, Panaitescu02}, the standard adiabatic synchrotron interpretation of GeV extended emission seems to indicate low values of order $\epsilon_B \sim 10^{-6}$~\cite{Kumar09, Kumar10, Barniol11, Barniol11b, He11, Liu11}. As discussed in \cite{Lemoine13, Lemoine13b}, this anomalous value of the magnetization may actually point to the decay of the shock-generated Weibel micro-turbulence, away from the shock front; this issue is also briefly discussed in \S \ref{S:RR}.

While our current understanding of shock formation indeed suggests the existence of intense turbulence with $\epsilon_B\,\sim\,10^{-2}$ behind the shock, such turbulence exists on plasma scales $\lambda\,\sim\,10\,c/\omega_{\rm pi}$ and as a consequence, it should decay on some multiples of $\lambda$~\cite{Gruzinov99}; this decay has been seen in state-of-the-art PIC simulations~\cite{Chang08, Keshet09, Medvedev11}. Given that the plasma is advected away from the shock at velocity $c/3$, the damping time $\tau$ is effectively a damping length $c\tau/3$ measured relatively to the location of the shock front. Then the question is how to sustain a magnetic turbulence with $\epsilon_B\,\sim\,10^{-2}$ on the whole width of the blast, which spans some $10^9$ skin depth scales at an observer timescale $t_{\rm obs}\,\sim\,10^5\,$s~\cite{Gruzinov99}? The ``decaying micro-turbulence'' interpretation of the low values of $\epsilon_B$ derived in GRB afterglows with extended GeV emission offers a simple solution to this problem~\cite{Lemoine13, Lemoine13b}. For an order of magnitude estimate, if $\epsilon_B$ decays from $10^{-2}$ in the shock vicinity down to $10^{-6}$ at the back of the blast, on some $10^8-10^9$ skin depths scales, this suggests that $\epsilon_B\,\propto\, (x\omega_{\rm pi}/c)^{\alpha_t}$ with $\alpha_t\,\sim\,-0.5$, $x$ denoting the distance to the shock front.

Interestingly, this scaling agrees with the PIC simulations of ~\cite{Keshet09}. Finally, as stressed in ~\cite{Lemoine13b}, the advantage of using GRBs with extended emission is to be able to determine the four afterglow parameters (blast energy, external density, $\epsilon_B$ and $\epsilon_e$) using four constraints with four wavebands (radio, optical, X-ray, $>100\,{\rm MeV}$); in contrast, earlier estimates have generally used only three wavebands, implying that the $\epsilon_B$ parameter was poorly constrained due to implicit degeneracies in the models.

Beyond shock-generated Weibel turbulence, additional sources may nevertheless be envisaged downstream of the shock: (1) the interaction of the shock with inhomogeneities in the circumburst medium may give rise to the Richtmyer-Meshkov instability, whereby a small scale dynamo effect amplifies the background magnetic field \cite{Sironi07, Inoue11}; (2) a large scale Rayleigh-Taylor instability at the contact discontinuity may also pollute the shocked circumburst medium with magnetized plasma from the shocked gamma-ray burst ejecta \cite{Levinson09}; (3) recent hydrodynamical simulations of a gamma-ray burst jet have shown that instabilities propagate in the blast from the boundaries of the jet, suggesting a possible new source of magnetic field amplification \cite{Meliani10}. Whether any of these instabilities can pump the magnetic field up to a persistent $\epsilon_B\,\sim\,10^{-2}$ on the scale of the blast remains an open issue.

\subsubsection{Trans-relativistic supernovae}
\label{S:Trel}
Finally, one should point out the recent measurement of the magnetic field strength in the blast of a trans-relativistic SNIbc supernova, SN2009bb~\cite{Soderberg10, Chakraborti11}, leading to values $B\,\sim\,1\,{\rm G}$ about 20 days after explosion! This estimate was obtained through the detection and follow-up of the synchrotron self-absorption frequency in the radio range, combined with equipartition arguments between the electrons and the magnetic fields \cite{Barniol13}.

This measurement is of particular importance in the context of the origin of ultra-high energy cosmic rays; the simultaneous estimate of the size of the blast and of the magnetic field indeed allows to estimate the confinement energy of particles in such trans-relativistic supernovae: $E_{\rm conf}\,\sim\, 6\times 10^{19}\,{\rm eV}\,(Z/26)$ \cite{Chakraborti11}, implying that Fe nuclei could in principle be accelerated to the relevant energy range if the blast wave is indeed able to accelerate particles at a Bohm rate (as is required to reach the confinement limit).

\subsection{Theory}
\label{S:RTh}

\subsubsection{Instabilities at relativistic shock waves}
\label{S:IRS}
Instabilities at relativistic shock waves enter the scene at two stages of the process: the shock formation, and the particle acceleration.

\paragraph{Shock formation:} Collisionless shock formation is likely to arise from the encounter of two plasma shells. In the fireball scenario for GRB for example, internal shocks results from the collisions of plasma shells ejected by a central engine. In shocks simulations, the shock is formed by launching two plasmas against each other \cite{Silva03, Spitkovsky08}. It is important to recognize that in the kind of collisionless environment occurring in astrophysical context the mean free paths involved are so large, that shells should pass through each other without anything happening.  The reason why something happens instead of nothing in reality or in simulations, is that such counter-streaming configurations are unstable.\\
For shocks in pair plasmas, the full unstable spectrum excited as the two plasma shells overlap, has been analyzed in the unmagnetized regime \cite{Bret08, Bret10}. A detailed investigation of the shock formation process, yielding for example to a theoretical estimate of the shock formation time from first principles, is in development \cite{Bret13, Bret14}. For a wide range of parameters, the filamentation instability governs the unstable spectrum (hence the term ``Weibel shocks'' as the filamentation instability is frequently called ``Weibel'' though notable differences exist between these two, see \cite{Bret10} and \S \ref{S:M-SF}). The technical challenge comes from the fact that the Vlasov/Maxwell kinetic calculations required to study these collisionless plasmas are quite involved, partly because the Lorentz factor couples integrations along the 3 momentum dimensions. Also, the leading instability is likely to be found for any orientation of the perturbation wave vector. Hence, the search for the most unstable mode demands a scanning of the full unstable spectrum. For the magnetized case, few such studies are available so far \cite{Godfrey75, Bret09, Timofeev09}.\\
The formation of a shock in an electron/ion plasma is expected to be qualitatively different. As the two shells come in contact and overlap, counter-streaming electrons from each shells go unstable. Electronic instabilities grow, stop and heat the electrons, and saturate before ions start to react. Therefore, the instabilities triggering the shock formation are the ones originating from the counter-streaming ions over a bath of hot electrons. For typical parameters involved in realistic scenarios, the resulting unstable spectrum has been investigated and gives the ions filamentation instability as the possible dominant instability \cite{Yalinewich10, Shaisultanov12}. Yet, the background electronic temperature remains elusive as simulations indicate it is much higher than a naive estimate assuming their initial kinetic energy has been transferred to heat \cite{Lyubarsky06}.

\paragraph{Particle acceleration:} Once the shock has been formed, it propagates and accelerates particle. To this aim, the Fermi mechanism requires an upstream turbulence to scatter particles back to the shock front. In the seminal papers as \cite{Blandford78}, such turbulence was simply {\it assumed}. It is now understood that this turbulence is generated by the unstable interaction of the accelerated, or reflected, particles at the shock front, with the upstream medium (see \S \ref{S:M-PAME}). The nature of the unstable modes generated ahead of the shock has been investigated by several authors \cite{Rabinak11}, yielding conditions for these modes to significantly back-react over the shock \cite{Nakar11} or specifying the parameter windows where Fermi cycles can be closed \cite{Lemoine10}.

\subsubsection{Particle transport in relativistic shock waves}
\label{S:RThPart}
\begin{figure}
\begin{center}
\includegraphics[width=0.3\textwidth]{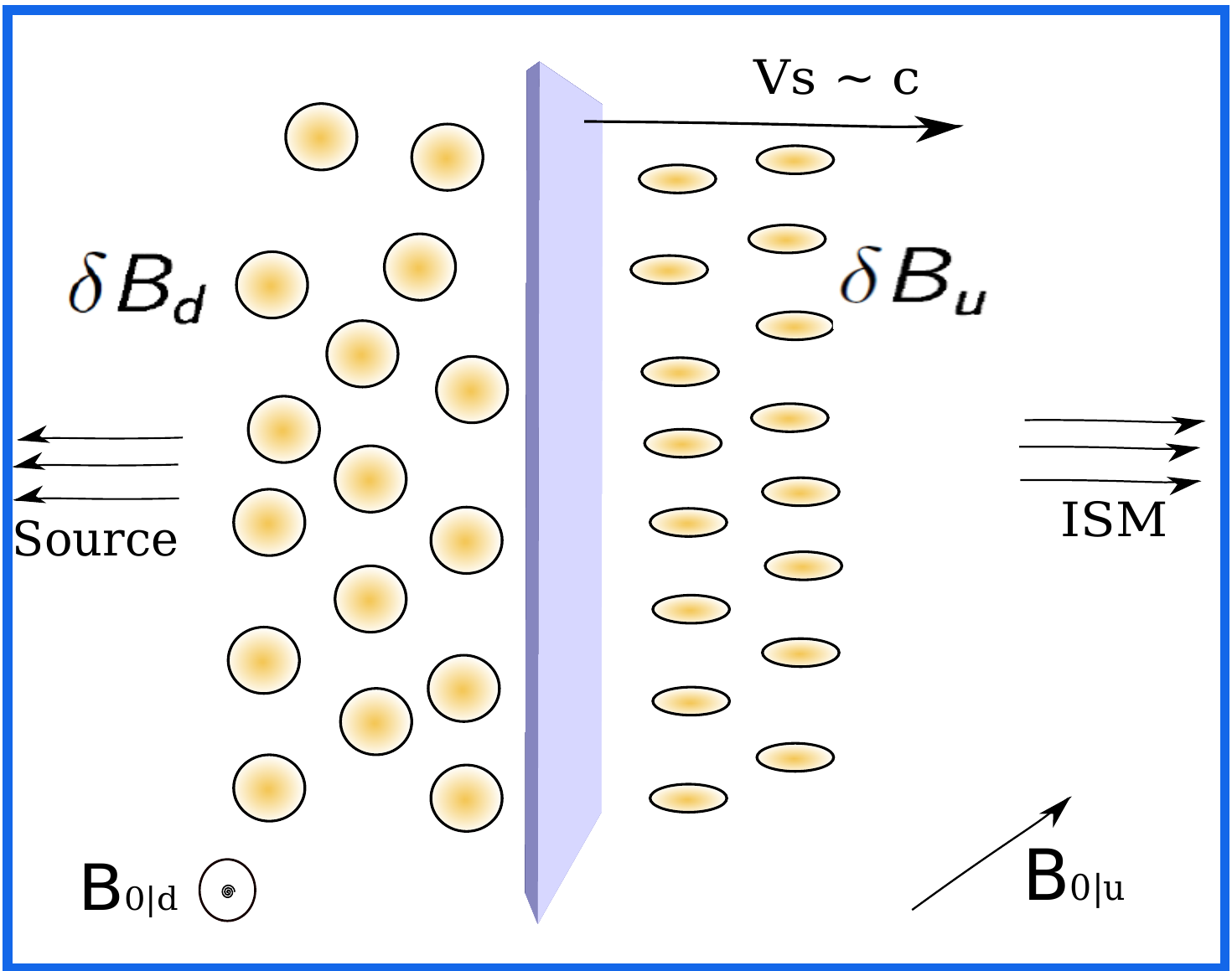}
\includegraphics[width=0.3\textwidth]{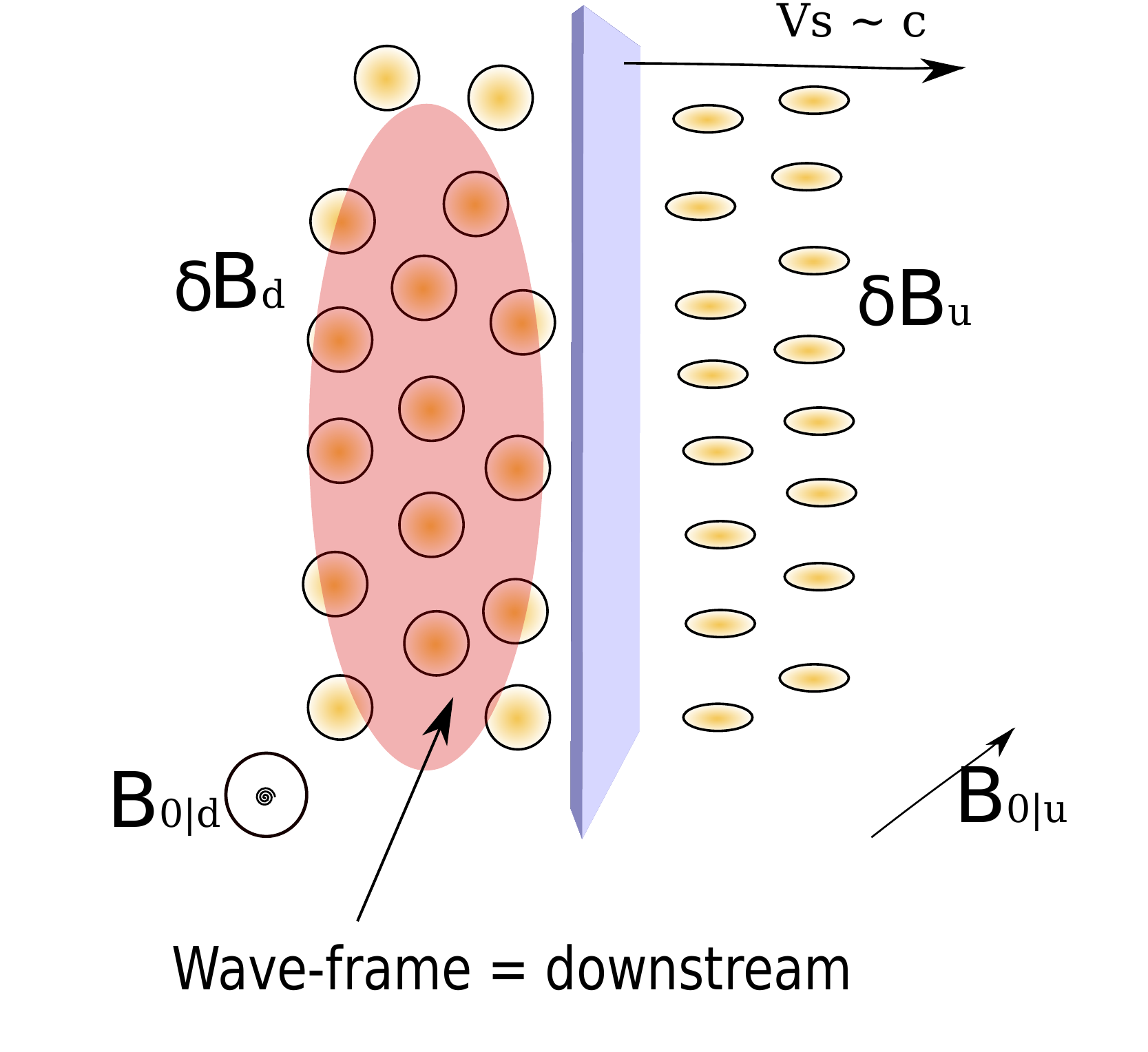}
\includegraphics[width=0.3\textwidth]{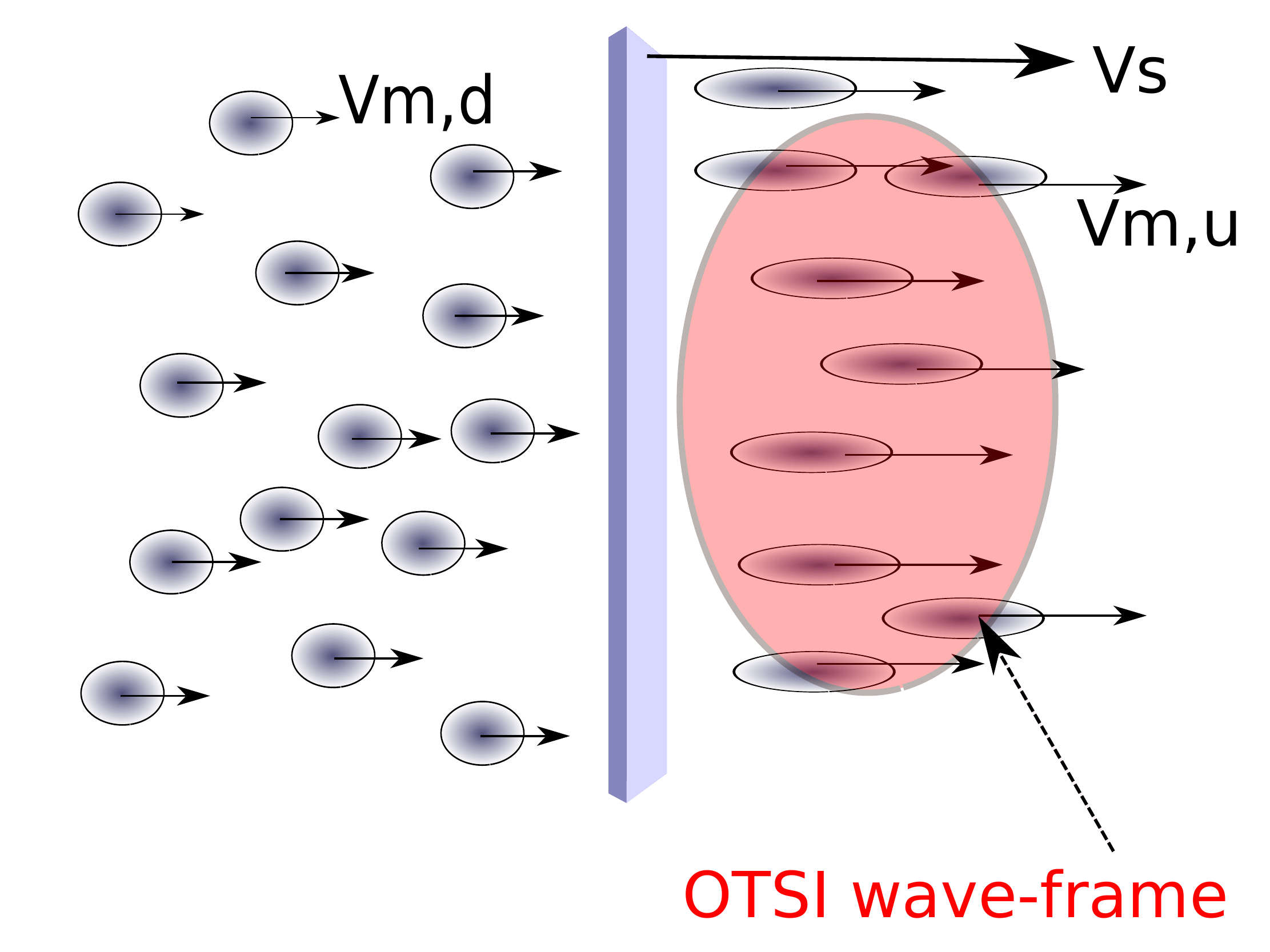} 
\caption{Left panel: Magnetic field structures around the shock front. The shock propagates from the left to the right. In the upstream region magnetic disturbances are anisotropic with coherence length along the shock normal being much larger than in the transverse one (along the plane of the shock front). Downstream, the turbulence is compressed at the shock front and magnetic disturbances should be roughly isotropic with a single coherence length in all directions. Middle panel: wave-frame of magnetic turbulence downstream. Right panel: wave-frame of upstream turbulence, with non-zero velocity relative to the shock (illustrated in the particular case of the oblique two-stream instability (OTSI), see text). Also, $v_{d,m}$ ($v_{u,m}$) is the velocity of downstream (upstream) electromagnetic irregularities relative to the upstream (lab) frame.}
\label{Fig:rel_frames}
\end{center} 
\end{figure}

 Particle transport in relativistic shock environment is strongly constrained by the Lorentz factor of the shock $\gamma_{\rm sh}$, especially in the ultra-relativistic limit ($\gamma_{\rm sh} >> 1$). Upstream of the shock reflected and first Fermi cycle accelerated particles get a typical energy boost $\gamma_{\rm sh}^2$. They move almost with the shock but by an angle more than $\simeq 1/\gamma_{\rm sh}$ they are caught up by the shock before to get isotropized in the upstream flow \cite{Achterberg01}. Consequently, supra-thermal particles in the shock precursor exhibit a very anisotropic distribution. Downstream, the flow moves away from the front with the velocity $V_{\rm sh, d} = c/3$ and it must carries intense turbulence in order to make the Fermi process operative through a very efficient particle scattering.\\
The presence of an uniform magnetic field $B_0$, frozen in the upstream flow, puts additional limitation on particle scattering. If strong micro-turbulence is not generated by plasma instabilities at she shock, particle scattering across the mean field is inhibited and supra-thermal particles are advected downstream \cite{Begelman90, Pelletier09}. Recent particle-in-cell simulations (PIC) \cite{Sironi11} and analytical works \cite{Lemoine10} demonstrate that strong micro-turbulence is generated only if the upstream magnetization is very weak, i.e $\sigma << 10^{-4}$ (see \S \ref{S:Rsim}).\\
An important point is the natural frame where the particle transport is to be considered as can be seen in figure \ref{Fig:rel_frames}. Since particles scatter off the magnetic irregularities, both transport and diffusion processes need to be investigated in the frame where these disturbances are at rest, i.e. the wave-frame of the turbulence. Downstream of the shock the turbulence appears to be quasi-static relatively to the shock front position. Therefore, downstream frame is equivalent to the wave-frame for particle scattering. Upstream, if intense micro-turbulence is generated, magnetic filaments are not frozen in the upstream flow and are able to move relative to the shock front along the shock normal \cite{Plotnikov13}. 

At this point we need to define some properties and notations that will be useful in particle transport studies.  
\paragraph{Particle transport generalities:} In order to study particle transport, one needs to know at each shock front side:
 \begin{itemize}
 \item The flow magnetization, i.e. the intensity of $B_0$.
 \item The magnetic turbulence {\it rms} strength $\sqrt{\langle \delta B^2 \rangle }$ and its coherence length $\ell_c$.
 \item At a given particle energy $E$, one defines an effective gyro-radius $R_g=E/(e \sqrt{B_0^2+ \langle \delta B^2 \rangle})$ and a 
 particle reduced rigidity $\rho=R_g/\ell_c$.
 \end{itemize}
Hence, two different regimes of particle scattering can be defined:
\begin{itemize}
\item A resonant scattering regime ($R_g < \ell_c$) with turbulence wave-modes that strongly depend on their turbulence spectrum. For instance, Kolmogorov-type turbulence leads to $\langle \delta B^2 \rangle \propto k^{-5/3}$; 
\item A non-resonant scattering regime ($R_g > \ell_c$) which is not sensitive to the mode spectrum, the scattering frequency depending on the inverse of the reduced rigidity only.  
\end{itemize} 
In the case of the weakly magnetized relativistic shock, scattering develops in an intense micro-turbulence such that the effective Larmor radii of the particles are always larger than the coherence length, thus only the non-resonant regime has to be considered. This fact results from both the existence of high-energy particles together with the nature of plasma instabilities that give rise to turbulent electromagnetic fields of small coherence length (see \S \ref{S:IRS}). Hence, we are mainly interested in the case $\rho>>1$.

\paragraph{Transport around relativistic shocks:}
We now consider in some detail how the transport process operates up- and downstream:
\begin{enumerate}
\item \underline{Upstream:}
\begin{itemize}
\item Supra-thermal particle precursor is highly anisotropic in momenta around the shock normal: $p_{t} \simeq p_{\ell}/\gamma_{\rm sh}^2$. Where ``t'' and ``$\ell$'' subscripts define transverse and longitudinal direction to the shock normal, respectively.
\item Generated turbulence is anisotropic too: the most unstable wave-modes have wave-numbers $k_{\ell} \ll k_{t}$, so the coherence length along the shock normal is much longer than in transverse direction ($\ell_{c, \ell} >> \ell_{c,t}$).
\item Upstream turbulent region exhibits filamentary structure resulting in non-zero phase velocity of electromagnetic disturbances along the shock normal \cite{Plotnikov13}. High Lorentz factors may be reached, comparable to $\gamma_{\rm sh}$ and the acceleration process is different from the case of frozen-in modes in the upstream flow.
\end{itemize}

Once the field components are expressed in the wave-frame of the turbulence, the spatial transport of supra-thermal behaves as a small-angle scattering process: at each coherence cell a particle is randomly deflected by an angle $\ell_{c,\ell} / R_g$ in the normal direction and by $\ell_{c,t} / R_g$ in the transverse one. The diffusion coefficients can be expressed as \cite{Plotnikov13}:
 \begin{eqnarray}
 D_{t}&=&{ \langle \Delta x_t^2 \rangle \over 4 \Delta t} =   {1\over 3} \ell_{t} c \left({p_{\ell,0} c
      \over \epsilon_\star}\right)^2 ,\\
 D_{\ell} &=&  {\langle \Delta x_\ell^2 \rangle \over 2 \Delta t}= {\ell_{c,\ell} \over \ell_{c,t}} D_t \ .
 \end{eqnarray}
Here $\epsilon_\star$ is the rms energy in the electromagnetic turbulence \cite{Plotnikov13}. All quantities with an index $0$ are calculated in the mean magnetic field alone. An important point is that such diffusion is operative when the scattering time is smaller than $t_{L,0}/\gamma_{\rm sh}$. Elsewhere, particles return downstream by regular gyration in the mean field.
  
\item \underline{Downstream:}
There the turbulence appears as static small-scale magnetic fluctuations (see the middle panel in Fig.\ref{Fig:rel_frames}). Its coherence scale $\ell_{c,d}$ is roughly isotropic because of shock compression. All particles have $R_g > \ell_c$ and one logically expects the diffusion coefficient to be $D = c^2 / (3 \nu_s) \propto E^2$. Where the scattering frequency is $\nu_s= c \ell_c / R_g^2$. As the downstream medium is magnetized by the external magnetic field of magnitude $\gamma_{\rm sh} B_0$ (generically perpendicular to the shock normal), then the diffusion coefficients become anisotropic relatively to the mean field orientation \cite{Plotnikov11}:
 \begin{eqnarray}
 D_{\parallel} &=& {c^2 \over 3 \nu_s} , \\
 D_{\perp} &=& {c^2 \over 3} {\nu_s \over \nu_s^2 + \omega_{L,0}^2}  \ .
 \end{eqnarray}
When $\nu_s < \omega_{L,0}$, $D_{\perp}$ saturates at a constant value and the cross-field diffusion becomes inefficient to transport the particles to the shock front. Acceleration process is locked by the effect of finite magnetization at the energy $E_{max}$ such that $\nu_s(E_{max}) = \omega_{L,0}(E_{max})$. In the shock rest frame this limit reads as:
\begin{equation}
\label{Eq:Emaxdr}
E_{max} = e{\delta B_{\rm rms}^2 \over B_0} \ell_c \ .
\end{equation}
If one consider for instance a GRB with a Lorentz factor $\gamma_{\rm sh} \sim 300$ with $\delta B_{\rm rms}^2/ 4\pi \sim 10^{-2} \rho c^2$ and a magnetization $\sigma \simeq 10^{-9}$, Eq.\ref{Eq:Emaxdr} leads to an energy limit, measured in the interstellar frame, of $10^{15} eV$. This suggests that, to get the highest possible energy, the intensity of the micro-turbulent field must compensate the smallness of the coherence length. The condition for the acceleration process to be efficient is $1 < \rho < \delta B / B_0$, $D_{\perp} \simeq D_{||}$. If $\rho >> \delta B / B_0$, particles are advected with the downstream flow and no further acceleration is possible. Evidences for such small-angle scattering was recently reported in PIC simulations \cite{Sironi13}, \cite{Stockem12} and a similar energy limitation was found (see \S \ref{S:PICR}).
\end{enumerate}

\subsection{Simulations of mildly and trans-relativistic shocks}
\label{S:Rsim}
We define in what follows a trans-relativistic shock as one where the typical ion speeds of the downstream and upstream plasma, which are measured in the reference frame of the downstream plasma, are a significant fraction of the light speed but where relativistic mass effects are not yet important for the ion dynamics. The temperatures of the inflowing upstream electrons are considered to be non-relativistic, but they are relativistic in the downstream region due to heating at the shock. Some of these hot downstream electrons may also escape into the upstream plasma and form a relativistic electron population ahead of the shock. 

\paragraph{Electrostatic shocks} 
The larger mobility of the electrons implies that they diffuse more rapidly than the ions from the dense downstream plasma into the dilute upstream plasma. A positive net charge develops in the dense plasma and a negative net charge in the dilute plasma, which are both located close the shock transition layer with its large plasma density gradient. This space charge results in an electrostatic field. The polarity of this unipolar electrostatic field is such that it counteracts the outflow of electrons from the dense plasma and drags electrons from the dilute plasma into the dense one. Since both processes take place simultaneously, there is a permanent exchange of downstream and upstream electrons across the shock. The positive potential of the downstream region relative to the upstream implies that it slows down the upstream ions in the downstream frame of reference. The potential associated with this electrostatic field is tied to the electron's thermal pressure gradient and it is thus determined by the density jump across the shock and by the electron temperature. A shock is only stable if this potential is sufficiently strong to slow down the inflowing upstream ions such that their velocities become comparable to those of the downstream ions after they have crossed the potential. Both populations can mix in this case and form a single hot ion population. Shock stability also requires that the excess thermal pressure of the downstream plasma is balanced by the ram pressure of the inflowing upstream medium so that the shock becomes stationary in its rest frame. Otherwise a double layer \cite{Langmuir29, Raadu88} or a rarefaction (expansion) wave (see \cite{Sack87} for a review) develops. Yet, a feature of fast electrostatic shocks is a dense shock-reflected ion beam \cite{Forslund70, Forslund71}. The source of this beam is the partial reflection of incoming upstream ions by the shock potential. The incoming upstream ions are not mono-energetic. The shock potential may, for example, adapt to a value that can be overcome by the ions that move towards the shock at the mean speed of the upstream ions. A thermal velocity spread of the upstream ions implies in this case that some ions can not cross the shock potential and are reflected back upstream. The density of the shock-reflected ion beam increases with the shock Mach number and practically all ions are reflected if the shock Mach number is close to the stability limit \cite{Forslund70}.\\

Nonrelativistic electrostatic shocks are frequently observed in the laboratory \cite{Taylor70, Romagnani08, Morita10}, in space \cite{Mozer77} and in particle-in-cell (PIC) and hybrid simulations (see \cite{Karimabadi91} for the latter). Shocks in simulations are generated either by letting plasma clouds collide with a wall or with a second plasma cloud \cite{Forslund70, Forslund71, Karimabadi91, Chen07, Kato10}, by the expansion of a dense into a dilute plasma \cite{Sarri11} or by employing an ion beam with a velocity modulation, which evolves into a shock \cite{Gohda04}. The speed of such shocks is limited to a few times the ion acoustic speed. Collisions of identical plasma clouds at a mildly relativistic speed do not result in electrostatic shocks if the plasma temperatures are non-relativistic to start with \cite{Dieckmann06}. The ambipolar electric field is in this case not strong enough to yield a shock. The result is at least initially a shock-less plasma thermalization through beam instabilities. \\

Electrostatic shocks with higher Mach numbers than a few that move at mildly relativistic speeds can not be ruled out altogether though. It has been shown (see \cite{Sorasio06}) that the maximum Mach number with respect to the ion acoustic speed is of the order of a few if the shock forms in a spatially uniform plasma. The maximum Mach number of electrostatic shocks can be raised if the shock forms as a result of the collision of two different plasma populations, for example between a supernova blast shell and the interstellar medium. It is in principle possible that stable mildly relativistic electrostatic shocks exist even if the upstream medium is cool. Electrostatic shocks with Mach numbers of the order of 100 and speeds of the order of 0.4\textit{c} have been observed \cite{Dieckmann09}, although under highly idealized simulation conditions.  

Unmagnetized shocks remain electrostatic only if the plasma flow speed is low, since instabilities that yield the growth of magnetic fields tend to grow slowly in this case and they often saturate at low magnetic amplitudes. Magnetic fields can be generated by a filamentation instability between the incoming upstream plasma and shock-reflected particles, provided that the flow speed is large enough to transform ion beam instabilities from being electrostatic \cite{Karimabadi91,Forslund70b} into being electromagnetic \cite{Yalinewich10}. The threshold speed is however not well known and it is likely to depend on details of the plasma phase space distribution \cite{Stockem14}. If the processes at the shock result in an anisotropic electron velocity distribution, then the Weibel instability can generate magnetic fields too. Unstable electron distributions develop in the shock transition layer due to the plasma density gradient-driven electrostatic field \cite{Thaury10, Quinn12}.

A recent PIC simulation study \cite{Niemiec12} confirms that magnetic effects are important for slow trans-relativistic and initially unmagnetized or parallel shocks. The shock forms only after the colliding plasmas have been heated up significantly by beam instabilities and after magnetic fields have grown. This suggests that the initial collision speed in that simulation has been too high for the formation of a strictly electrostatic shock and that this shock is mediated by a combination of electrostatic and electromagnetic forces. Increasing the collision speed to a moderately relativistic $\approx 0.9c$ \cite{Kazimura98, Frederiksen04} already results in a shock transition layer that is magnetically dominated and filamentary. The shock structure should thus be strongly dependent on the flow speed in the trans-relativistic regime. Until now, no systematic laboratory or simulation studies of trans-relativistic electron-ion shocks in the intermediate velocity regime and in initially unmagnetized plasmas exist, primarily because of the long shock formation time and the resulting high computational cost.
 
\paragraph{Perpendicular and quasi-perpendicular mildly relativistic shocks} 
Most simulations have considered a background magnetic field that is stronger than what we find in most relevant astrophysical regimes such as the interstellar medium. A strong magnetic field implies that the fast electron plasma oscillations and the slow ion cyclotron oscillations can be resolved simultaneously at a reasonable computational cost. This strong background magnetic field typically preserves its spatially uniform structure perpendicularly to the shock normal during the simulations, because the comparatively weak magnetic fields generated by plasma instabilities can not modulate it. 

As we go to trans-relativistic shocks, the relative speed between the incoming upstream plasma and the shock-reflected ion beam becomes relativistic. Consequently, the character of the instability between both counter-streaming ion beams in the foreshock becomes increasingly magnetic. This has multiple consequences. Electron surfing acceleration, which requires the transport of electrostatically trapped electrons across a magnetic field, can be robust against non-planar electric field structures if electrons have multiple encounters with patchy electrostatic structures \cite{Matsumoto12}. However if the ion beam instability becomes primarily magnetic this mechanism is supressed \cite{Dieckmann08}. Another consequence of the development of the filamentation instability upstream of the shock is the growth of filamentary magnetic fields with amplitudes that exceed those of the uniform background magnetic field \cite{Hededal05}. It follows that  the shock propagates into a filamentary rather than into a spatially uniform foreshock magnetic field. In particular the ion acceleration mechanisms such as the SDA and SSA will be modified if the shock is trans-relativistic since both require a magnetic field that is uniform perpendicular to the shock normal. It is up to now unclear for which shock parameter regimes these ion acceleration mechanisms are efficient.
 
As we go to mildly relativistic perpendicular shocks, a type of electron acceleration mechanism sets in, which is not observed in this form and has not been reported to occur at non-relativistic shocks. This process is known in its basic form as electron acceleration by a magnetic wall \cite{Smolsky96}. Consider an electron-ion plasma, which is initially free of any net charge and current. This plasma moves at a relativistic speed towards a uniform magnetic field, which is oriented perpendicularly to the plasma flow. The penetration depth of ions and electrons into this magnetic field differs due to their different Larmor radii. The charge separation results in the growth of a strong electrostatic field. Electrons are dragged by this field across the magnetic field and accelerated to a speed, for which their Larmor radius becomes comparable to that of the ions. This type of acceleration is also observed when two plasmas, which carry a perpendicular magnetic field, collide at a moderately relativistic speed \cite{Hededal05}.

\paragraph{Quasi-parallel mildly relativistic shocks}
Trans-relativistic shocks, which are quasi-parallel relative to the magnetic field far upstream of the shock, have also been investigated in PIC \cite{Shikii10, Dieckmann10, Murphy10} and hybrid simulation studies \cite{Gargate12}. Circularly polarized electromagnetic waves have been observed with magnetic amplitudes, which exceed those of the background field by more than one order of magnitude. The similarity of the wave properties in the PIC and hybrid simulations demonstrates that these waves are robust against changes in the electron distribution, which are represented differently in both types of codes. These waves rotate the upstream magnetic field vector from a quasi-parallel to a quasi-perpendicular configuration close to the shock. The circular polarization of the wave implies that its magnetic field vector rotates around the shock normal. The simulation results are supported by analogous observations of such magnetic structures at the oblique Earth's bow shock \cite{Mann94}. These structures are known as Short Large Amplitude Magnetic Structures (SLAMS) and they are supported by the ions. The ions form a compact beam in phase phase space with a trajectory that resembles a corkscrew. The axis of the corkscrew is aligned with the direction of the wave vector and the corkscrew motion involves the two velocity components orthogonal to that along this axis. A prototype of such a non-linear wave is described in \cite{Krasovsky10}. 

The emergence of these waves typically coincides in the PIC simulations with the acceleration of electrons to very high energies. The simultaneous generation of strong magnetic fields and acceleration of electrons to highly relativistic speeds is important for astrophysics, because such shocks should emit strong electromagnetic radiation. Two mechanisms and, possibly, their combination can explain the coincidence of the wave generation and electron acceleration. The first mechanism attributes the growth of the strong magneto-wave to a streaming instability. Particles are accelerated by repeated shock crossings and they reach highly relativistic speeds. Unless the upstream magnetic field is almost perpendicular, some of these particles can move upstream where they undergo a non-resonant magnetic instability with the incoming upstream plasma (see section \S \ref{S:LWNR}). The second mechanism \cite{Dieckmann10} attributes the growth of the magnetic field to plasma processes at the boundary between the colliding plasmas. The shock formation and the particle acceleration are here a consequence and not the cause of the strong initial magneto-wave. Consider two colliding plasma clouds that transport a magnetic field, which is oriented quasi-parallel to their collision velocity vector. Both clouds are initially separated in space and share a boundary. The magnetic field is initially frozen-in in each cloud and the magnetic field is continuous across both clouds. The relative motion of both clouds implies that they carry a different convective electric field and $\nabla \times \mathbf{E} \neq 0$ at the contact boundary. A magnetic field thus grows in the interval where both clouds overlap after the simulation started and it amplifies the perpendicular component of the magnetic field. This magnetic field is frozen-in in the overlap layer with a mean speed that is determined by the speed and densities of both plasma clouds through momentum conservation. The ion beams from both clouds move at a high speed relative to this magnetic field and are deflected away from the initial collision velocity vector. The resulting net ion current amplifies the perpendicular magnetic field. The unperturbed ion motion along the collision direction together with the rotation of the orthogonal velocity vector by the magneto-wave implies that this localized deflection results in a corkscrew orbit of the ions. The magnetic amplitude grows to a value that results in the formation of a shock. The now quasi-perpendicular magnetic field of this wave acts on the incoming upstream plasma like a magnetic wall \cite{Trakhtengerts66}. The different penetration lengths of electrons and ions in the wave field introduces a space charge, which accelerates the electrons to highly relativistic speeds \cite{Lembege89, Bessho99} almost perpendicularly to the collision- and magnetic field direction. This electron beam can in turn drive instabilities that result in magnetic vortices \cite{Murphy10}.

\begin{figure}
\centering
\includegraphics[width=\textwidth]{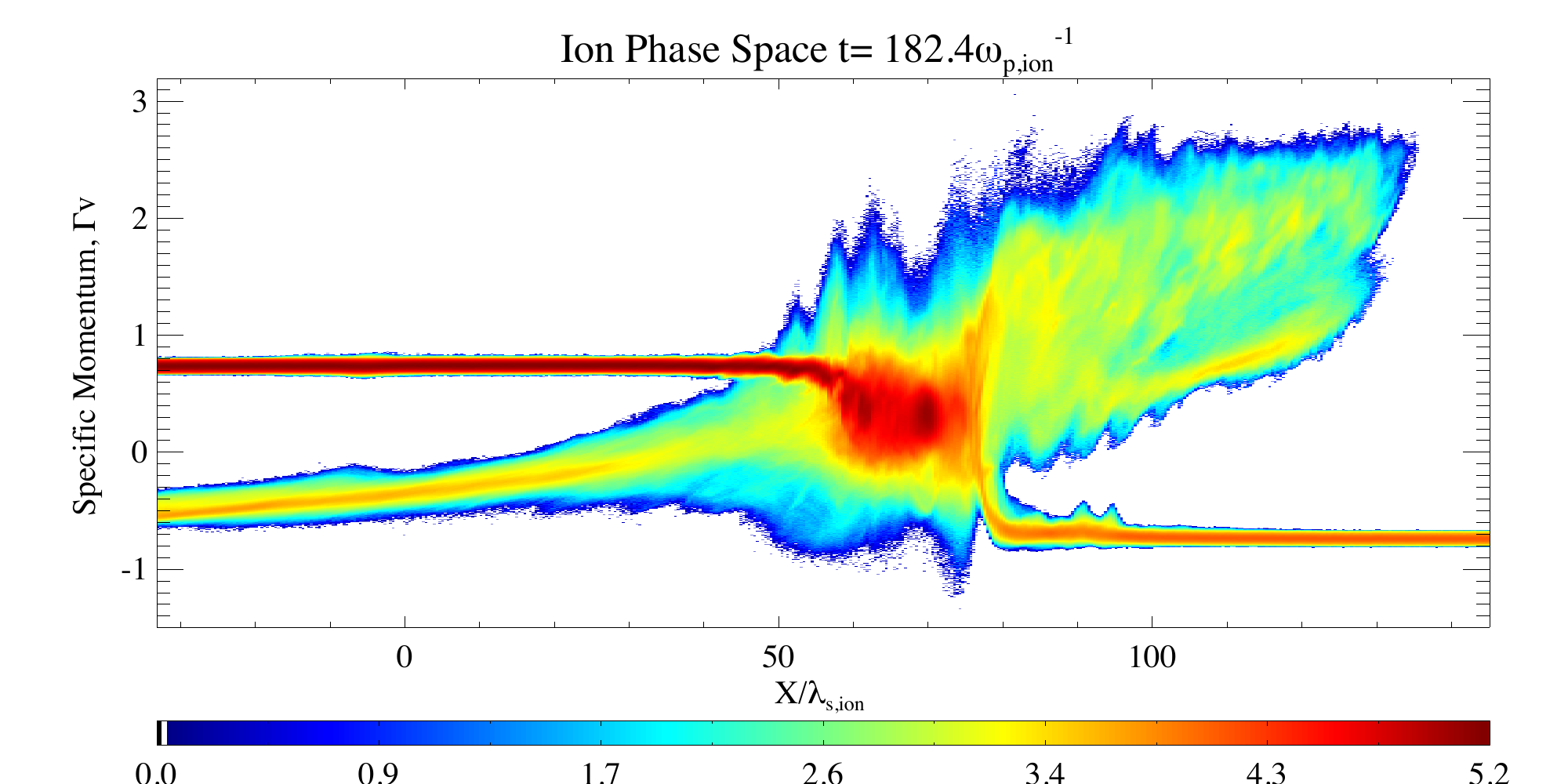}
\caption{Ion phase space distribution in a 0.9c mildly relativistic shock: Logarithm of electron density as a function of specific x-momentum $\Gamma v_x$ and x at $t=T_2$. The forward shock is forming at x=80. Between x=80 and x=130, the incoming ions are reflected, this region is the foreshock.
Between x=60 and x=80, the shock forms a thermalised downstream region. At this time, the reverse shock is still forming at x=60 (see \cite{Murphy10} for details).}
\label{FigIonPST2}
\end{figure}

For simplicity, many shocks are modeled as symmetric collisions, between identical plasma species. This assumption is likely to be violated for real interstellar shocks. Overdense shocks have different properties to shocks with unity density ratio. In particular, increased particle acceleration has been observed and increased magnetic field amplification \cite{Murphy10}, in comparison to shocks in uniform media. A prototypical asymmetric density plasma protoshock at mildly relativistic speeds is shown in Figure \ref{FigIonPST2}.
There is forming forward shock at X=80, hot ionized downstream ($ 60<X<80$), a shock reflected ion beam ($X > 80$), and the density ratio is 10. 
Notably, the reverse shock is not yet visible in the region X=60. 

\subsection{Particle-in-cell simulations: Electron-Positron and Electron-Proton Relativistic Magnetized Shocks}
\label{S:PICR}
\subsubsection{Particle Acceleration in Relativistic Magnetized Shocks}
\label{S:PARMS}
The internal structure of relativistic shocks and the efficiency of particle acceleration depend on the conditions of the pre-shock flow, such as bulk velocity, magnetic field strength and orientation. As found by \cite{Sironi11, Sironi13}, for highly relativistic flows, the main parameter that controls the shock physics is the magnetization $\sigma$ (again, the ratio of electromagnetic to kinetic energy density of the pre-shock medium, see \S \ref{S:Def}).

\paragraph{Weak magnetization shocks:} If $\sigma \le 10^{-3}$, shocks are governed by electromagnetic plasma instabilities (the filamentation or Weibel instability), that generate magnetic fields stronger than the background field. Such shocks do accelerate particles self-consistently up to non-thermal energies, and the accelerated particles populate a power-law tail $dN/dE\propto E^{-p}$ with slope $p\simeq2.5$, that contains 3\% of particles and 10\% of flow energy \cite{Sironi13}. In electron-proton shocks, the acceleration process proceeds similarly for the two species, since the electrons enter the shock nearly in equipartition with the ions, as a result of strong pre-heating in the self-generated upstream turbulence. In both electron-positron and electron-ion shocks, the maximum energy of the accelerated particles scales in time as $\varepsilon_{max}\propto t^{1/2}$ (\cite{Sironi13}, see discussion below). This scaling is shallower than the so-called (and commonly assumed) Bohm limit $\varepsilon_{max}\propto t$, and it naturally results from the small-scale nature of the Weibel turbulence generated in the shock layer. In magnetized plasmas (yet with $\sigma\lesssim10^{-3}$), the energy of the accelerated particles increases until it reaches a saturation value $\varepsilon_{sat}/\gamma_0 m_i c^2\sim\sigma^{-1/4}$, where $\gamma_0 m_i c^2$ is the mean energy per particle in the upstream bulk flow (so, $\gamma_0$ is the bulk Lorentz factor of the upstream flow, as measured in the downstream frame). Further energization is prevented by the fact that the self-generated turbulence is confined within a finite region of thickness $\propto \sigma^{-1/2}$ around the shock \cite{Sironi13}.

\begin{figure}[!bp]
\begin{center}
\includegraphics[width=\textwidth]{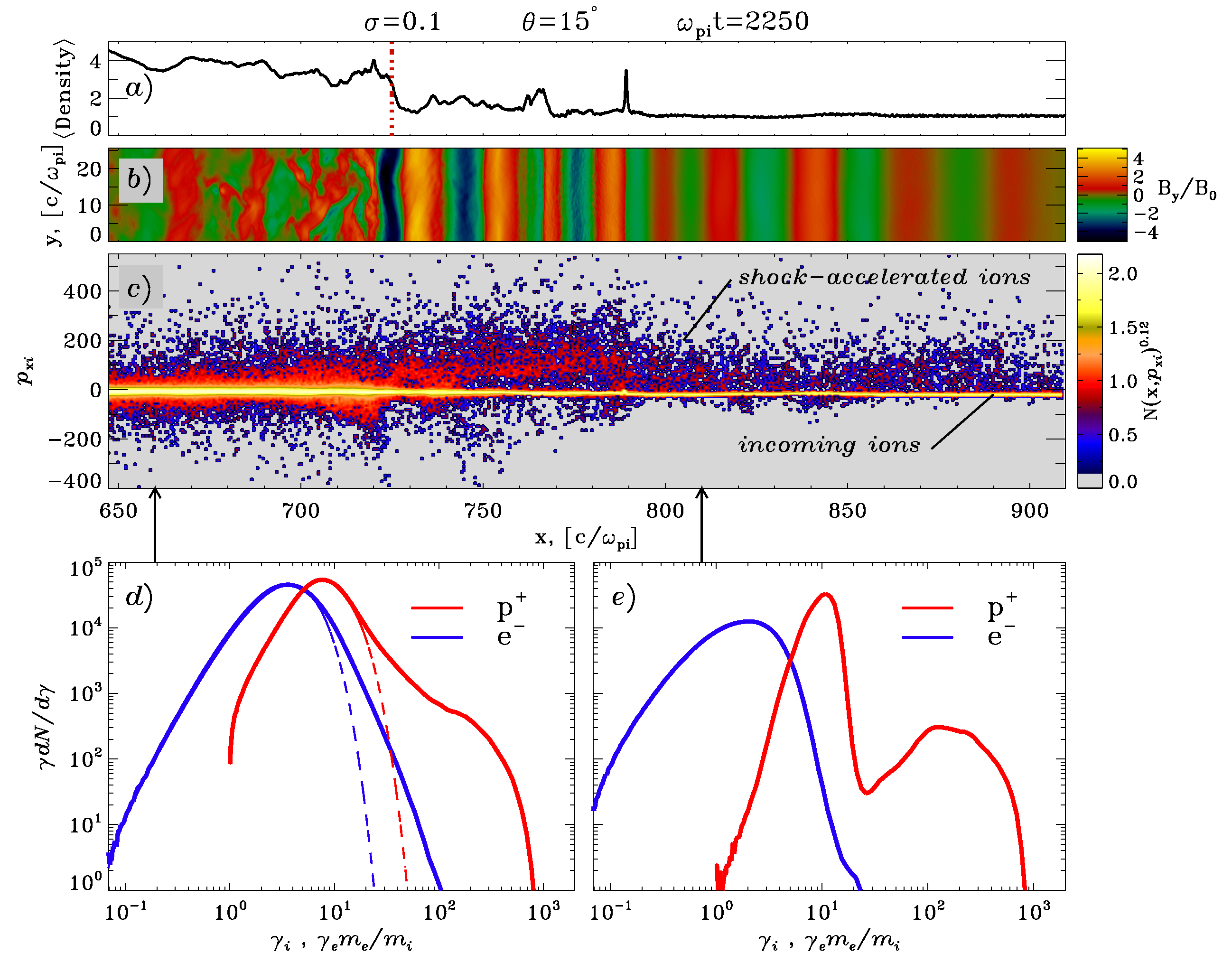}
\caption{\footnotesize{Internal structure of an electron-ion subluminal shock with $\sigma=0.1$ and $\theta=15^\circ$ \cite{Sironi11}. The shock front is located at $x\sim725\,c/\omega_{\rm pi}$ (vertical dotted red line in panel (a)), and it separates the pre-shock region (to its right) from the compressed post-shock region (to its left). A stream of shock-accelerated ions propagates ahead of the shock (see the diffuse cloud in panel (c) to the right of the shock, at $x\gtrsim725\,c/\omega_{\rm pi}$). Their interaction with the pre-shock flow (narrow beam to the right of the shock in panel (c)) generates magnetic turbulence ahead of the shock (see the transverse non-resonant waves in panel (b), to the right of the shock). In turn, such waves govern the process of particle acceleration. In fact, the particle spectrum behind the shock (solid lines in panel (d); red for ions, blue for electrons) is not compatible with a simple thermal distribution (dashed lines), but it shows a clear non-thermal tail of high-energy particles, most notably for ions (red solid line).} }
\label{fig:shock1}
\end{center}
\end{figure}

\paragraph{High magnetization shocks:} If now $\sigma\gtrsim10^{-3}$, the shock structure and acceleration properties depend critically on the inclination angle $\theta$ between the pre-shock field and the shock direction of propagation. If the magnetic obliquity is larger than a critical angle $\theta_{\rm crit}\simeq34^\circ$ (as measured in the post-shock frame), charged particles would need to move along the field faster than the speed of light in order to outrun the shock (``superluminal'' configurations). In that view, only ``subluminal'' shocks ($\theta\lesssim\theta_{\rm crit}$) are efficient particle accelerators. As illustrated in figure~\ref{fig:shock1}, a stream of shock-accelerated particles propagates ahead of the shock (panel (c), for $x\gtrsim725\,c/\omega_{\rm pi}$), and their counter-streaming with the incoming flow generates magnetic turbulence in the pre-shock region (panel (b)). In turn, such waves govern the acceleration process, by providing the turbulence required for the Fermi mechanism. 

The post-shock particle spectrum in subluminal shocks shows a pronounced non-thermal tail of shock-accelerated particles with a power-law index $2\lesssim p\lesssim 3$ (panel (d)). The tail contains 5\% of particles and 20\% of flow energy \cite{Sironi11}. In contrast, superluminal shocks ($\theta\gtrsim\theta_{\rm crit}$) show negligible particle acceleration. Here, due to the lack of significant self-generated turbulence, charged particles are forced to slide along the background field lines, whose orientation prohibits repeated crossings of the shock. This inhibits the Fermi process, and in fact the particle distribution behind superluminal shocks is purely thermal \cite{Sironi11}. The same conclusion holds for both electron-positron and electron-ion flows. In electron-ion shocks, the incoming electrons are heated up to the ion energy, due to powerful electromagnetic waves emitted by the shock into the pre-shock medium, as a result of the synchrotron maser instability \cite{Gallant92, Lyubarsky06, Sironi11}. Yet, such heating is not powerful enough to permit efficient injection of electrons into the Fermi acceleration process at superluminal electron-ion shocks.

\begin{figure}[!bp]
\begin{center}
\includegraphics[width=\textwidth]{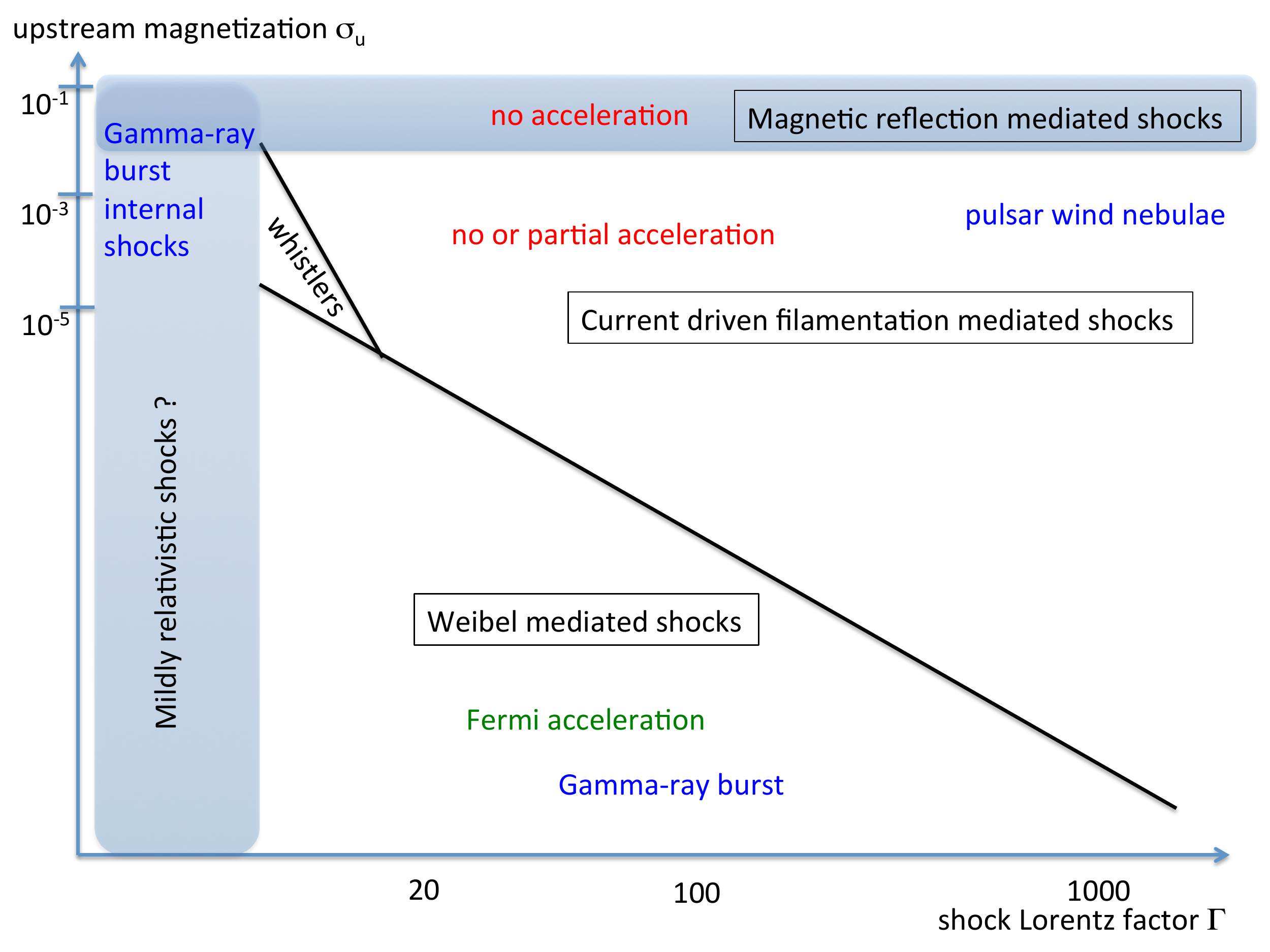}
\caption{\footnotesize{Summary plot of the dominance zone of the main instabilities in relativistic shocks in terms of upstream magnetization and shock Lorentz factor. Three zones are clearly identified: a low-magnetization case where Fermi acceleration operates, an intermediate-magnetization case where no or partial acceleration occurs, and  a high-magnetization case with no Fermi acceleration; see \cite{Lemoine10}.}}
\label{fig:Sketch}
\end{center}
\end{figure}

Overall, the results of PIC simulations of relativistic shocks imply that non-thermal particle acceleration only occurs if the pre-shock magnetization is weak ($\sigma\lesssim10^{-3}$, see the summary sketch in figure \ref{fig:Sketch}), or if the upstream field is nearly aligned with the shock direction of propagation (i.e., in subluminal shocks). Polarization measurements of PWNe \cite{Wilson72, Schmidt79}, GRBs and AGN jets \cite{Gabuzda04} suggest that the shocks in these systems should be appreciably magnetized ($\sigma\gtrsim0.01$) and superluminal, yet they need to be efficient particle accelerators, in order to explain the prominent non-thermal signatures of these sources. A possible solution to this  discrepancy is proposed below for PWNe in \S \ref{S:Rpuls}.

\paragraph{Discussion: issues on acceleration in high magnetization shocks:} Despite the apparent lack of non-thermal particles in PIC simulations of superluminal highly magnetized shocks, it is still possible to put firm constraints on models of non-thermal emission. In particular, the strong electron heating observed in electron-ion shocks implies that a hypothetical power-law tail in the electron spectrum should start from energies higher than the ion bulk kinetic energy. For models of GRBs and AGN jets that require a power-law distribution extending down to smaller energies, this would suggest that electron-positron pairs may be a major component of the flow. Ejections from synchrotron sources, like the termination shocks of pulsar winds or the jets of active galactic nuclei, consist mainly of pair plasmas, with a small fraction of protons or heavier ions \cite{Bucciantini10, Kino12}. It has been shown above that in collisionless shocks, an increase of the ambient magnetic field changes the dominant acceleration/heating process from Fermi acceleration in the unmagnetized or weakly magnetized case to magnetic reflection in the case of a strong perpendicular magnetic field \cite{Sironi11b}. In a strongly magnetized perpendicular pair shock non-thermal acceleration is almost completely suppressed \cite{Langdon88, Gallant92}. However, if a small fraction of ions is present and dominant in terms of kinetic energy \cite{Blasi11}, the lighter particles can be accelerated efficiently \cite{Hoshino92, Amato06, Stockem12}.

Positrons and electrons thermalize rapidly during the early stages of shock formation process, while the ions preserve their beam character over tens of ion plasma frequencies and are reflected from the shock front. While in non-relativistic shocks the electrostatic field is responsible for ion reflection \cite{Hasegawa05, Kato10}, in relativistic scenarios the electromagnetic fields are more important due to the increased Larmor radius. The upstream ions can penetrate the shock region and enter the downstream, but they will eventually be reflected back into the upstream due to the strong field. These oscillations in the shock front region lead to a compression of particles and a strong overshoot in the magnetic field \cite{Hoshino92}.

The acceleration process in electron-positron-proton shocks is based on the so-called synchrotron maser instability \cite{Hoshino91,Treumann11}. The gyro motion of the ejected plasma particles in the strong ambient field gives rise to wave emittance and subsequent resonant absorption. The left-handed polarized waves emitted by the gyrating ions are resonantly absorbed by the positrons due to the matched rotation direction. Nevertheless, the circular polarization of the waves also leads to an excitation of the electrons. Up to 20\% of the entire positron population was observed to be contained in a non-thermal tail and the most energetic particles can reach Larmor radii of the order of the shock thickness. The fraction of wave absorption by ions is small due to their high mass, so that the ion spectrum stays thermal. The acceleration efficiency and particle spectra are determined by the ion properties. The fraction of energy in the non-thermal part of the particle spectrum increases with the ion to electron density and mass ratios, and the cutoff energy scales with the ion mass ratio \cite{Amato06}. The temporal evolution of the non-thermal tail behaves according to a power-law, $\gamma_{max} \propto t^{\alpha}$ with $1/3 < \alpha < 1$ \cite{Stockem12}. The acceleration time can be estimated by (see \S \ref{S:NRth}):
\begin{equation}
t_{acc} = \frac{3}{v_u-v_d} \int_{p_0}^{p} \frac{dp'}{p'} \left( \frac{\kappa_u(p')}{v_u} + \frac{\kappa_d(p')}{v_d} \right) \ ,
\end{equation} 
From Bohm diffusion as a standard transport model, a linear scaling $\gamma \propto t_{acc}$ is expected \cite{Gargate12}. The reduced powers $\alpha < 1$ indicate that the process is slowed down by small wavelength scattering \cite{Kirk10, Sironi13}, which for highly relativistic velocities scales as $\gamma \propto t_{acc}^{1/2}$.

The shock formation process and relevant scales are also determined by the actual constitution of the unperturbed plasma. In the case of highly relativistic upstream fluid velocities, the shock speed is approximated well by:
\begin{equation}\label{eq:shockspeedapprox}
\left( 1 + \frac{1}{\sigma} \right) \beta_{\rm sh}^2 - \left[ \frac{\Gamma }{2} + \frac{1}{\sigma} (\Gamma - 1) \right] \beta_{\rm sh} - \left( 1 - \frac{\Gamma}{2}\right) = 0 \ ,
\end{equation}
assuming that the species are all thermalized, obeying the pressure-energy relation $p = (\Gamma-1) e$ with the adiabatic constant $\Gamma$, and characterized only by the total magnetization, which is given by $\sigma = \sigma_e \, \left[ 2 + \frac{m_i n_i}{m_e n_e} \left( 1- \frac{m_e}{m_i}\right) \right]^{-1}$ with electron upstream magnetization $\sigma_e = B^2/(4\pi n_e m_e c^2 \gamma)$ \cite{Gallant92}. An exact expression for the calculation of the shock speed can be found in \cite{Stockem12}, taking into account the real shapes of the non-thermal spectrum and an arbitrary upstream fluid velocity. The deviations from Eq.\ref{eq:shockspeedapprox} become stronger the lower the upstream magnetic field, but always stay below 10\% for standard collisionless shock scenarios \cite{Stockem12}. The shock speed increases with the magnetization and decreases with the ion to electron mass and density ratios due to their reciprocal dependence on $\sigma$. At the same time, an increase of the mass and density ratios leads to larger scales. The shock forms on a larger temporal scale and also the spatial scales are increased and strong wave generation in the downstream on the scale of the ion Larmor radius appears, making the shock profile become less smooth \cite{Hoshino92}.

Most of the simulations have been performed in a 1D setup and the question was raised if the acceleration efficiency was overestimated due to the artificially increased heating in longitudinal direction \cite{Hoshino92}. 2D effects were found to play a minor role if the ambient magnetic field is perpendicular and strong $(\sigma \gtrsim 0.1)$, and the final acceleration spectra show only marginal differences. In parallel and oblique shocks, however, at least two-dimensional simulations are required to capture the physics. While the positron spectra are not affected much by the geometry of the magnetic field, the electron spectra and energy cutoff are very sensitive to the obliquity angle of the field.

\subsubsection{Radiation at relativistic shock waves}
\label{S:RR}
\paragraph{Radiation Signatures of Relativistic Unmagnetized Shocks} As mentioned above, relativistic unmagnetized shocks (i.e., with $\sigma=0$) are governed by the Weibel instability, which generates small-scale magnetic fluctuations. It has been speculated that emission from these shocks may occur in the so-called ``jitter'' regime \cite{Bogovalov99, Sironi11}, if the scale $\lambda_B$ of the turbulence is so small that the wiggler parameter $a\equiv q\,\delta B\lambda_B/mc^2\ll1$, with synchrotron radiation occurring for $a\gg1$. Here, $\delta B$ is the field strength, $q$ and $m$ the electron charge and mass, and $c$ the speed of light. Jitter radiation has also been proposed as a solution for the fact that GRB prompt emission spectra below the peak frequency are not compatible with the predictions of synchrotron emission at mildly relativistic shocks (the so-called ``line of death'' puzzle \cite{Bietenholz97}).

\begin{figure}[!tbp]
\begin{center}
\includegraphics[width=\textwidth]{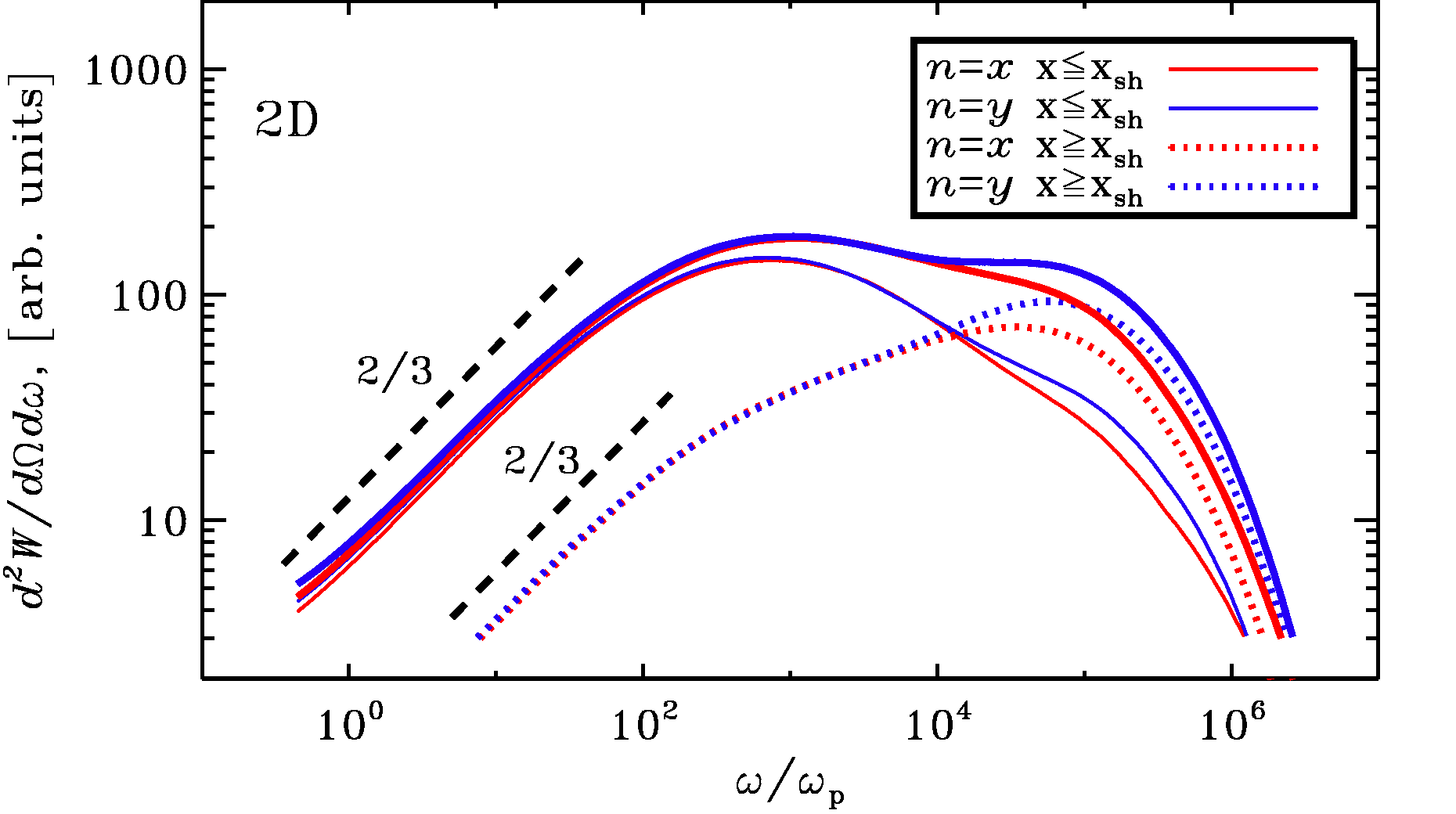}
\caption{\footnotesize{
\textit{Ab initio} photon spectrum (thick solid lines) from the 2D PIC simulation of an unmagnetized (i.e., $\sigma=0$) shock \cite{Sironi09}. Red lines are for head-on emission (along the shock direction of propagation), blue lines for edge-on emission (along the shock front). The slope at low frequencies is $2/3$ (black long-dashed lines), proving that the spectra are consistent with synchrotron radiation from a 2D particle distribution (in 3D, the predicted slope of 1/3 is obtained). By separating the relative contribution of post-shock (thin solid lines) and pre-shock (dotted lines) particles, one sees that pre-shock particles contribute significantly to the total emission (thick solid lines), especially at high frequencies. Frequencies are in units of the plasma frequency $\omega_{\rm p}$.}}
\label{fig:radiation1}
\end{center}
\end{figure}

However, recent PIC simulations, which includes an algorithm to extract \textit{ab initio} photon spectra, have revealed synthetic spectra that are entirely consistent with synchrotron radiation in the fields generated by the Weibel instability \cite{Gabuzda04}, see Fig.~\ref{fig:radiation1} for an illustration. The so-called ``jitter'' regime is recovered only by artificially reducing the strength of the fields, such that the parameter $a$ becomes much smaller than unity. So, if the GRB prompt emission results from relativistic unmagnetized shocks, it seems that resorting to the jitter regime is not a viable solution for the ``line of death'' puzzle. This can be easily understood on analytical grounds, because $a\,\sim\, 3.6\times 10^2\gamma_{\rm sh} \epsilon_{B,-2}^{1/2}\left(\lambda_{B}\omega_{\rm pi}/c\right)\,\gg\,1$ in terms of the shock Lorentz factor $\gamma_{\rm sh}$; one even expects $a\,\gg\,\gamma$ for all supra-thermal electrons, corresponding to the standard synchrotron regime \cite{Kirk10, Reville10, Plotnikov13}.

The small scale nature of the turbulence nevertheless brings in some interesting radiative signatures. First of all, particle scattering in small scale turbulence is slow, hence the maximal energy is limited: comparing the scattering timescale $t_{\rm scatt}\,\sim\, r_{\rm g}^2/\lambda_{\delta B}c$ (in the SRF) to the synchrotron energy loss timescale for electrons leads to a maximal Lorentz factor $\gamma_{\rm max}\,\sim\,\left(n r_e^3m_e/m_p\right)^{-1/6}$, $r_e$ denoting the electron classical radius~\cite{Kirk10, Plotnikov13, Sironi13}, or to a maximal synchrotron photon energy
\begin{equation}
\epsilon_\gamma\,\sim\, 3\,{\rm GeV}\,\epsilon_{B,-2}^{-1/2}\gamma_{\rm sh,2.5}^2n_{0}^{1/2} \ .
\end{equation}
The fiducial numerical values refer to the external shock wave of a GRB propagating in the ISM. Indeed, most high energy photons observed in the extended emission phase of GRBs have a rest frame energy below the above cut-off; in the present scenario, the few photons above this limit must therefore originate from IC interactions, see the discussion in \cite{Wang13}.

Secondly, small scale turbulence is expected to decay fast, on multiples of the plasma skin depth \cite{Gruzinov99, Lemoine15}; the PIC simulations of \cite{Chang08, Keshet09} do confirm this decay, with a roughly power-law behavior on $\gtrsim\,10^{2-3}\omega_{\rm p}^{-1}$ timescales. In a decaying turbulence, electrons of different energies cool in regions of different magnetic field strengths, which has direct implications for the afterglow spectrum and light curves of gamma-ray bursts in particular \cite{Rossi03, Derishev07, Lemoine13}; detailed synchrotron spectra and light curves are provided in the latter paper for a generic power-law decay and various cooling scenarios. Recent work \cite{Lemoine13b} further argues that the late time extended emission seen in several gamma-ray bursts from the radio to the GeV band do point to the decay of micro-turbulence behind the shock front, with a decay index $-0.5\,\lesssim\,\alpha_B\,\lesssim\,-0.4$, for $\epsilon_B \,\propto\, t^{\alpha_B}$ in terms of comoving time $t$ since injection through the shock.

\subsection{The particular case of relativistic shocks in striped winds}
\label{S:Rpuls}
Polarization measurements of PWNe suggest that the pulsar wind termination shock is highly magnetized and perpendicular (so, superluminal). Based on the PIC results shown above, particle acceleration via the Fermi process should be inhibited, in contradiction with the clear non-thermal signatures of these sources. In the attempt to tackle this apparent discrepancy, 2D and 3D PIC simulations have recently been performed to investigate the acceleration efficiency of perpendicular shocks that propagate in high-$\sigma$ flows with alternating magnetic fields \cite{Sironi11b}. Here, the assumption is that the pulsar wind ahead of the termination shock consists of alternating stripes of opposite magnetic polarity (hereafter, a ``striped wind''). For PWNe, this is the configuration expected around the equatorial plane of the wind, where the sign of the magnetic field alternates with the pulsar period \cite{Bogovalov99}.\\

\begin{figure}[!tbp]
\centering 
\includegraphics[width=\textwidth]{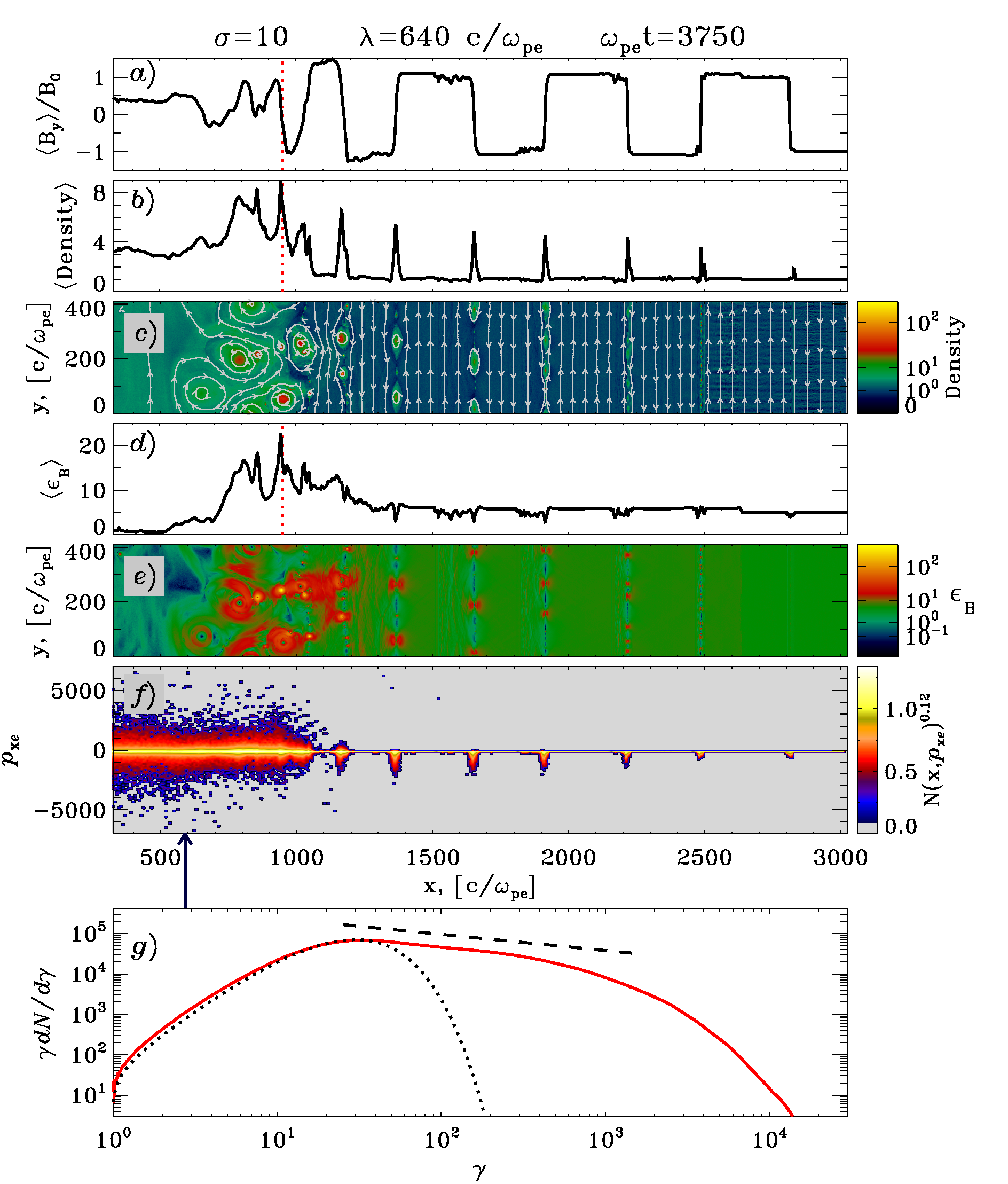}
\caption{\footnotesize{2D PIC simulation of a relativistic shock propagating in a striped flow with magnetization $\sigma=10$ and stripe wavelength $\lambda=640 d_e$, ($d_e$ is the electron plasma skin depth, see \S \ref{S:Def}); see \cite{Sironi11b} for details. The shock is located at  $x\sim 950\, d_e$ (vertical dotted red line), and the incoming flow moves from right to left. At the shock, the striped structure of the magnetic field is erased (panel (a)), the flow compresses (density in panel (b)), and the field energy (panel (d)) is transferred to the particles (phase space  in panel (f)). The microphysics of magnetic reconnection is revealed by the islands seen in the 2D plots of density and magnetic energy (panels (c) and (e), respectively) in a region around the shock. As a result of magnetic reconnection, the post-shock particle spectrum (red line in panel (g)) is much broader than a thermal distribution (dotted line), and it approaches a power-law tail with flat slope $p\simeq1.5$ (dashed line).}}
\label{Fig:fluid}
\end{figure}

At the termination shock, the compression of the flow forces the annihilation of nearby field lines, a process known as magnetic reconnection. As shown in Fig.~\ref{Fig:fluid}, magnetic reconnection erases the striped structure of the flow (panel (a)), and transfers all the energy stored in the magnetic fields (panel (d)) to the particles, whose distribution becomes much hotter behind the shock (see panel (f), for $x \le 1000$). As a result of field dissipation, the average particle energy increases by a factor of $\sigma$ across the shock, regardless of the stripe width $\lambda$ or the wind magnetization $\sigma$. \\

The reconnection process manifests itself as characteristic islands in density (panel (c)) and magnetic energy (panel (e)), separated by X-points where the magnetic field lines tear and reconnect. The incoming particles are accelerated by the reconnection electric field at the X-points and, in the post-shock spectrum, they populate a broad distribution (red line in panel (g)),  extending to much higher energies than expected in thermal equilibrium (dotted line). The acceleration efficiency reaches nearly 100\%, and the slope of the non-thermal tail is $p\simeq1.5$ (dashed line in panel (g)), flatter than what the Fermi process normally gives in relativistic shocks. The particles are accelerated primarily by the reconnection electric field at the X-points, rather than by bouncing back and forth across the shock, as in the standard Fermi mechanism \cite{Sironi14}. Quite surprisingly, the Fermi process can still operate along the equatorial plane of the wind, where the stripes are quasi-symmetric. Here, the highest energy particles accelerated by the reconnection electric field can escape ahead of the shock, and be injected into a Fermi-like acceleration cycle. In the post-shock spectrum, they populate a power-law tail with slope $p\simeq2.5$, that extends beyond the flat component produced by reconnection (This additional population of Fermi-accelerated particles is not present in panel (g) of \ref{Fig:fluid}, since the figure focuses on the characteristic shock structure at intermediate latitudes away from  the wind midplane.). 
The particle energy spectra extracted from the simulations can be used directly to interpret the radiative signatures of PWNe. The radio spectrum of the Crab Nebula, the prototype of the class of PWNe, requires a population of non-thermal particles with a flat spectral slope ($p\simeq1.5$), extending at least across three decades in energy \cite{Bietenholz97}. The particle spectrum should be steeper at higher energies, with slope $p\simeq2.5$, to explain the optical and X-ray flux \cite{Mori04}. 
One could interpret the optical and X-ray signatures of the Crab, which require a particle spectrum with $p\simeq2.5$, as synchrotron emission from the particles that are Fermi-accelerated close to the  equatorial plane of the  wind. At face value, the spectral index required for the radio spectrum of the Crab ($p\simeq1.5$) could naturally result from the broad flat component of particles accelerated by the reconnection electric field. However, the particle spectrum in the simulations approaches the flat tail required by the observations only when the combination $\lambda/(R_g\sigma)$ exceeds a few tens (for smaller values, the spectrum is a narrow thermal-like distribution). Here, $R_g$ is the particle gyration radius taken in the pre-shock magnetic field. Most theoretical models of the Crab predict a value of $\lambda/(R_g\sigma)$ that is too small to produce a broad flat tail in the spectrum. If radio-emitting electrons are accelerated at the termination shock of pulsar winds via magnetic reconnection, a revision of the existing theories of pulsar magnetospheres is required \cite{Arons07}. In any case, first-principles PIC simulations provide physically-grounded inputs for models of non-thermal emission in PWNe.

\subsection{Magnetohydrodynamics}
Magnetohydrodynamics (MHD) numerical simulations are the domain of the investigation of large scales (scales comparable to the shock radius) instabilities. These instabilities are relevant for the confinement of the highest particle energies (see also the corresponding \S \ref{S:NRsim} in the non-relativistic shock case). MHD perturbations participate also to the dynamics of the global flow \cite{Meliani10}. Here, more specifically, we discuss the contribution of MHD instabilities to the production of magnetic field in relativistic shocks. This aspect is of particular importance to explain values of $\epsilon_B$ in GRB (see the discussion in \S \ref{S:Robs}). We first discuss MHD instabilities associated with the presence of CRs and then pure MHD instabilities.\\
\cite{Pelletier09} considered a 1 and 2D analytical analysis of MHD waves evolution in a coupled MHD and CR fluid system in the CR precursor of highly relativistic perpendicular shocks. The source of free energy is the charge of CR that stands ahead the shock front, hence this instability is a particular case of a streaming instability. In this configuration, the CR charge destabilizes upstream waves through the effect of the electromotive force. Alfv\'en waves have been found stable whereas magneto-sonic waves are destabilized. This solution raises an issue concerning the possibility to reach turbulence levels $\delta B \gg B_0$ ($B_0$ is the background magnetic field) in the upstream medium as magnetic-sonic instabilities are expected to saturate at a level $\delta B \sim B_0$ at least in the linear analysis. \cite{Casse13} performed 1D relativistic MHD-CR numerical simulations to test the previous analytical calculations. The authors confirmed the stability of Alfv\'en waves and found that the magneto-acoustic instability saturates at a moderate level of $\delta B/B_0 \sim 4-5$ in the non-linear regime. The density fluctuations ahead the shock front also saturate in the non-linear regime at a level of $\delta \rho/\rho_0 \sim 4-5$ ($\rho_0$ is the ambient upstream density). This effect can lead to the production of a series of shocks in the CR precursor that participate to a pre-heating of the medium and a pre-acceleration of the particles. The instability is also active in the mildly relativistic regime although the growth rate scaling as $(1-1/\gamma_{\rm sh})$ is reduced in that case.\\

In the downstream medium \cite{Inoue11} have investigated the amplification and the decay of MHD turbulence excited by the Richtmyer-Meshkov instability (a Rayleigh-Taylor-type instability) using 3D simulations. This instability is produced by an interaction between a highly relativistic shock front and ambient density fluctuations. The impulsive acceleration of the shock passage in a density clump induces the instability. High values for $\epsilon_B$ up to 0.1 have been found but they are dependent on the ambient magnetization. In some cases, the magnetic energy density can grow by at least two orders of magnitude compared with the magnetic energy just behind the shock independently of the magnetization. However in the context of gamma-ray bursts a supplementary mechanism (for instance the above streaming instability) is necessary for this fluid instability to amplify the magnetic fluctuations with $\epsilon_B$ as high as $0.1-10^{-2}$. The turbulence decays with time following a power-law at late stages. \cite{Mizuno14} investigated the instability over longer timescales using 3D axisymmetric simulations and were able to catch the saturation of the magnetic field. The saturation level happens to depend strongly on the orientation and the strength of the background magnetic field; the perpendicular configuration with a weak field produce larger enhancements. The saturated magnetic field energy in the postshock region is found to be in equipartition with the flow kinetic energy. The fluid instability may also trigger magnetic reconnection in striped pulsar winds (see \S \ref{S:Rpuls}). Hence turbulent motions induced by the Richtmeyer-Meshkov instability may help to dissipate the magnetic field in striped pulsar winds as been shown by 2D MHD simulations \cite{Takamoto12}.

\subsection{On the origin of ultra high energy cosmic rays}
\label{S:Ruhecr}
The physics of particle acceleration in powerful astrophysical sources is of course central to the problem of the origin of ultra-high energy cosmic rays (UHECRs). The near isotropy of the arrival directions of UHECRs, the difficulty of confining $10^{20}\,$eV nuclei in the Galactic magnetic field, the challenge of accelerating particles up to such extreme energies and the detection of a high energy cut-off at $6\times 10^{19}\,$eV \cite{Abbasi08, Abraham08, Sagawa11} -- the expected location for the Greisen-Zatsepin-Kuzmin cut-off \cite{Greisen66, Zatsepin66} -- all point towards rare and powerful extra-galactic sources.

Recent data have indicated the presence of weak anisotropies in the arrival directions of the highest energy cosmic rays, although the statistical confidence level is not yet conclusive \cite{Abreu10}. Such anisotropies are indeed expected if the nuclei are light, due to the small angular deflection in intervening magnetic fields. However, the most extensive dataset of the Pierre Auger Observatory also suggests that the composition tends to be dominated by heavier nuclei at energies $\gtrsim10^{19}\,$eV~\cite{Abraham10}. This result is still disputed by other experiments \cite{Abbasi10, Sagawa11}, and as of now, the situation remains unclear.

In direct inverse proportion to the extent of firmly established experimental results, the number of theoretical models is rather large. These models will not be reviewed in detail here due to lack of space, the reader is referred to \cite{Lemoine01, Kotera11} for reviews, or \cite{Berezinskii90, Norman95, Henri99, Pelletier01, Lemoine13c} for discussions on the acceleration issue. A direct application of the Hillas bound \cite{Hillas84} underlines neutron stars \cite{Gunn69, Venkatesan97, Blasi00, Rudak01, Arons03, Fang12}, gamma-ray bursts \cite{Milgrom95, Vietri95, Waxman95, Milgrom96, Dermer01, Dermer02, Gialis03, Gialis04, Rieger05}, powerful AGN \cite{Dermer10, Murase12} or radio-galaxies \cite{Takahara90, Rachen93, Ostrowski02, Casse05} and large scale structure accretion shock waves \cite{Ostrowski02b, Inoue07}. In order to go beyond the Hillas bound, one may assume that the accelerator is embedded in an outflow of bulk Lorentz $\Gamma$ and that in the comoving frame, acceleration proceeds at a fraction of the Bohm rate, meaning an acceleration timescale $t_{\rm acc}\sim {\cal A}t_{\rm L}$, with ${\cal A}\gtrsim 1$. Then, one can derive a lower bound on the magnetic luminosity of the source, assuming that acceleration indeed produces $10^{20}\,$eV particles~\cite{Norman95, Waxman05, Lyutikov07, Lemoine09}: $L_B\,\gtrsim\,10^{45}Z^{-2}{\cal A}^2\,$erg/s. This bound limits the number of possible sources to fast spinning neutron stars, gamma-ray bursts and the most powerful radio-galaxies (Faranoff-Riley II) for $10^{20}\,$eV protons. For iron-like nuclei, the number of possible sources extends significantly, down to moderate luminosity AGN and FR~I radio-galaxies (i.e. BL Lacs and TeV blazars when seen head-on), possibly large scale structure shock waves.

Among the possible acceleration mechanisms, shock acceleration plays a special role: it emerges as a generic and natural consequence of astrophysical outflows, it produces power laws with differential energy spectrum indices close to $-2$ and it allows to extract a significant fraction of the kinetic energy flux flowing into the shock into the non-thermal hadronic power law, as demonstrated repeatedly in this report. The latter point is of particular interest, because the UHECR energy input rate $\dot\epsilon\approx 0.5\times 10^{44}\,$erg/Mpc$^3$/yr above $10^{19}\,$eV \cite{Katz09}, which corresponds to a substantial fraction ($\gtrsim\,$a few \%) of the total source luminosity/output energy. We thus focus on shock acceleration in the rest of this discussion, and more particularly on relativistic shock acceleration, given that the acceleration in the non-relativistic limit is suppressed by $(c/v_{\rm sh})^2$ and that most candidate sources involve relativistic outflows.

In the ultra-relativistic limit, $\gamma_{\rm sh}\beta_{\rm sh}\,\gg\,1$, the physics of shock acceleration is known to depend sensitively on the degree of magnetization of the ambient medium (see \S \ref{S:IRS}). In ideal conditions, one understands that the relativistic Fermi process takes place provided intense turbulence on spatial scales shorter than a gyration radius has been excited in the shock precursor. PIC simulations have demonstrated that in the unmagnetized, or weakly magnetized limit, relativistic shock waves do excite the filamentation instability, which gives rise to intense turbulence with strength $\epsilon_B\sim 0.1$ at the shock front, on scales $\lambda_{\delta B} \sim 10\,c/\omega_{\rm pi}$ (see \S \ref{S:PARMS}). However, particle transport in short scale turbulence takes place at a slow rate, with typical scattering time $t_{\rm scatt}\,\sim\,r_{\rm g}^2/\lambda_{\delta B}c$ (see \S \ref {S:RThPart}). Consequently, Bohm scaling does not apply at high energies and the ratio of acceleration timescale to gyro-time ${\cal A}\,\sim\,r_{\rm g}/\lambda_{\delta B}$ becomes much larger than unity as energy increases. The maximal energy for nuclei accelerated at such ultra-relativistic shock waves is then the best of the following two estimates, depending on how transport operates upstream of the shock front: $E_{\rm max}\,\approx\,Z e B_0 R \gamma_{\rm sh}\sim 10^{16}\,Z\,B_{0,-6}R_{17}\gamma_{\rm sh,2.5}\,$eV, corresponding to rotation in the background magnetic field $B_0$; or $E_{\rm max}\,\approx\,Z \gamma_{\rm sh} \epsilon_B^{1/2}(R/\lambda)^{1/2}m_pc^2\,\sim\,4\times 10^{15}\,Z\,R_{17}^{1/2}\gamma_{\rm sh,2.5}n_0^{1/4}\,$eV, corresponding to small angle deflection in the self-generated turbulence~\cite{Achterberg01, Bykov12b, Plotnikov13, Reville14}. The fiducial values chosen are representative of the external shocks of gamma-ray bursts.

Ultra-relativistic external shocks of gamma-ray bursts cannot push particles up to $10^{20}\,$eV, mostly because of the low magnetization of the circumburst medium. At large magnetizations however, acceleration is likely suppressed, at least in ideal conditions. The termination shock of the Crab pulsar apparently violates this rule, since it appears to accelerate electrons up to PeV energies at a Bohm rate. The spectral energy distribution of the Crab nebula at frequencies above the ultra-violet range indeed corresponds to an electron injection spectrum $\propto E^{-2.2}$, in remarkable agreement with the predictions of test particle Fermi acceleration with isotropic scattering \cite{Bednarz98, Kirk00, Achterberg01, Lemoine03, Keshet05}. What triggers Fermi acceleration at this termination shock, whose Lorentz factor $\gamma_{\rm sh}\sim 10^3 -10^6$ and magnetization $\sigma\sim 10^{-3}-10^{-2}$~\cite{Kennel84} -- see also the review~\cite{Kirk09} -- remains the subject of debate. Nevertheless, assuming that nuclei are injected along with the pairs in the pulsar wind, one could expect such nuclei to be accelerated up to the confinement energy in the nebula, $E\,\sim\, Z e B R\,\sim\, 3\times10^{17}\,Z\,$eV for an estimated magnetic field $B\sim 300\,\mu$G and nebula size $R\sim 1\,$pc. This remains insufficient and, actually, the energy output of Crab-like pulsars is too small to explain the UHECR flux. If, however, a fraction of neutron stars are born with high angular velocity, corresponding to $1-10\,$ms periods, their rotational energy reservoir could be sufficient to account for the observed flux, provided the ions tap a sizable fraction of it~\cite{Venkatesan97}; furthermore, ion acceleration at the termination shock could operate up to confinement energies of order $\sim 10^{19}-10^{20}\,$eV for protons in such young nebulae \cite{Lemoine14c}.

Most scenarios of UHECR acceleration have focused on mildly relativistic shock waves, with $\gamma_{\rm sh}\beta_{\rm sh}\sim 1$, e.g. in gamma-ray burst internal shocks \cite{Waxman95, Gialis03, Gialis04}, at the reverse shock \cite{Waxman01, Dermer10}, in blazar internal shocks \cite{Dermer09}, in radio-galaxy outflows \cite{Rachen93, Casse05} and more recently at the external shock of trans-relativistic supernovae \cite{Wang07, Budnik08, Liu11b, Chakraborti11, Liu12}, which are generally associated with low luminosity gamma-ray bursts. The latter form an interesting class of objects, because their high occurence rate, compared to the long gamma-ray bursts, makes it easier to accomodate the cosmic ray flux above $10^{19}\,$eV. However, their restricted acceleration capabilities imply that only heavy nuclei (oxygen to iron) could be produced at ultra-high energies; acceleration of protons should stop short of $10^{19}\,$eV; see also the discussion in \S.~\ref{S:Trel} for an estimate of the magnetic field and radius of the blast wave in those objects.

A common and crucial assumption in studies in this field of research is that acceleration proceeds according to a Bohm scaling in the self-amplified field, i.e. ${\cal A}\,\sim\,1$. Whether this applies remains to be demonstrated by PIC simulations, which so far indicate relatively inefficient acceleration at mildly relativistic shocks \cite{Sironi13}. There are reasons however to be optimistic \cite{Lemoine12, Plotnikov13}; in particular, the precursor of mildly relativistic shock waves should extend on long spatial scales (as compared to ultra-relativistic shock waves), and thus possibly allow the development of new instabilities. The investigation of such issues may require the development of new numerical tools beyond PIC simulations in order to probe the physics of the shock on long spatial and temporal scales.

\section{Laboratory experiments in connection to astrophysics}
\label{S:LS}
\subsection{Introduction}
Advances in the field of high energy-density physics (HEDP) have marched in parallel with recent progress in astrophysical plasma theory discussed in this review. The increase in both the experimental capability and availability of modern high-power plasma devices, has been successfully exploited to tackle problems of astrophysical relevance, an area commonly referred to as {\it laboratory astrophysics}. While this field has been around for many years, see for example the many excellent reviews \cite{SagdeevPodgorny, Remington, Lebedev, Zakharov, Savinetal, ZweibelYamada}, its over-lap with the interests of the high-energy astrophysics community has seen particular rapid growth. Laboratory experiments continue to provide unique insight into several the physical mechanisms discussed in this review.

Many fields of experimental physics, such as atomic, nuclear physics, high-energy particle physics, etc.  are directly applicable to astrophysical processes. However, much of what is done in the field of laboratory astrophysics, involves the {\it scaling} of laboratory experiments to astrophysical phenomena. This relies on the similarity of physical processes occurring on vastly different scales. For example, in the case of a plasma jet produced using a Z-pinch device, a typical jet structure might have a spatial extent of several centimeters \cite{Suzuki}, as compared with a Herbig-Haro jet from a young stellar object, which are observed to reach a fraction of a parsec ($\approx 3\times10^{18}$~cm) or more in length \cite{Reipurth}. The fact that any similarity between these two systems, operating on such different scales can be considered at all, may seem quite surprising to some, but a simple comparison of the equations that govern their dynamics shows that, subject to certain conditions, this can be achieved. If both systems are considered from a hydrodynamical viewpoint, i.e. neglecting magnetic fields for the moment, then writing these equations in dimensionless form, it follows that the two systems differ only in their numerical values for the dimensionless quantities, the Reynolds number: $$\frac{1}{Re} = \frac{\nu}{U\ell}$$ and the P\'eclet number: $$\frac{1}{Pe} = \frac{\chi}{U\ell}$$ where $U$ and $\ell$ are characteristic velocity and length scales of the system, $\nu$ the kinematic viscosity, and $\chi$ the thermal diffusivity \cite{Zeldovich}. If both these quantities are large compared to unity $Pe, Re \gg 1$, in the two systems being compared, then they are said to be hydro-dynamically similar \cite{Ryutov, Remington, Cross}, and a simple transformation can be used to relate the dynamical variables.

The condition, $Pe, Re \gg 1$, is easily achieved in astrophysical flows, due to the low collision rate and large length scales involved. It is also possible to satisfy this condition in the laboratory using laser produced plasmas or pinch devices. For example, typical Reynolds numbers in a laser ablated plasma, using 100 Joules of laser energy on a metal target, can be of the order $10^4$ or larger. It should be noted that, neglecting other effects this is comparable, or superior, to current state of the art numerical simulations. Hydrodynamic similarity has been demonstrated in a number of experiments, looking at the early stages of type II supernovae \cite{Kane, Remington}.

Achieving magneto-hydrodynamic similarity, from laboratory to astrophysical scales, often proves to be more challenging. With the inclusion of magnetic fields, similarity requires in addition, that the magnetic Reynolds number, $Rm$, be considerably larger than unity $$\frac{1}{Rm} =  \frac{\eta}{U\ell}$$ where $\eta$ is the magnetic diffusivity. Again, this number is typically extremely large in astrophysical systems, but  achieving very large magnetic Reynolds numbers in laser/Z-pinch plasmas is more challenging, with typical values ranging from less than unity, to a few tens. (We note that considerably higher magnetic Reynolds numbers can be achieved in magnetic confinement fusion devices, however, since the focus is on shocks, we do not discuss these further.) The issue of achieving large magnetic Reynolds numbers, is closely related to the production of a collisionless plasma, since $\eta \sim (c/\omega_{\rm pe})^2/\tau_e$, where $\tau_e$ is the Coulomb collision time for electrons. Collisionless plasma physics has been central to much of this review, and in particular, the physics of collisionless shocks. The ability to reproduce collisionless shocks under controlled laboratory conditions is highly desirable, and provides a novel platform for the study of collective plasma effects, and the interplay between microscopic and macroscopic plasma processes. 

Many aspects of such shocks remain to be fully understood, and as discussed in this review, it continues to be an active field of study. While theory, satellite observations and numerical simulations have proved invaluable, the advantages of laboratory plasma experiments  are self evident, being unavoidably multi-dimensional, and multi-scale. In the following, we give a flavour of some of the experiments that are pertinent to the review. This is by no means an exhaustive review of the field. We hope, nevertheless, that it gives an introduction to some experiments of interest to non-experts in the high energy astrophysics community. 

\subsection{Collisionless Shock Experiments}
The first experiments investigating collisionless shocks in laboratory plasmas dates back to sixties, using electromagnetic shock-tube or pinch devices, and are reviewed in \cite{SagdeevPodgorny, Eselevich}.  While these experiments provided valuable information, their relevance to astrophysical shocks were limited \cite{SagdeevPodgorny, Drake}. Recent years, have seen the use of high-power lasers, become the primary field for such investigations. This may be largely attributed to the increase in academic access to high-power laser facilities, such as the LULI laboratory at \'Ecole Polytechnique, France, the Vulcan laser at Rutherford Appleton Laboratories in Oxfordshire, the Jupiter Facility at Lawrence Livermore National Laboratory or the GEKKO XII laser at Osaka University, Japan. All of these facilities are capable of providing a total energy of a few hundred to a few thousand Joules of energy onto a sub mm target, on nanosecond or sub-nanosecond timescales. Larger facilities, such as Omega ($30$ kJ) or the National Ignition Facility ($\gtrsim 1$~MJ) also provide limited academic access, with astrophysically relevant investigations making up a sizeable fraction of such experiments. There are also continuing efforts to exploit pinch machines, as is being done on MAGPIE \cite{Lebedev}, as well and large plasma devices (LAPD) \cite{Schaeffer}, and plasma gun devices \cite{Merritt}.

Collisionless shock experiments using laser produced plasmas has itself a long history. In \cite{Antonov}, a collisionless shock was produced by irradiating a solid target mounted inside a low-density plasma filled pinch device. In \cite{BellChoi}, the supersonic flow of a laser ablated plasma past a stationary obstacle was investigated. Both these experiments showed good agreement between numerical simulations and experimental results. These pioneering experiments, largely overlooked for more than a decade, are at last being developed further, exploiting the increased availability of laser power and improved diagnostics, provided at the above mentioned facilities. Several recent experiments \cite{Romagnani08, Ahmed, Kuramitsuetal} have detailed the formation of collisionless electrostatic shocks (see \S \ref{S:Rsim}), using a laser driven ablation flow in a low density gas filled target chamber. An impressive suite of diagnostics provide detailed information on the shock evolution. Proton imaging \cite{BorghesiPP, KuglandPP} is used to image the electromagnetic field structure, and magnetic induction probes can provide high time resolution measurements of magnetic fields at a fixed location \cite{Gregori}. At sufficiently high densities, interferometry can be used to measure density, while Schlieren imaging (a standard optical technique used in laboratory plasmas to measure variations in the refractive index of the gas). and streaked optical pyrometry (SOP) can be used to track the shock motion. In addition, spectrometers can be used to measure the electron temperature. A recent experiment \cite{Niemann} has coupled a high power laser with the LAPD device, which contains a low density, magnetised plasma. Exploiting the high reproducibility of the system, this group has provided convincing evidence of a collisionless magnetosonic shock. Although the Alf\'enic Mach numbers of these shocks remain quite modest, they certainly offer an exciting avenue for astrophysically relevant investigations. The successful generation of a supercritical collisionless shock opens the possibililty for laboratory investigations of diffusive shock acceleration \cite{RevilleNJP}, particularly with application to supernova studies. 

Experiments addressing the formation of unmagnetized shocks have also been performed. These experimental set-ups for such investigations typically involve the interpenetration of two ablation flows in vacuum. The first such investigations were performed by \cite{Woolseyetal, Courtois}. Two parallel foils were irradiated with 60J of laser energy, generating two oppositely directed ablation flows, capturing the early stages of shock formation. Schlieren images showed the formation of small scale features, indicating the early stages of shock formation. The inclusion of an external magnetic field, was also demonstrated to introduce strong sub-Larmor scale structure, again indicative of shock formation. Variations on this method have also shown interesting results. \cite{Kuramitsuetal} use a similar set-up, of two oppositely facing foils, and irradiate one of them. The resulting fast-moving ablated plasma then \lq reflects\rq\, off the opposite foil and a feature is observed to propagate in the flow. 

More recent experiments have taken advantage of the sizeable increase in laser power, as compared with these early investigations \cite{Parketal, Kugland, Fox}. Using a similar double foil set-up \cite{Woolseyetal} on the OMEGA and OMEGA-EP lasers, irradiating the foils with more than a kJ of laser energy, the interpenetration of two ablation flows was studied. A Thomson scattering probe was used to determine the plasma conditions in the interaction region, which clearly show the formation of a hot ($>$keV), dense ($\sim10^{18} ~{\rm cm}^{-3}$) plasma \cite{Rossetal}. The proton radiography images \cite{Kugland}, show the development of non-linear structures in the interaction region, however, convincing evidence for shock formation has not been found. The Weibel instability, operating between the two interpenetrating plasma has been suggested as the dominant mechanism, in generating the features observed in the proton images \cite{Fox, Huntington}. These experiments are ongoing. The successful generation of a collisionless shock via this avenue is eagerly anticipated by the whole community. It has been suggested that NIF scale laser is required to produce a fully formed unmagnetized collisionless shock \cite{Sakawa14}. 

Until then, there is much that can be studied with magnetized shocks, or indeed weakly collisional shocks. Recent work by \cite{Meineckeetal}, have demonstrated the mechanism of magnetic field amplification by shock interaction with a clumpy medium, one of the mechanisms suggested for producing the strong fields at the outer shocks of young supernova remnants. We are also on the verge of generating neutral electron-positron pair plasmas in the laboratory. The generation of a neutral pair beam has already been reported \cite{Sarrietal}, (see also \cite{Chen}) while the next generation of high intensity lasers, ELI \cite{Mourou}, have already been suggested as potential sources of copious pair production \cite{BellKirk, Ridgers}. Advances such as these will open many opportunities to study such plasmas, which to date are exclusively found in astrophysical sources. 

\section{Summary and conclusions}
\label{S:Sum}

This review provides the reader with an instantaneous view of the field of collisionless shock microphysics. We first considered the case of non-relativistic shocks. These are well probed by in-situ satellite measurements in the environment of magnetospheric and heliospheric shock waves. We discussed several important aspects which control the shock phenomenon at the smallest scales: shock formation, shock non-stationarity and their associated instabilities. The non-stationarity has basically two different origins: self-reformation and emission of non-linear waves that do not include any micro-instabilities and non-stationary effects connected with micro-instabilities. In natural plasmas both are inter-connected and can be explored with the help of particle-in-cell and hybrid numerical techniques. Among the large number of micro-instabilities the Weibel instability is likely the dominant instability which mediates the shock formation in an unmagnetized background. We also detailed the non-resonant streaming instability relevant for astrophysical shocks triggered by a population of energetic particles (that may become cosmic rays). This instability (both linear and non-linear saturation stages) has been recently widely investigated by different numerical techniques including hybrid and PIC-MHD simulations. We discussed three types of shock acceleration mechanisms: shock drift, shock surfing and diffusive shock acceleration. Especially SDA has been shown to be potentially active in the process injection of moderately energetic particles at astrophysical shock fronts. We gave an overview of the multi-wavelength (radio, optical, X-rays and gamma-rays) observational probes of particle acceleration in supernova remnant shocks, the favorite sources of galactic cosmic rays. A particular emphasize has been made on the recent high-angular resolution structures observed by the X-ray satellite Chandra: X-ray filaments and X-ray stripes. These structures can be interpreted as evidences of turbulent motions and magnetic field amplification. MFA phenomenon is thought to be closely connected to the development of streaming instabilities even if other ways to amplify the magnetic field exist without invoking any particle acceleration. If substantial progresses have been made in the theory of cosmic ray acceleration at supernova remnant shocks yet no definite proof exists that SNR are the sources of the galactic CRs. Turning to relativistic shocks, we first gave several observational hints that MFA is also occurring there especially in shocks associated with the gamma-ray bursts phenomenon. We provided details on the transport of energetic particles in the turbulence that develops in relativistic flows. This turbulence is likely different from the NR case because it is anisotropic and restricted to scales smaller than the particle Larmor radius; it is non-resonant. This produces strong constraints on the maximum energies we may expect from acceleration in R shocks and question the sources of extragalactic CRs. Relativistic and mildly-relativistic shocks have been widely investigated recently in both electron-positron and electron-proton plasma configurations using PIC simulations. The results point out the importance of the local magnetization and magnetic field obliquity for particle acceleration efficiency. In particular high-magnetization shocks do not permit an efficient Fermi acceleration and alternative scenarios are required to explain the radiation detected from pulsars and their nebul\ae. Finally R-MHD simulations are starting to complement the global picture on the magnetic field expected in R shocks. The last chapter of this review discussed the importance of the developments realized in laboratory astrophysics which have investigated the conditions the shock physics can be scaled down to laboratory scales.  

\section{Acknowledgments}
This review is issued from a working group hosted by the international space science institute (ISSI) in Bern Switzerland. The authors acknowledge supports by ISSI. A.M. acknowledges the support by the french ANR MACH project. A.C.B. acknowledges the support by Grant ENE2013-45661-C2-1-P from the Ministerio de Educacion y Ciencia, Spain and Grant PEII-2014-008-P from the Junta de Comunidades de Castilla-La Mancha, Spain.

\end{document}